\documentclass[useAMS,usenatbib]{mn2e}
\usepackage{graphicx,color}
\usepackage{MnSymbol,gensymb}
\usepackage{journal_names}
\usepackage{fixltx2e}
\usepackage{lscape}
\usepackage[compatibility=false]{caption}
\usepackage{subfig}
\usepackage{threeparttable}
\title[Millimetre southern SFR survey]{Grain growth signatures in the protoplanetary discs of Chamaeleon and Lupus}
\author[Ubach et al.]
{C.~Ubach$^{1,5}$\thanks{E-mail: cubach@astro.swin.edu.au},
S.~T.~Maddison$^{1}$, C.~M.~Wright$^{2}$, D.~J.~Wilner$^{3}$, D.~J.~P.~Lommen$^{4}$,
\newauthor B.~Koribalski$^{5}$\\
$^{1}$Centre for Astrophysics \& Supercomputing, Swinburne University of Technology, H39, PO Box 218, Hawthorn, VIC 3122, Australia\\
$^{2}$School of Physical, Environmental and Mathematical Sciences, UNSW@ADFA, Canberra ACT 2600, Australia\\
$^{3}$Harvard-Smithsonian Center for Astrophysics, 60 Garden Street, 02138 Cambridge, MA, USA\\
$^{4}$Raffles Institute, 1 Raffles Institution Lane, Singapore, 575954 \\
$^{5}$CSIRO Astronomy \& Space Science, Australia Telescope National Facility (ATNF), PO Box 76, Epping NSW 1710, Australia\\
}

\begin{document}


\pagerange{\pageref{firstpage}--\pageref{lastpage}} \pubyear{2012}
\label{lastpage}

\maketitle

\label{firstpage}

\begin{abstract}

{{We present ATCA results of a 3 and 7~mm continuum survey of 20 T Tauri stars in the Chamaeleon and Lupus star forming regions. This survey aims to identify protoplanetary discs with signs of grain growth.
We detected 90\% of the sources at 3 and 7~mm, and determined the spectral slopes, dust opacity indices and dust disc masses.
We also present temporal monitoring results of a small sub-set of sources at 7, 15~mm and 3+6~cm to investigate grain growth to cm sizes and constrain emission mechanisms in these sources.
Additionally, we investigated the potential correlation between grain growth signatures in the infrared (10~$\mu$m silicate feature) and millimetre (1--3~mm spectral slope, $\alpha$).}}

{{Eleven sources at 3 and 7~mm have dominant thermal dust emission up to 7~mm, with 7 of these having a 1--3~mm dust opacity index less than unity, suggesting grain growth up to at least mm sizes.
The Chamaeleon sources observed at 15~mm and beyond show the presence of excess emission from an ionised wind and/or chromospheric emission.}}
Long-timescale monitoring at 7~mm indicated that cm-sized pebbles are present in at least four sources.
Short-timescale monitoring at 15~mm suggests the excess emission is from thermal free-free emission.
Finally, a weak correlation was found between the strength of the 10~$\mu$m feature and $\alpha$, suggesting simultaneous dust evolution of the inner and outer parts of the disc.
{{This survey shows that grain growth up to cm-sized pebbles and the presence of excess emission at 15~mm and beyond are common in these systems, and that temporal monitoring is required to disentangle these emission mechanisms.}}

\end{abstract}



\section{Introduction}
\label{sec-introduction}

\begin{figure}
\centering
	\includegraphics[width=.97\columnwidth]{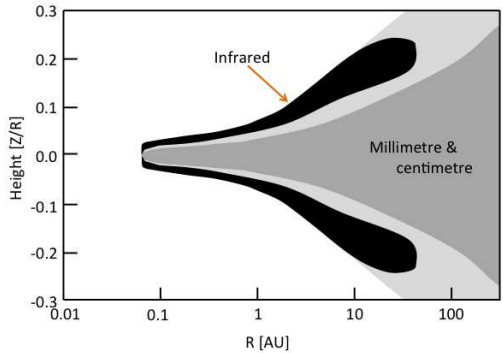}
	\caption{{{Schematic of a protoplanetary disc, adapted from \citet{2010ARA&A..48..205D} and \citet{2012A&A...539A...9M}. The grey area shows the gas within one (dark) and two (light) pressure scale heights. The black areas show the approximate region of the infrared emission. The millimetre and centimetre emission, tracing the larger dust grains, originates mainly from the mid-plane of the outer disc. Note that the origin of the observed infrared emission strongly depends on the orientation of the disc.}}}
	\label{fig-ppdPictogram}
\end{figure}

Observational signatures of grain growth {{in protoplanetary discs}} can be obtained from both the infrared (IR) and millimetre (mm) bands, which probe different regions of the disc and are sensitive to different grain sizes. IR emission comes from the warm inner {{($\sim$1--5~AU)}} and upper layers of the disc, while the mm emission comes from the cooler outer {{disc}} regions ${{(>10~\rm{AU})}}$ and mid-plane where the bulk of the dust resides -- see Fig.~\ref{fig-ppdPictogram}. The smallest submicron-sized grains are strongly coupled to the gas, but as they grow they begin to decouple from the gas and settle to the mid-plane, resulting in grain size sorting \citep{Chiang01, D&D04}. Since thermal dust emission is dominated by dust grains whose size is similar to the observing wavelength \citep{draine06}, multi-wavelength observations are therefore needed to understand the distribution of the dust phase in protoplanetary discs.

The strength and shape of the 10~$\mu$m silicate feature can be used as a grain growth indicator.
\citet{Przy} found that low mass T Tauri stars with submicron-sized grains typically have a strong triangular shaped 10~$\mu$m feature, while discs with micron-sized grains typically have weaker and broader 10~$\mu$m features. Similar results were found by \citet{Boekel} for their survey of intermediate mass Herbig Ae/Be stars. {{An equivalent effect was suggested to be caused by crystallisation \citep[e.g.,][]{2003ApJ...585L..59H,2006ApJ...639..275K}. \citet{2006ApJ...639..275K} found an increase in flux near 11.3~$\mu$m can be caused by polycyclic aromatic hydrocarbon and/or crystalline forsterite, which can mislead the grain growth interpretation.}}

The 1-3~mm spectral slope{{, $\alpha_{\rm{1-3mm}}$ (hereafter $\alpha$),}} where the flux F $\varpropto \nu^{\alpha}$ and $\nu$ is the frequency, can be used to estimate the dust opacity index $\beta$ where {{dust opacity}} $\kappa_{\nu} \varpropto \nu^{\beta}$. Assuming the emission is optically thin, the dust opacity index can be written as $\beta \sim \alpha-2$ and an opacity index $\beta \approx 1$ indicates grain growth up to mm sizes \citep{draine06}.
This approach has been used to detect mm-sized grains in $\sim$50 discs to date \citep{Natta04,Andrews05,Lommen07,Lommen09,Lommen10,Ricci10a,Ricci10b}.
However, care must be taken when relating $\alpha$ and $\beta$, as shallow spectral slopes can result from an extended optically thin disc or a compact optically thick disc \citep{Beckwith1991}. To break this degeneracy the discs have to be spatially resolved at mm wavelengths.
Recently \citet{2012A&A...540A...6R} determined that although only a small fraction of optically thick material is required to create a shallow spectral slope in the disc where the maximum grain size {{$\ll$\,0.1~mm,}} the required dust {{overdensities}} are much larger than the physical processes which concentrate dust (e.g., streaming instabilities, gravitational instabilities) {{can achieve}}. This strengthens the {{notion}} that shallow spectral slopes at long wavelengths are due to large mm/cm-sized grains in the outer parts of the disc.

By extending the dust opacity index relationship to 7~mm and beyond, it is possible to determine if grains up to cm sizes are present in discs \citep{2003A&A...403..323T,2005ApJ...626L.109W,Rod06,Lommen09}. However for wavelengths of 7~mm and longer, the emission can result from a range of physical processes in young stellar objects, including thermal emission from dust, free-free emission from an ionised wind, and non-thermal chromospheric emission \citep[e.g.,][]{2007prpl.conf..555D,2007prpl.conf..539M}.  While chromospheric emission will be unresolved without the use of very long baseline interferometry, both thermal dust emission from the disc and free-free emission from a wind may be extended, and so simply resolving the emission is not enough to determine the dominant emission process. 

{{The spectral slope from 7~mm and beyond can be used to break this degeneracy. Depending on optical depth, two spectral slopes are known for free-free emission from spherically symmetric, constant velocity, ionised wind. For an opaque wind the spectral slope, $\alpha_{\rm{ff}} = 0.6$ \citep{1975A&A....39....1P}, while for optically thin wind, $\alpha_{\rm{ff}} = -0.1$~\citep{1967ApJ...150..807M}.}} Using these {{definitions, \citet{Rod06}}} used the cm spectral slope from 2 to 3.6~cm {{of four T Tauri stars}} to determine that $\sim$20\% of the 7~mm flux was due to free-free emission. 

Another way to determine the emission mechanisms present in a disc is by temporal monitoring {{of flux}} variability at 7 and 15~mm. Thermal dust emission is expected to be constant over time, while thermal free-free emission from an ionised wind is expected to vary by $\sim$20--40$\%$ on a timescale of years \citep{2002ApJ...580..459G,2007ApJ...657..916L}. Non-thermal emission varies over a timescale of minutes to hours by an order of magnitude or more \citep{Kutner86, Chiang96}. 

The results from IR and mm surveys have been used to study a potential correlation between the 10~$\mu$m silicate feature and the millimetre spectral slope \citep{Lommen07, Lommen10, Ricci10b}. \citet{Lommen07,Lommen10} suggested a tentative correlation between the strength and shape of the 10~$\mu$m silicate feature and $\alpha$ for a sample of Chamaeleon, Lupus, and Taurus sources exists. 
A correlation between these grain growth signatures would suggest simultaneous dust evolution of the inner and outer parts of the disc. The submicron-sized grains would no longer be replenished as grains grow to mm to cm sizes, which flattens the 10~$\mu$m silicate feature, while {{the}} simultaneous growth of larger particles in the midplane leads to a shallower mm spectral slope \citep{Lommen07, Lommen10}. 
However, \citet{Ricci10b} found no such correlation in their sample of $\rho$ Ophiuchus and Taurus sources, and suggested that the small sample size and some inconsistency between the spectral slopes of specific {{sources could}} have lead to the discrepancy between these two results. 

In this {{work we}} identify discs with signs of grain growth from observations at 3, 7, 15~mm, 3+6~cm {{of}} 20 T Tauri stars in the Chamaeleon and Lupus star forming regions conducted using the Australia Telescope Compact Array (ATCA)\footnote{The Australia Telescope Compact Array is operated by the Australia Telescope National Facility (ATNF) which is a division of CSIRO.}, extending the 3 and 7~mm work of \citet{Lommen07,Lommen10}. {An introduction to the survey and the data reduction process can be found in Section~\ref{sec-ob}{{, with the}} results of the {{survey}} presented in Section~\ref{sec-results}. In Section~\ref{sec-discussion} the}
~dust opacity indices, the dust disc masses, and dominant emission mechanism at the longer wavelengths are determined, and the potential correlation between the IR and mm grain growth signatures is explored, {with the conclusions presented in Section~\ref{sec-summary}.}


\section{Observations and data reduction}
\label{sec-ob}

\begin{table*}
\caption{List of sources observed with ATCA {{in this survey}}.}  
 \begin{tabular}{llllcccccc}
	 \hline \hline
	& Source	&	RA	&	DEC	&	Spectral Type	&	Cloud	&  Distances & T$_{\rm eff}$ &	Wavelengths & Comments	\\
	&	&	(J2000)	&	(J2000)	&		&		& (pc)  &	 (K) & (mm) & \\
	\hline \hline									
	 \multicolumn{9}{c}{Chamaeleon}											\\
	\hline	
	1 & SY Cha		&		10 56 30.5		&		-77 11 39.3		&		M0.5		&		Cha~I		&	160	&	3778	&	7	&			\\										
	2 & CR Cha		&		10 59 06.9		&		-77 01 39.7		&		K0--K2		&		Cha~I 	&	160	&	4900	&	7,15	&			\\
	3 & CS Cha		&		11 02 24.9		&		-77 33 35.9		&		K4		&		Cha~I		&	160	&	4205	&		3,7,15 	& Binary$^a$		\\
	4 & DI Cha		&		11 07 21.6 		&		-77 38 12.0		&		{{G1--G2	}}	&		Cha~I		&	160	&	5860	&		3,7,15 	&	Binary$^a$	\\
	5 & T Cha		&		11 57 13.6 		&		-79 21 31.7		&		G2		&		Isolated		&	 100	 	&	5600	&	 7,15,30,60 	&		\\
	6 & Glass I		&		11 08 15.1		&		-77 33 59.0		&	         K4		&		Cha~I		&		160	&	5630	&	 3,7 	&	Binary$^a$	\\
	7 & SZ Cha		&		10 58 16.7		&		-77 17 17.1		&		K0 G		&		Cha~I		&		160 	&	5250	&	7	&		\\
	8 & Sz 32		&		11 09 53.4		&		-76 34 25.5		&		K4.7		&		Cha~I		&		160 	&	4350	&	 3,7,15,30,60		&	HH jet$^a$ \\
	9 & DK Cha		&		12 53 17.2 		&		-77 07 10.7		&		F0 D		&		Cha~II		&		178 	&	7200	&	 3,7,15,30,60 	&	Outflow$^a$	\\	
	\hline							
	 \multicolumn{9}{c}{Lupus}											\\
	\hline	
	10 & IK Lup       	&		15 39 27.8         	&	      -34 46 17.2 		&	       K7 D 	&	      Lupus  1    	&	150	&	3802	&	7	&	Binary$^{b}$\\								
	11 & Sz 66		&		15 39 28.3		&		-34 46 18.0		&		M2 D 	&		Lupus 1		&		150 	&	3350	&	7	&	Binary$^{b}$	\\
	12 & HT Lup		&		15 45 12.9		&		-34 17 30.8		&		K2		&		Lupus 1		&		150 	&	4898	&	7	&	Triple system, HH jet$^b$	\\
	13 & GW Lup		&		15 46 44.7		&		-34 30 36.0		&		M2		&	Lupus 1		&		150 &	3499	&	7	&		\\
	14 & GQ Lup	&		15 49 12.1		&		-35 39 03.9		&		K7V D		&		Lupus 1		&		150 	&	3899	&	  3,7 	&	 Planet$^{b}$	\\
	15 & RY Lup		&		15 59 28.4		&		-40 21 51.2		&		G0V: C 	&		Lupus 3		&	200 	&	4592	&	7	&		\\
	16 & HK Lup		&		16 08 22.5		&		-39 04 46.3		&	    K5	&		Lupus 3	&		200 	&	4350	&	7	&		\\
	17 & Sz 111	&		16 08 54.7		&		-39 37 43.1		&		M1.5	&		Lupus 3		&		200 &	3573	&	7	&	 Cold disc	\\
	18 & EX Lup		&	16 03 05.5	&		-40 18 25.3		&		M0		&		Lupus 3		&		200 	&	3802	&	  3,7 	&	\\
	19 & MY Lup		&		16 00 44.6		&		-41 55 29.6		&				&		Lupus 4		&		165 	&	5248	&	7	&		\\
	20 & RXJ1615.3-3255	&		16 15 20.2 		&		-32 55 05.0 		&		K5		&		Isolated		&		184 	&	4590	&	7	&	\\
	\hline	
 \end{tabular}
   	\begin{tablenotes} 
	   	\item[1] $^{a}$ See Section~\ref{app-sub-cham} for further details {{on Chamaeleon sources}}.
		\item[2] $^{b}$ See Section~\ref{app-sub-lup} for further details {{on Lupus sources}}. 
		\item[5] References: Chamaeleon: \citet{1991AJ....101.1013H, 1997A&A...327.1194W, 1998A&A...330..145V, 2004ApJ...602..816L, 2007ApJ...664L.107B, 2008ApJ...680.1295S}. Lupus: \citet{2008hsf2.book..295C, 2007ApJ...658..480M, 2011MNRAS.418.1194M, 2010A&A...518L.128V, 1997ApJ...478..295C, 1998A&A...332..849F}		
	\end{tablenotes}
 \label{tab-sources}
 \end{table*}	
\subsection{Sample}
\label{sub-sample}

	This survey targets 20 T Tauri stars in the Chamaeleon and Lupus star forming regions at a range of wavelengths {{presented}} in Table~\ref{tab-sources}. 
	The primary aim was to obtain 3 and 7~mm fluxes for all sources. The sources were selected to overlap with the sample observed by \citet{Lommen07,Lommen10} and the Spitzer c2d programme \citep{2005ApJ...628..283Y,2008ApJ...676..427A,2008ApJS..177..551M}, with additional sources {{with strong}} 1.3~mm fluxes from \citet{H1993} and \citet{Dai10} to  {{improve}} the chance of detection at 3~mm. The sample {includes 7 Chamaeleon I}
~cloud sources, the isolated Chamaeleon source T Cha, the Chamaeleon II source DK Cha, 10 sources from the Lupus 1, 3, and 4 clouds, and the isolated Lupus source RXJ1615.3-3255. Note {{that}} Sz 32 and WW Cha were observed simultaneously at 7~mm and beyond, where Sz 32 was the primary target and WW Cha was in the field of view of the observations. 

A subsets of Chamaeleon sources with the highest expected centimetre fluxes were observed at 15 mm {{(6 sources)}} and 3+6 cm {{(3 sources)}} in order to {{determine}} the emission mechanisms. {{The expected cm fluxes were estimated}} by extrapolating the {{spectral slope from 3 to 7~mm, $\alpha_{3-7}$, using the 3 and 7~mm fluxes, where {{$\alpha =$~d(log F$_{\nu}$)/d(log$\nu)$}}. No Lupus sources were available during the scheduled observing time for long wavelength observations.

\subsection{Observations and data calibration}
\label{sub-ob-sample}

	{{From}} May 2009 to August 2011 we performed continuum observations with ATCA in the mm and cm bands.  ATCA has an east-west track and northern spur with five movable antennas with a wavelength range from 3~mm to 30~cm, and one fixed antenna (CA06) at 3~km on the eastern end of the track with no 3~mm receiver\footnote{Additional information about ATCA can be found at \small{http://www.narrabri.atnf.csiro.au/observing/\\users\_guide/html/atug.html}}. These observations used ATCA in compact hybrid configurations -- observations details and specific array configurations can be found in Table~\ref{tab-obdetails}. {{{Compact}} hybrid configurations were chosen to ensure good u-v plane coverage and {{hence maximum}} detection sensitivity in relatively short integration times.}

	These observations used the new Compact Array Broadband Backend (CABB)\footnote{For more information about CABB see: http://www.narrabri.atnf.csiro.au/observing/CABB.html} which has a maximum bandwidth of 2~GHz (a factor 16 improvement), higher level of data sampling, and improved continuum sensitivity by a factor of four from the previous backend {{\citep{CABB}}}. For these {{continuum}} observations we used {{the 2048 channels of 1~MHz width}}. {{The}} observations were conducted in dual side band mode, with frequency pairs centred at 93$+$95~GHz (3~mm band), 43$+$45~GHz (7~mm band), 17$+$19~GHz (15~mm band), and simultaneous observations at 9.9+5.5~GHz (3 and 6~cm bands respectively).

	The gain calibrator was observed between each target scan of 10~minutes length at 3 and 7~mm, of 3 or 5~minutes at 15~mm (see Table~\ref{tab-15mm}) and of 15~minutes at 3+6~cm. QSO~B1057-79 and QSO~B1600-44 were used as gain calibrators for Chamaeleon and Lupus respectively, Uranus was used as the primary flux calibrator at 3 and 7~mm, and the quasar QSO~B1934-638 was used as the primary flux calibrator from 15~mm to 6~cm.

	The weather was generally good throughout the observations. We reached a sensitivity of $\sim$0.1~mJy/beam required to detect 90\% of the sources at better than 5$\sigma$ for the 3 and 7~mm band observations, and detected {{4/6} sources at 15~mm and  {1/3} at 3+6~cm.}
~The two 7~mm non-detections had the lowest 1.3 and 3~mm fluxes and the highest RMS during the observations, and the non-detections at {{longer wavelengths had}} the lowest extrapolated fluxes of the targeted sample.
	 
	 The data calibration followed the standard CABB procedure described in the ATCA user guide\footnote{\small{http://www.narrabri.atnf.csiro.au/observing/users\_guide/\\html/atug.html}} for all {{wavelengths using}} the software package \texttt{MIRIAD} version 1.5 {{\citep{1995ASPC...77..433S}}}. 

\section{Analysis and results}
\label{sec-results}

The flux and RMS values at 3 and 7~mm were extracted from the image plane using \texttt{IMFIT} and \texttt{IMSTAT}. At the longer wavelengths, values were extracted from the u-v plane using \texttt{UVFIT} and \texttt{UVRMS} due to the short {{3--15}} minute scans. A summary of {{the fluxes}} are presented in Table~\ref{tab-results2}.

The resulting source maps {{were}} created {{by Fourier-transforming the visibility data using a specific weighting (here ``natural"), then the \texttt{CLEAN} and \texttt{RESTORE} algorithms}}. For sources observed over several epochs, data from each day {{was}} calibrated separately and combined with \texttt{INVERT}.

\subsection{{{Which sources}} are resolved at 3 and 7~mm?}
\label{subsec-resolved}

We determined if a source was extended by looking at the visibilities. As the resolution of the array increases, the flux will decrease for a resolved source and remain constant for an unresolved source. Plots of the u-v distance as a function of amplitude created using the \texttt{UVAMP} task in \texttt{MIRIAD} are presented in Fig.~\ref{fig-uvamps-3mm} for 3~mm and Fig.~\ref{fig-uvamp-lup-7mm} for 7~mm. These plots suggest DK Cha at 3 and 7~mm and CS Cha at 3~mm have extended emission, while marginally extended emission is observed at 7~mm for CR Cha, CS Cha, {{SZ Cha}} and RXJ1615.3-3255. For these marginally resolved sources, we will take the conservative approach and assume the {{emission}} were not resolved.

{RXJ1615.3-3255 was resolved at 1.3~mm with the SMA, with an estimated Gaussian size of 1.53 arcsec ($\sim$28 AU assuming a distance of 184~pc) \citep{Lommen10}. Since the H168 {{array}} has a resolution of 1753 $\times$ 763 AU {{(beam size of 9.5$\times$4.1 arcsec)}} and the map in Fig.~\ref{fig-rxj1615map} does not appear to have extended emission, it is unlikely the emission was resolved here at 7~mm.}

{{The visibilities for Sz 32 and MY Lup presented in Fig.~\ref{fig-uvamp-lup-7mm} suggest the presence of a contributing source near each targeted source location.}}
{{Following the \citet{1974gegr.book..256F} approach of using the visibility function to determine the distance between double sources,}} we find {{that the}} Sz 32 and MY Lup visibilities are consistent with the location of WW Cha at $\sim35^{\prime\prime}$ to the {{south-west}} of Sz 32 and an unidentified source at $\sim50^{\prime\prime}$ to the {{north}} of MY Lup -- maps are presented in Fig.~\ref{fig-maps}. The smaller field of view at 3~mm ($30^{\prime\prime}.2$ versus $65^{\prime\prime}.4$ at 7~mm) did not allow for either {WW Cha or the unidentified source}
~to be detected at 3~mm. This is consistent with \citet{Lommen09}, who previously {{observed}} and resolved WW Cha at 3 and 7~mm and only detected Sz 32 {{in the field of their 7~mm observations}}.

Although looking at the visibility {{data}} is commonly used in literature to determine source extension \citep[e.g.][]{2007ApJ...659..705A}, this approach could suffer from phase decorrelation. Phase decorrelation degrades the signal as the baseline increases, causing a similar decrease in amplitude on the longer baselines. {{Given the generally}} good weather {{conditions}}, the use of compact {{array}} configurations, and reasonable RMS values from the ATCA seeing monitor\footnote{The ATCA seeing monitor is a fixed interferometer tracking the 30.48~GHz  beacon on a geostationary communication satellite at elevation of 60$^\degree$ \citep{2006PASA...23..147M}.}{{, phase decorrelation was not substantial in our observations.}} 

{{To ensure that phase decorrrelation was not effecting our visibilities amplitudes, we used three other}} methods to investigate source extension at 3 and 7~mm. If {{a}} source is extended, {a point source fit will exclude some emission, whilst a Gaussian fit should include most of the emission.}
~A source could be considered extended when {{(1) the point source}} fit is less than the Gaussian fit (after accounting for uncertainties), {{(2)}} the Gaussian fit is greater than the {{point source}} fit plus $n$ times the RMS, where $n$ is a predetermined value (\citet{Lommen07, Lommen09} used $n=2$ for pre-CABB {{ATCA}} data, and assuming a factor of four improvement with CABB, {{here we use }}$n=8$) and {{(3)}} when the synthesised beam is smaller than the Gaussian size obtained from the Gaussian fit. These {{three}} alternative approaches were considered (see Table~\ref{tab-results}) and the {{results are consistent with the sources considered resolved when using the visibility data (CS Cha and DK Cha), while marginally resolved sources were considered unresolved with these three alternative approaches.}}

	Thus from this analysis we have determined that the emission of CS Cha was resolved at 3~mm, and the emission of DK Cha was resolved at 3 and 7~mm. In Table~\ref{tab-results2} sources in \textbf{boldface} were resolved at 3~mm with ATCA in this work or in the literature. {{\citet{Lommen07,Lommen10} resolved SZ Cha, HT Lup, RY Lup and Sz 111 at 3~mm with ATCA.}} 

\begin{figure*}
	\subfloat{\label{fig:cscha.93969.2}\includegraphics[width=.68\columnwidth]{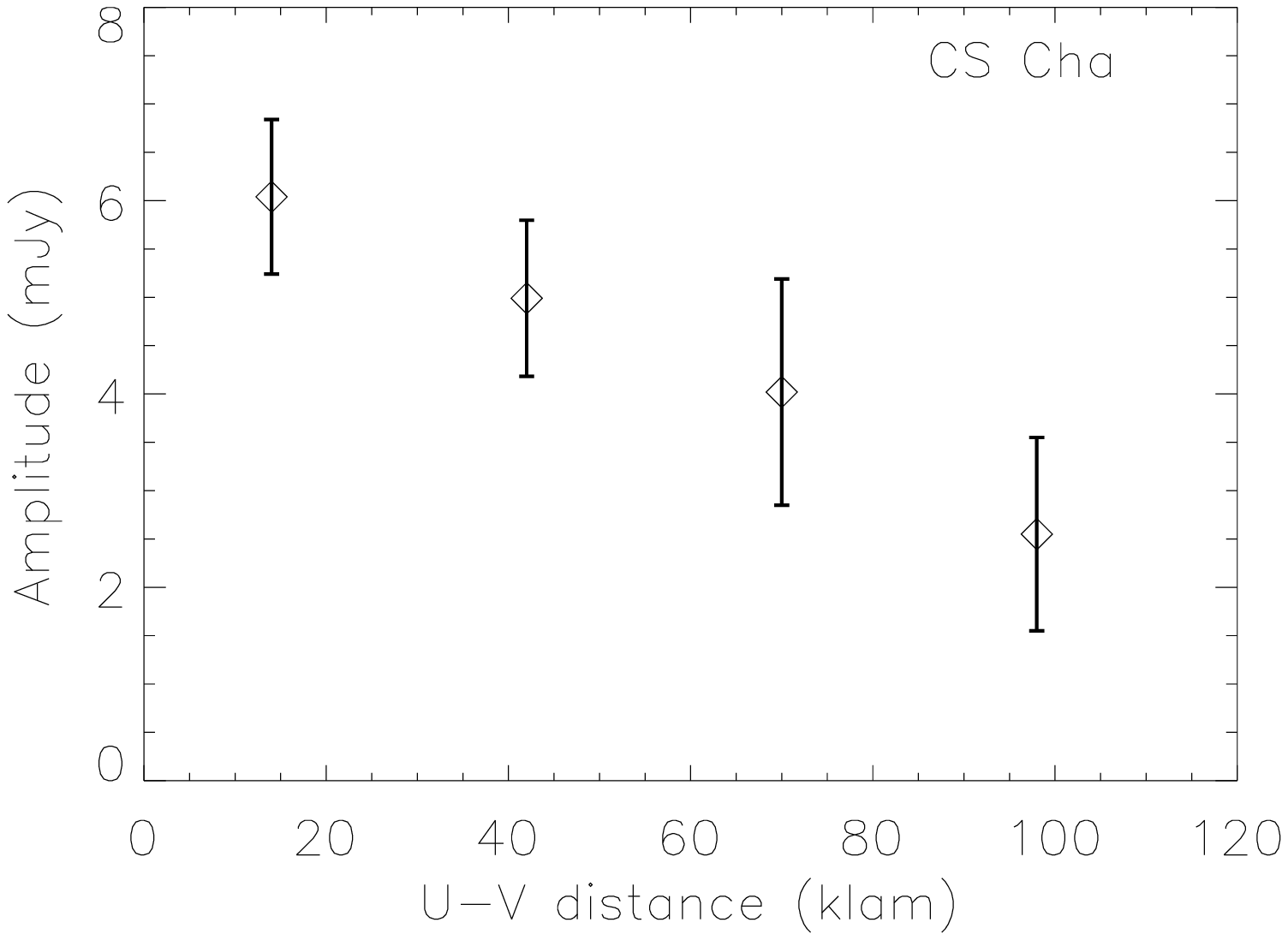}}
	\subfloat{\label{fig:dicha.93969.2}\includegraphics[width=.68\columnwidth]{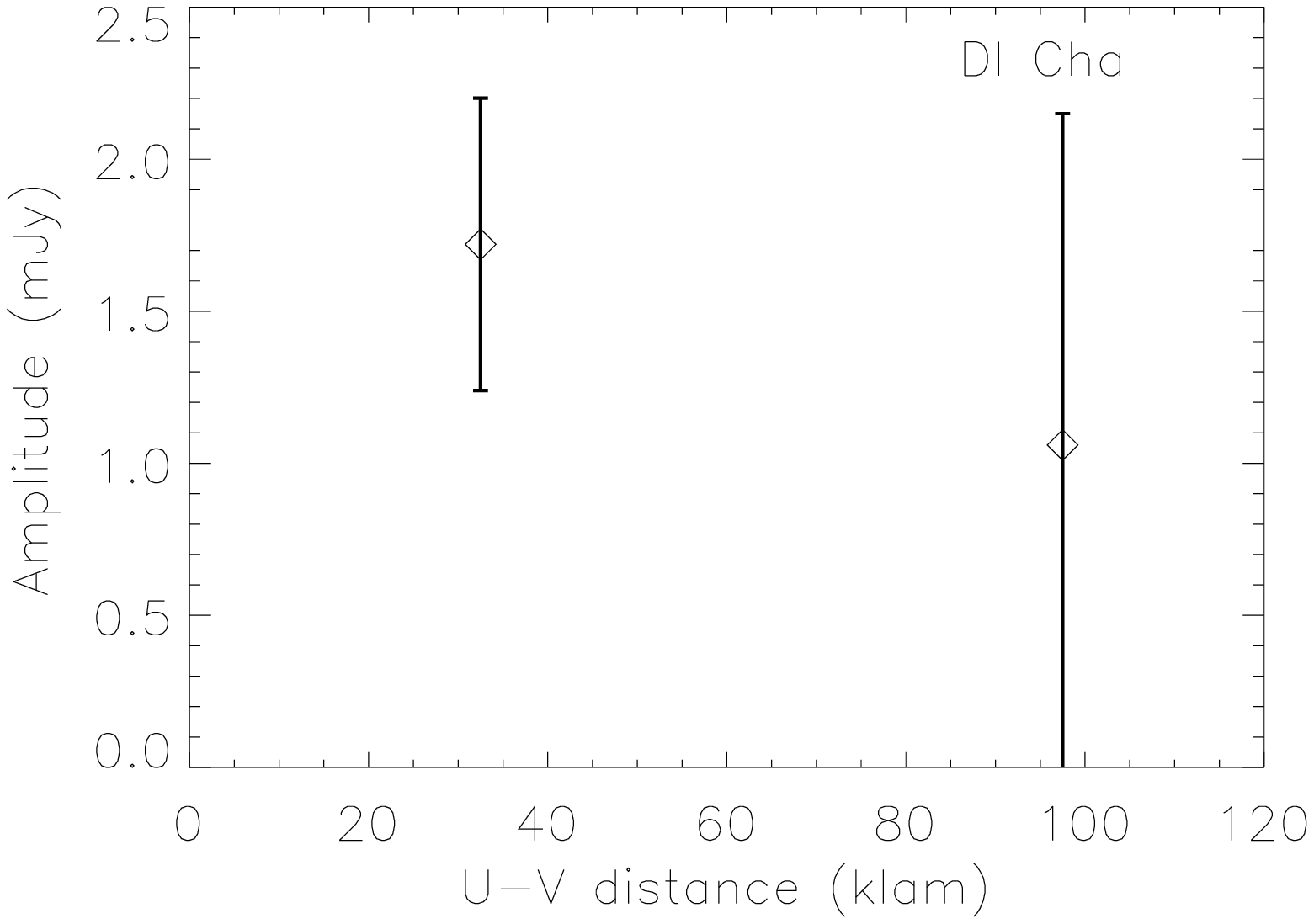}}
	\subfloat{\label{fig:glassi.93969.2}\includegraphics[width=.68\columnwidth]{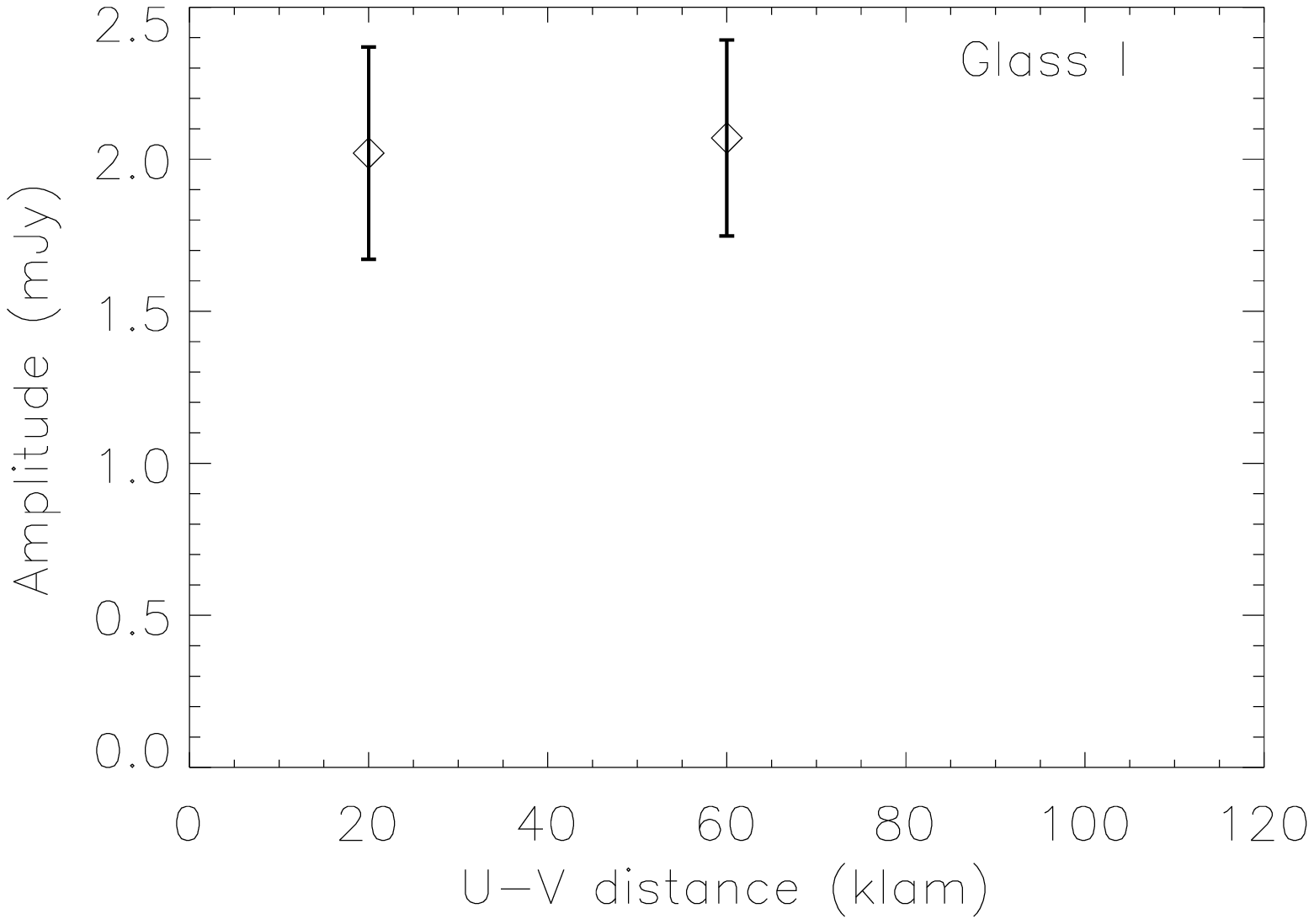}}\\
	\subfloat{\label{fig:sz32.93969.2}\includegraphics[width=.68\columnwidth]{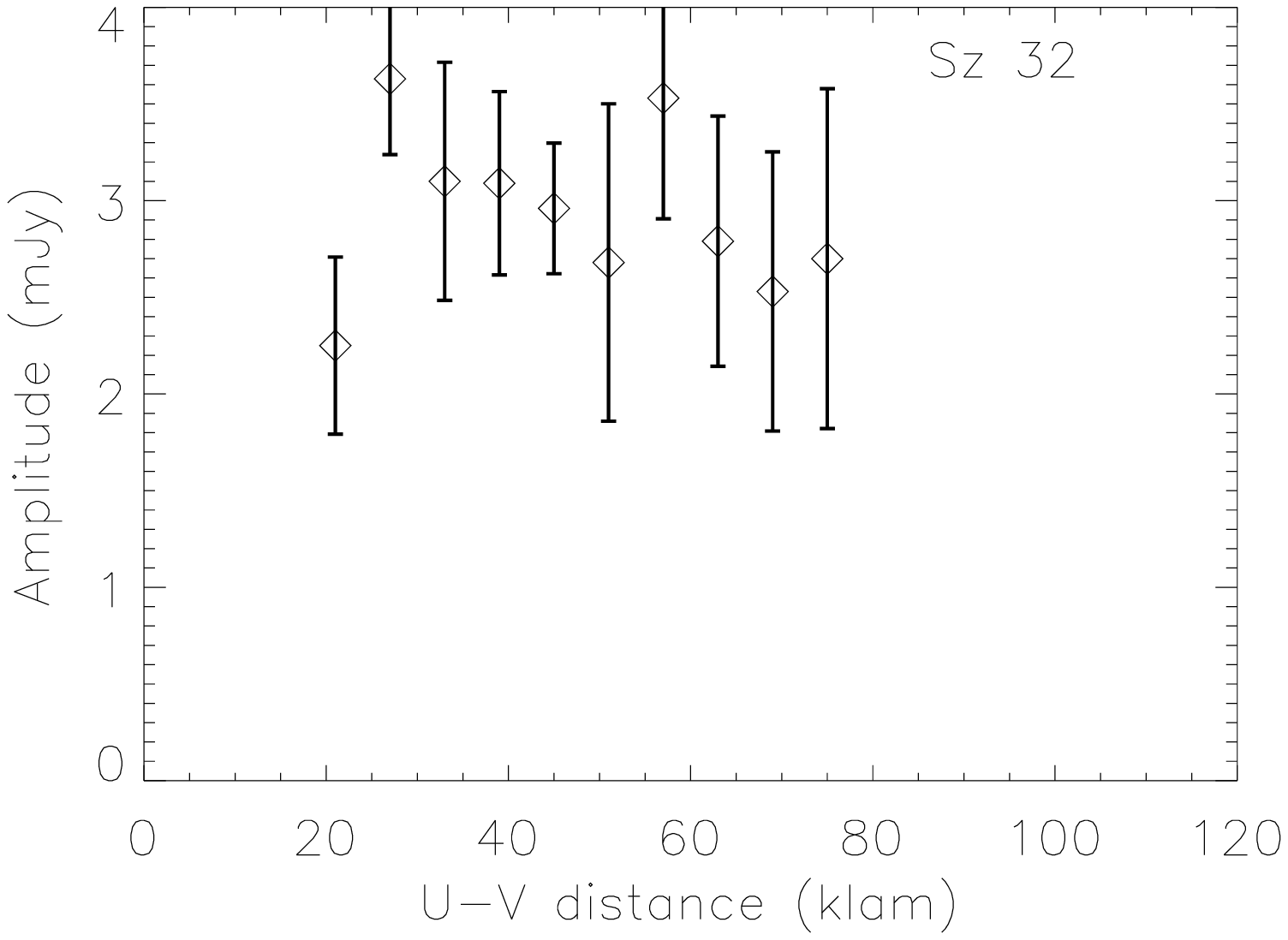}}
	\subfloat{\label{fig:dkcha.93969.2}\includegraphics[width=.68\columnwidth]{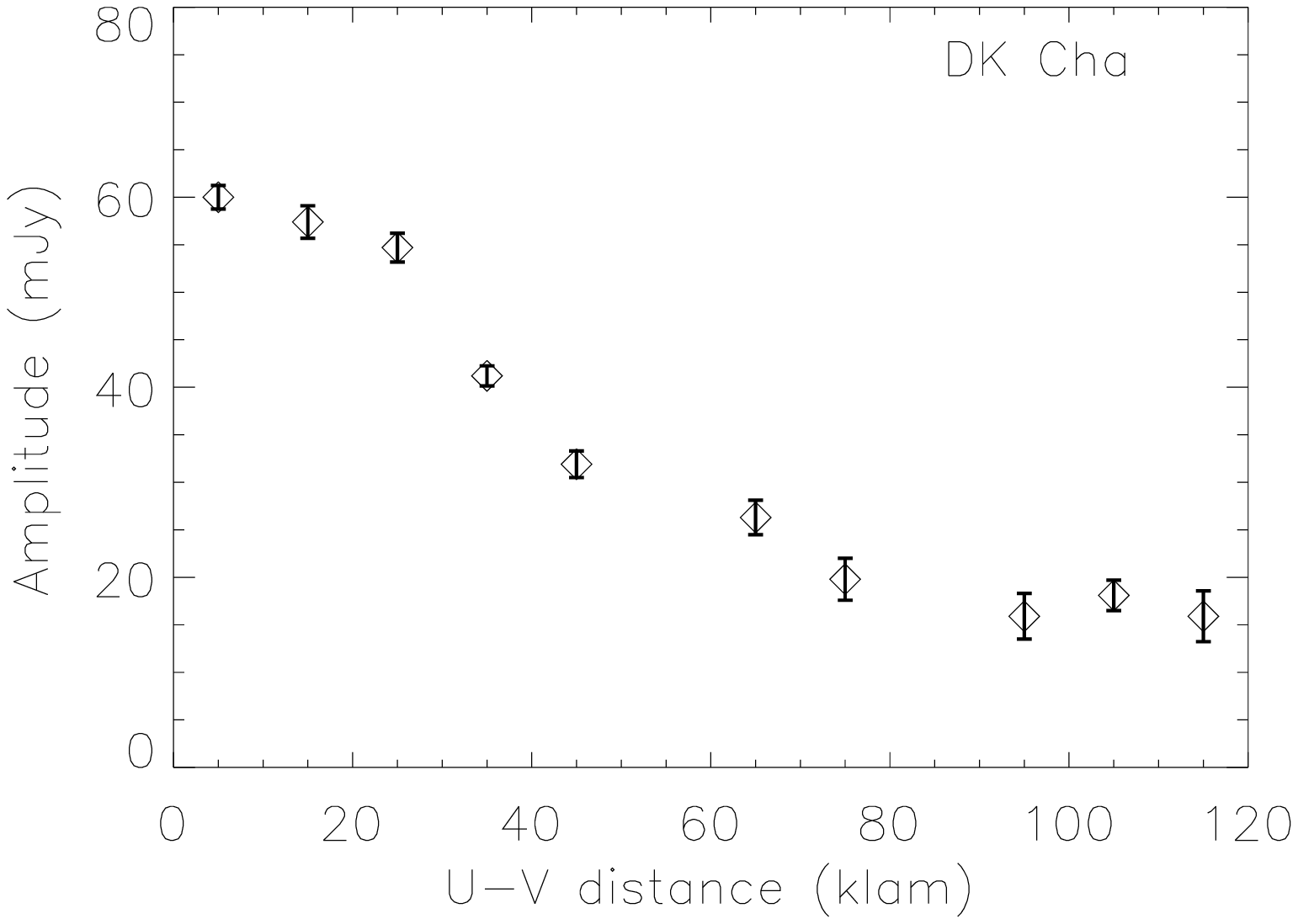}}
	\subfloat{\label{fig:gqlup.93969.2}\includegraphics[width=.68\columnwidth]{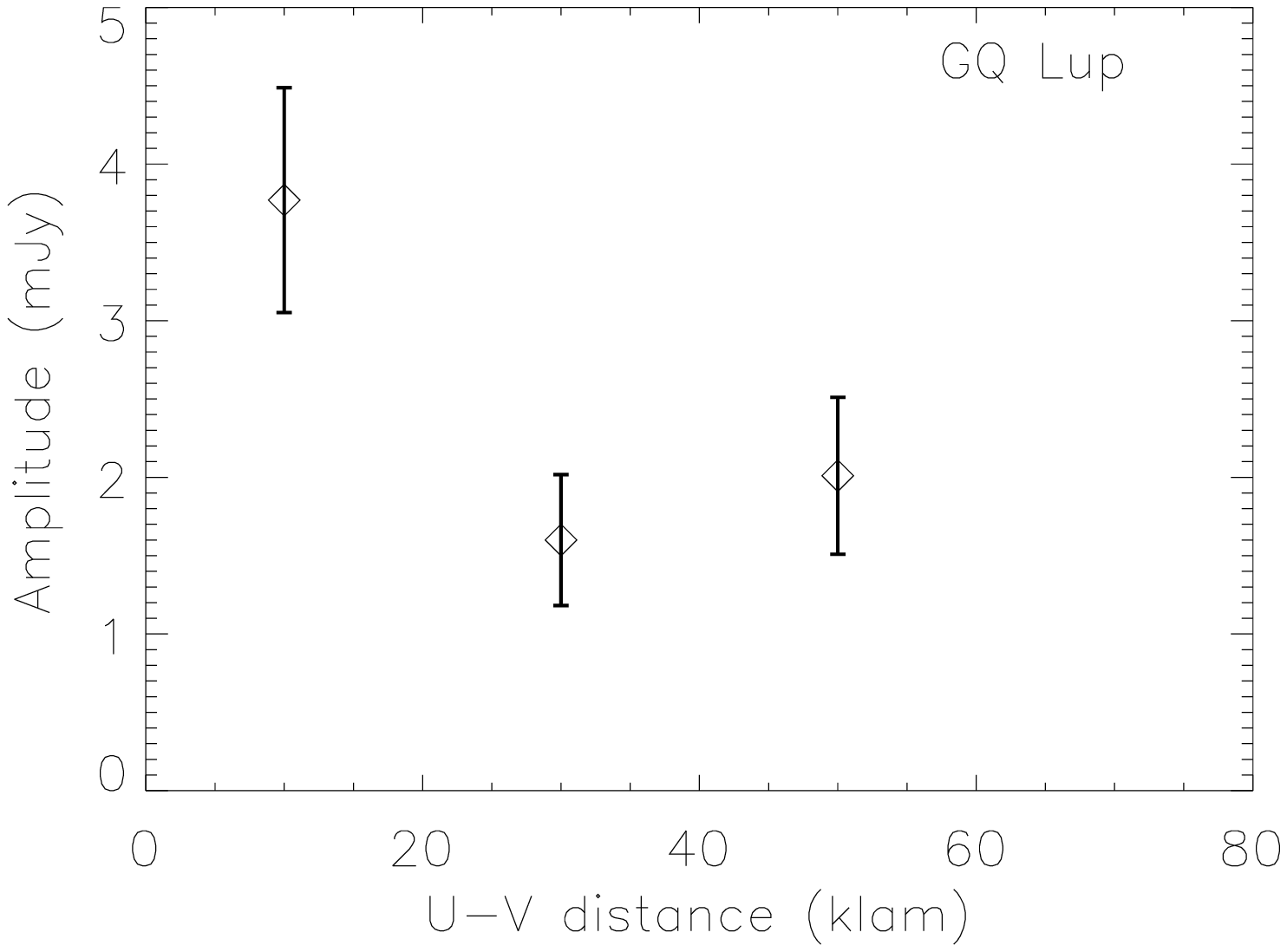}}\\
	\subfloat{\label{fig:exlup.93969.2}\includegraphics[width=.68\columnwidth]{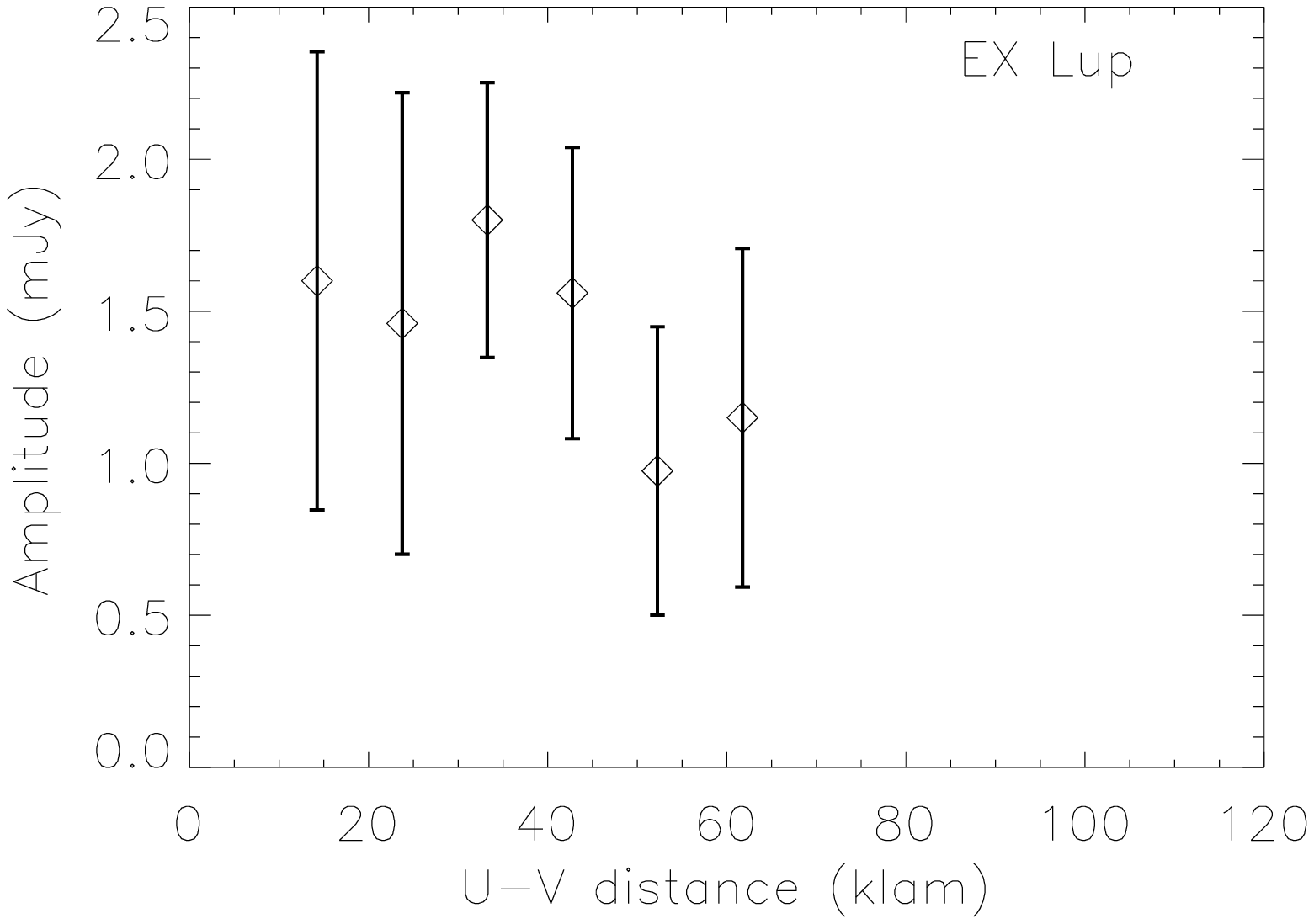}}
	 \caption{Visibility amplitude versus baseline length {{(or u-v distance)}} for Chamaeleon and Lupus sources at 3~mm, with the  {{1$\sigma$}} statistical error bars for each bin.}
  \label{fig-uvamps-3mm}
\end{figure*}

\begin{figure*}
	\subfloat{\label{fig:cr-cha.44030.2}\includegraphics[width=.68\columnwidth]{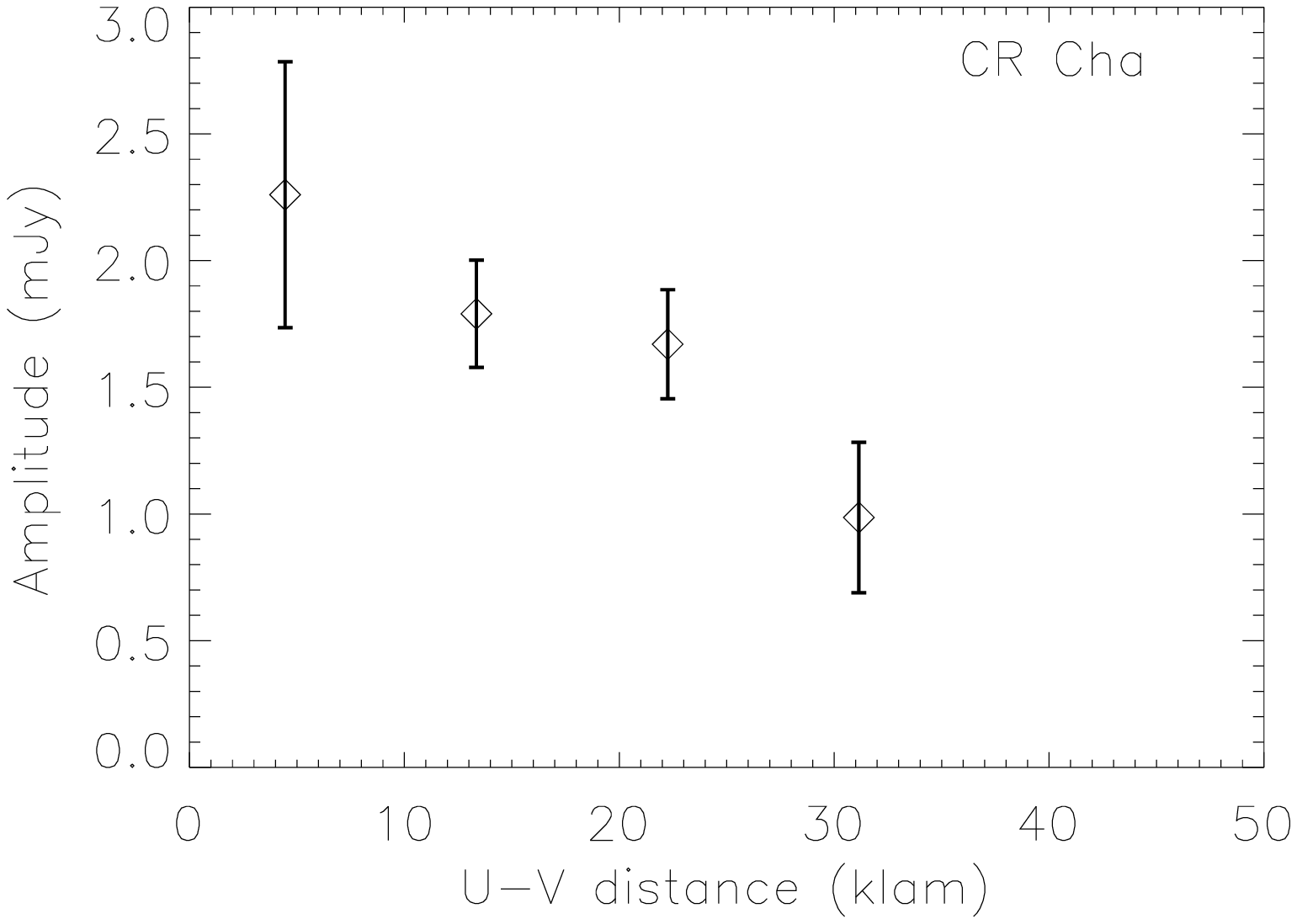}} 
	\subfloat{\label{fig:cs-cha.44030.2}\includegraphics[width=.68\columnwidth]{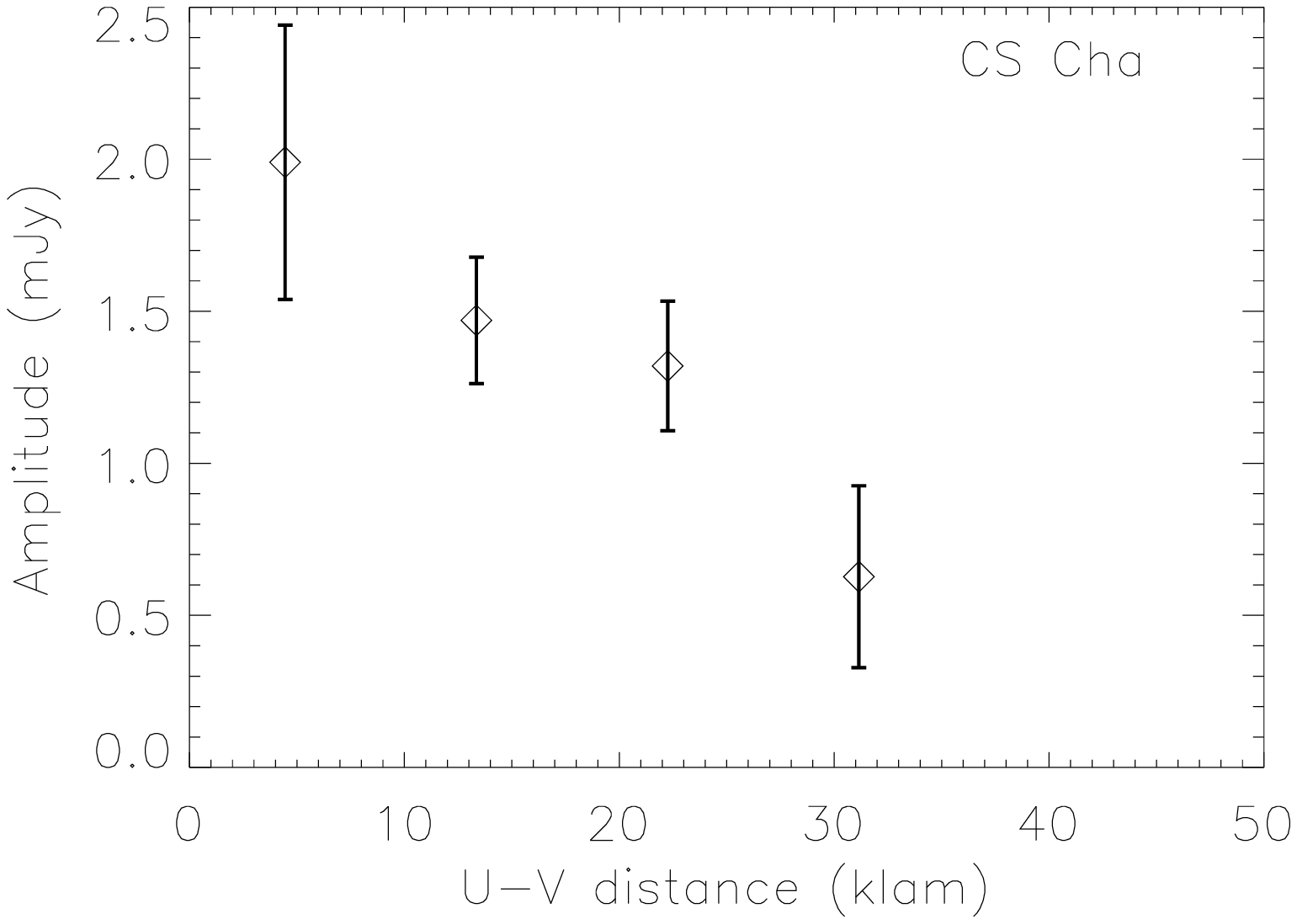}} 
	\subfloat{\label{fig:di-cha.44030.2}\includegraphics[width=.68\columnwidth]{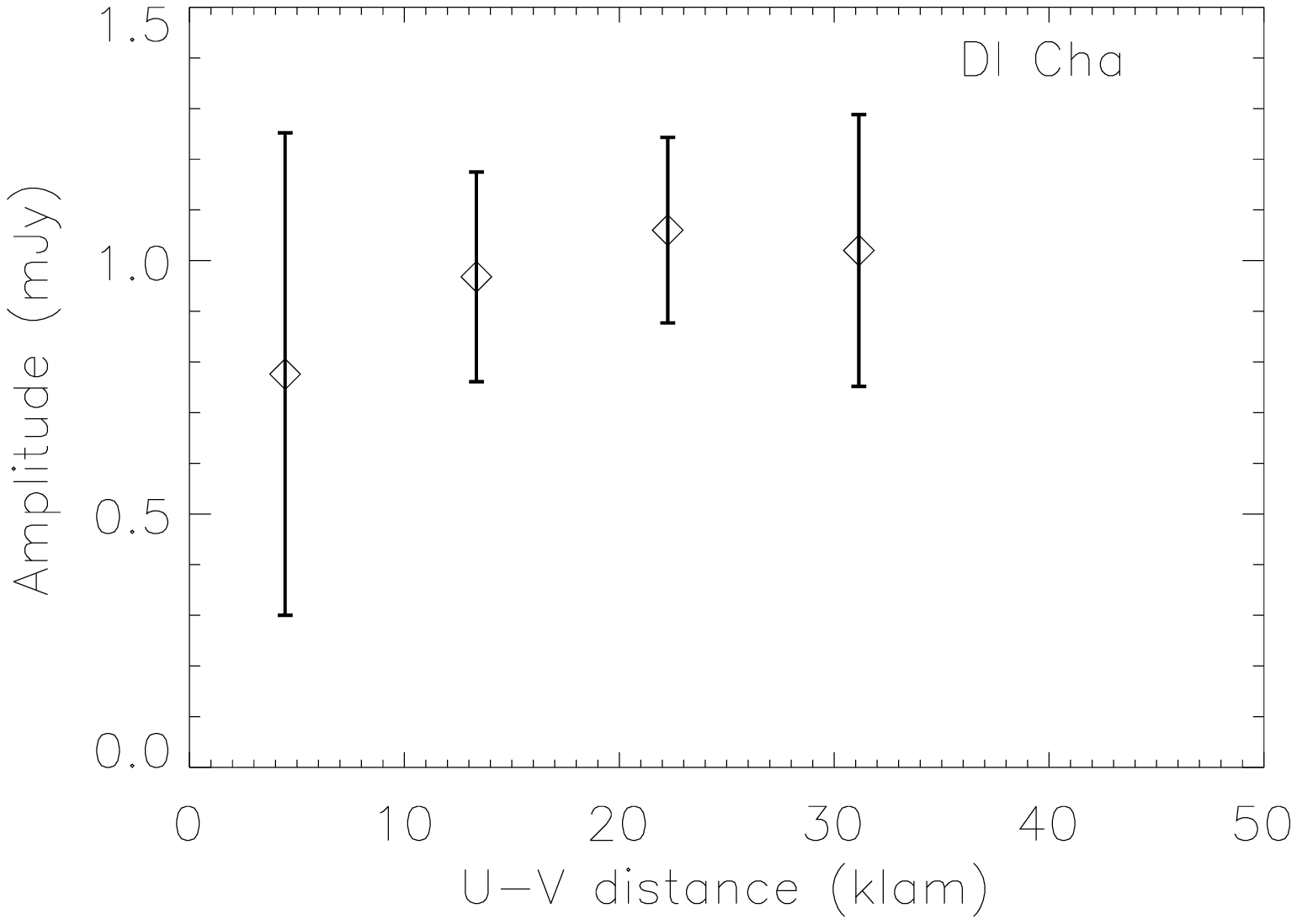}}\\ 
	\subfloat{\label{fig:t-cha.44030.2}\includegraphics[width=.68\columnwidth]{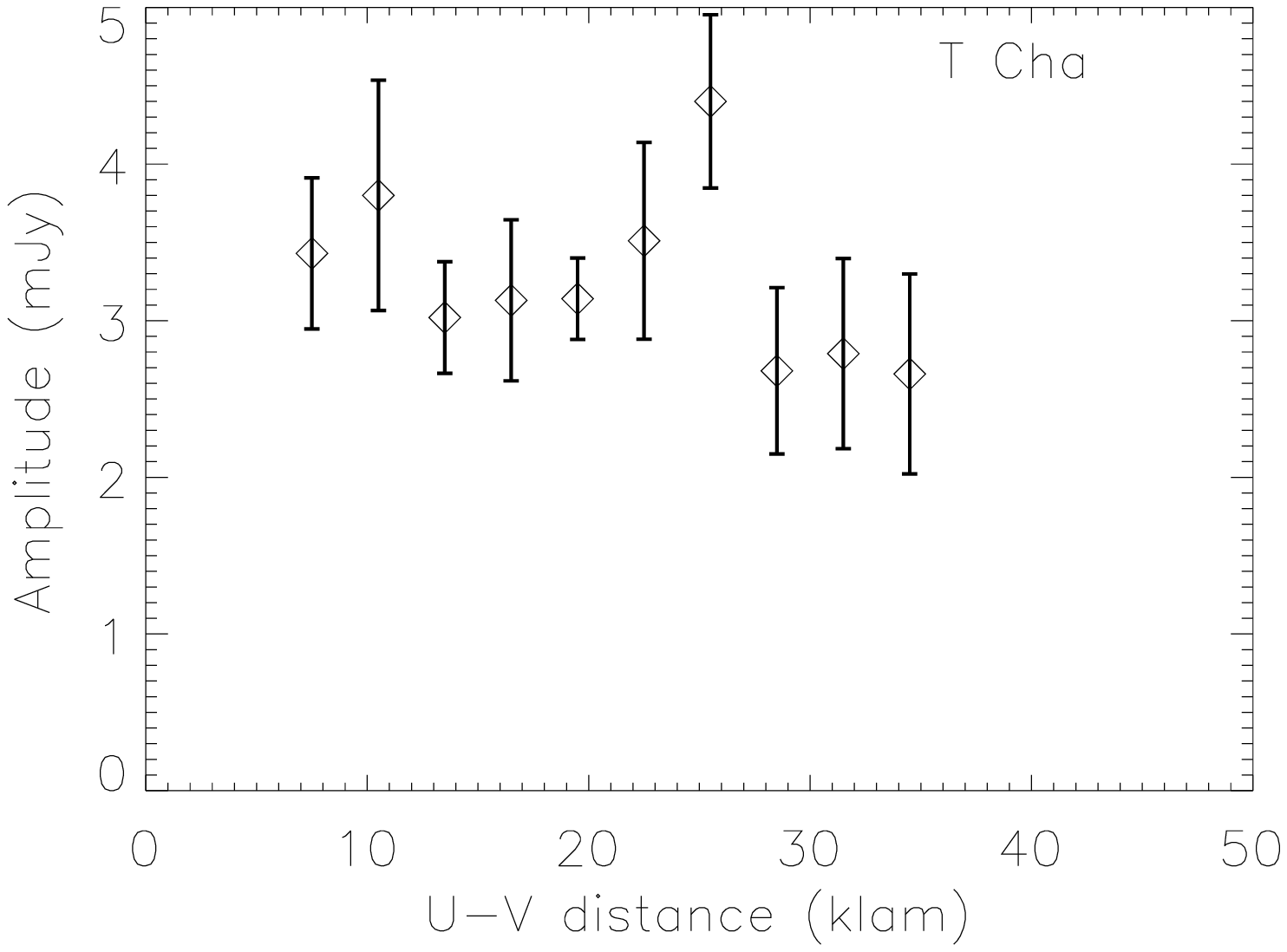}} 
	\subfloat{\label{fig:glassi.44030.2}\includegraphics[width=.68\columnwidth]{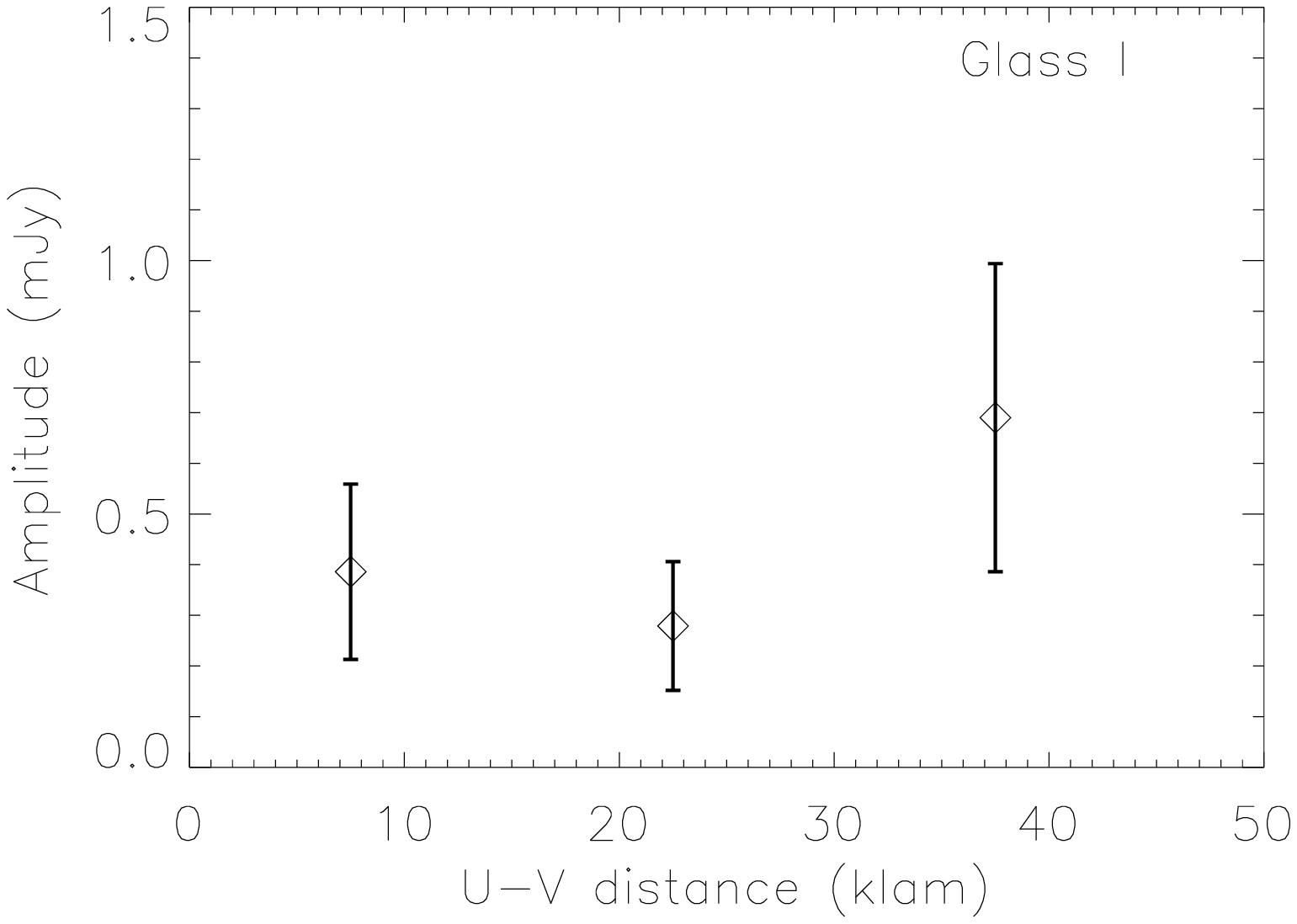}} 
	\subfloat{\label{fig:sz-cha.44030.2}\includegraphics[width=.68\columnwidth]{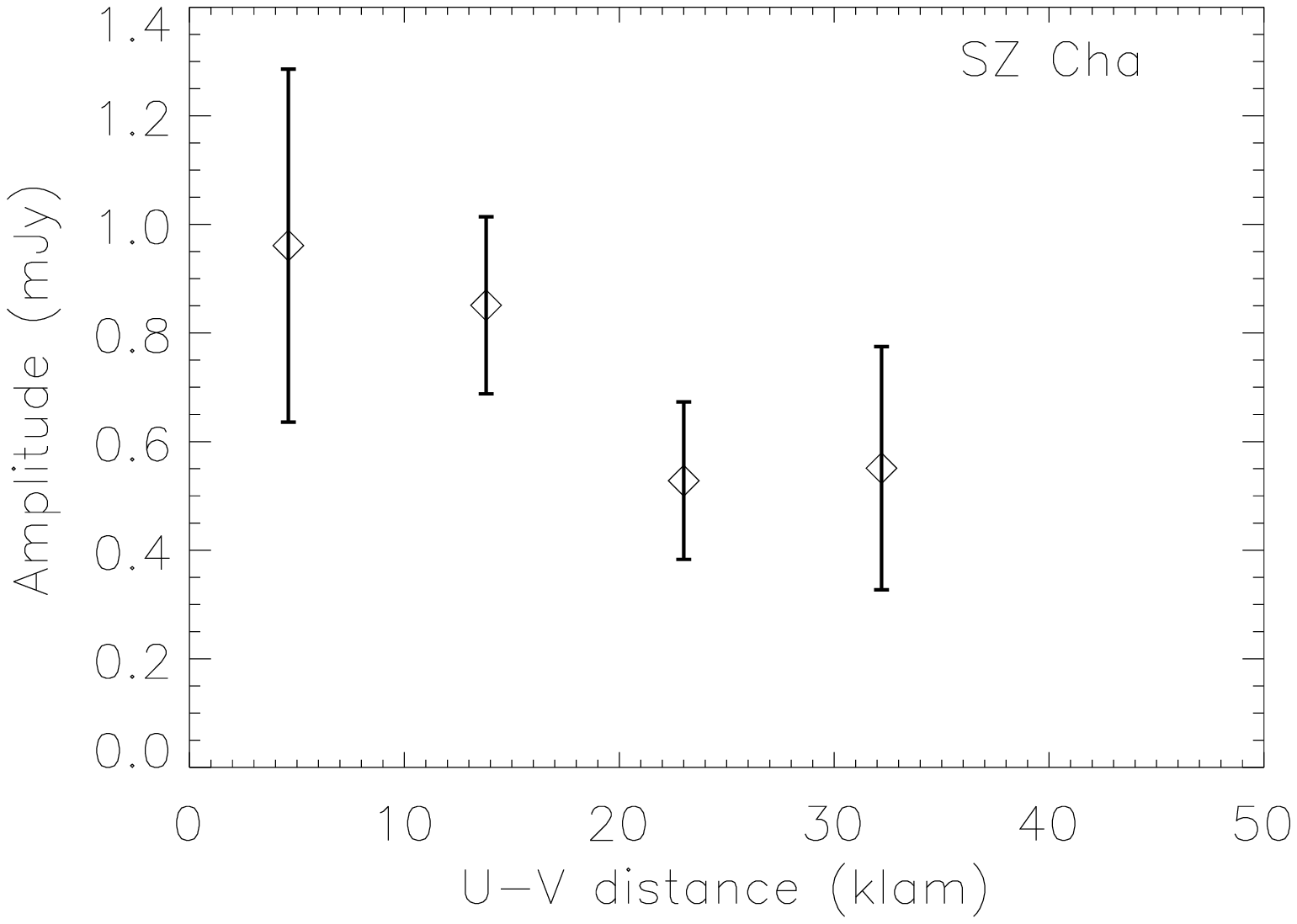}}\\ 
	\subfloat{\label{fig:sz-32.44030.2}\includegraphics[width=.68\columnwidth]{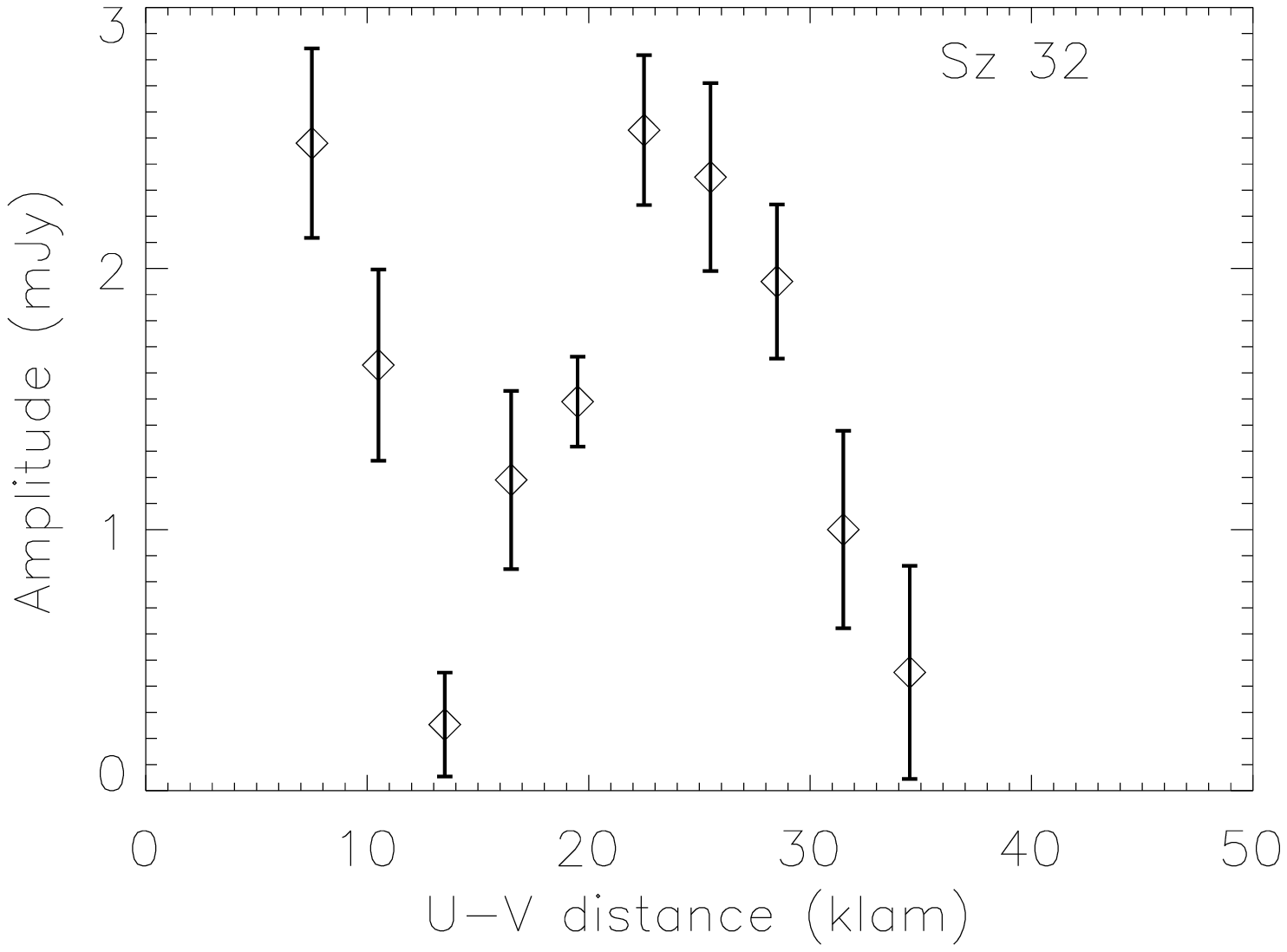}} 
	\subfloat{\label{fig:dkcha.44030.2}\includegraphics[width=.68\columnwidth]{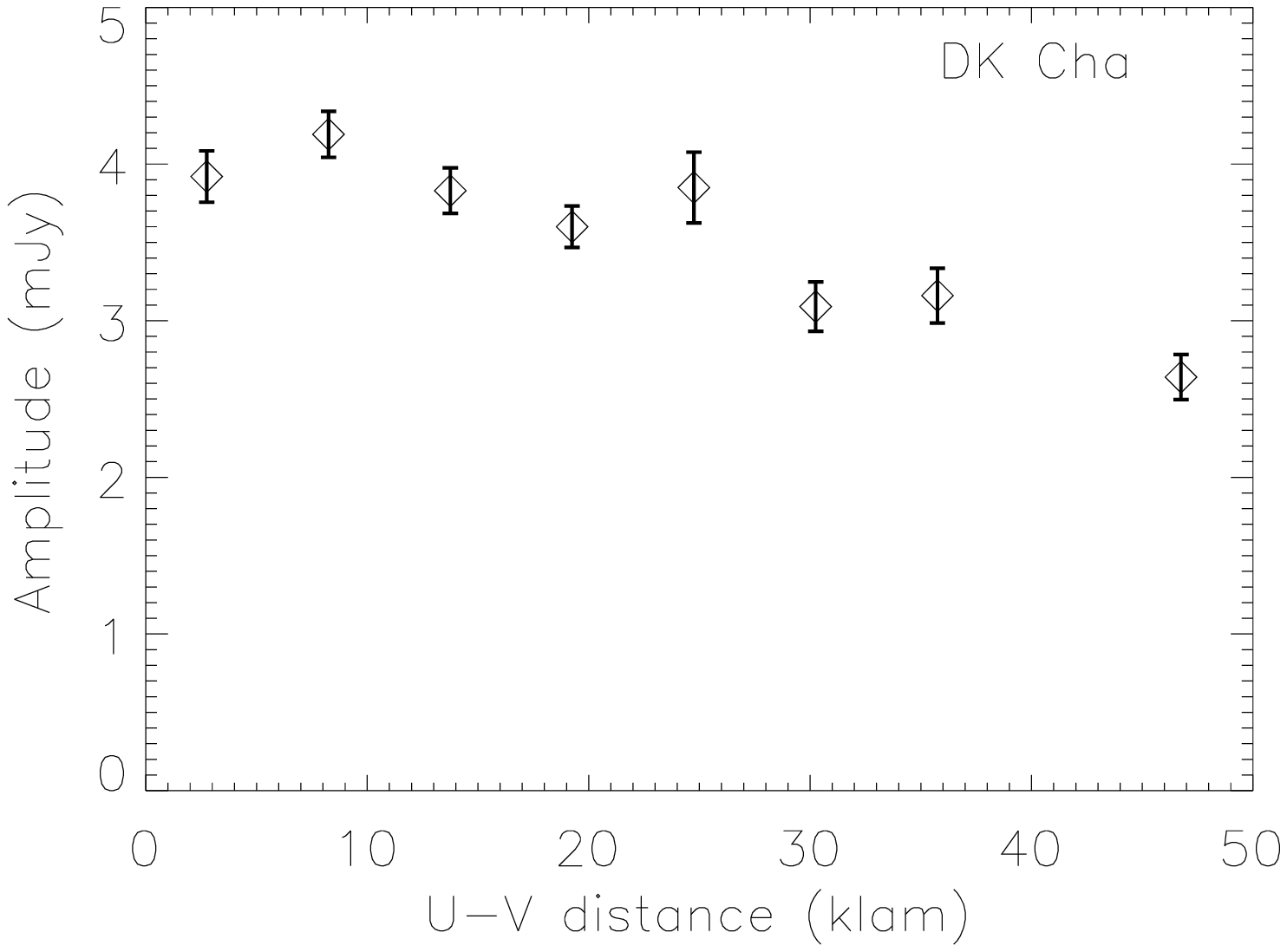}} 
	 \caption{Visibility amplitude versus baseline length {{(or u-v distance)}} for Chamaeleon sources at 7~mm, with the  {{1$\sigma$}} statistical error bars for each bin.}
\end{figure*}

\begin{figure*}
	\ContinuedFloat
	\subfloat{\label{fig:sz65-66.44030.2}\includegraphics[width=.68\columnwidth]{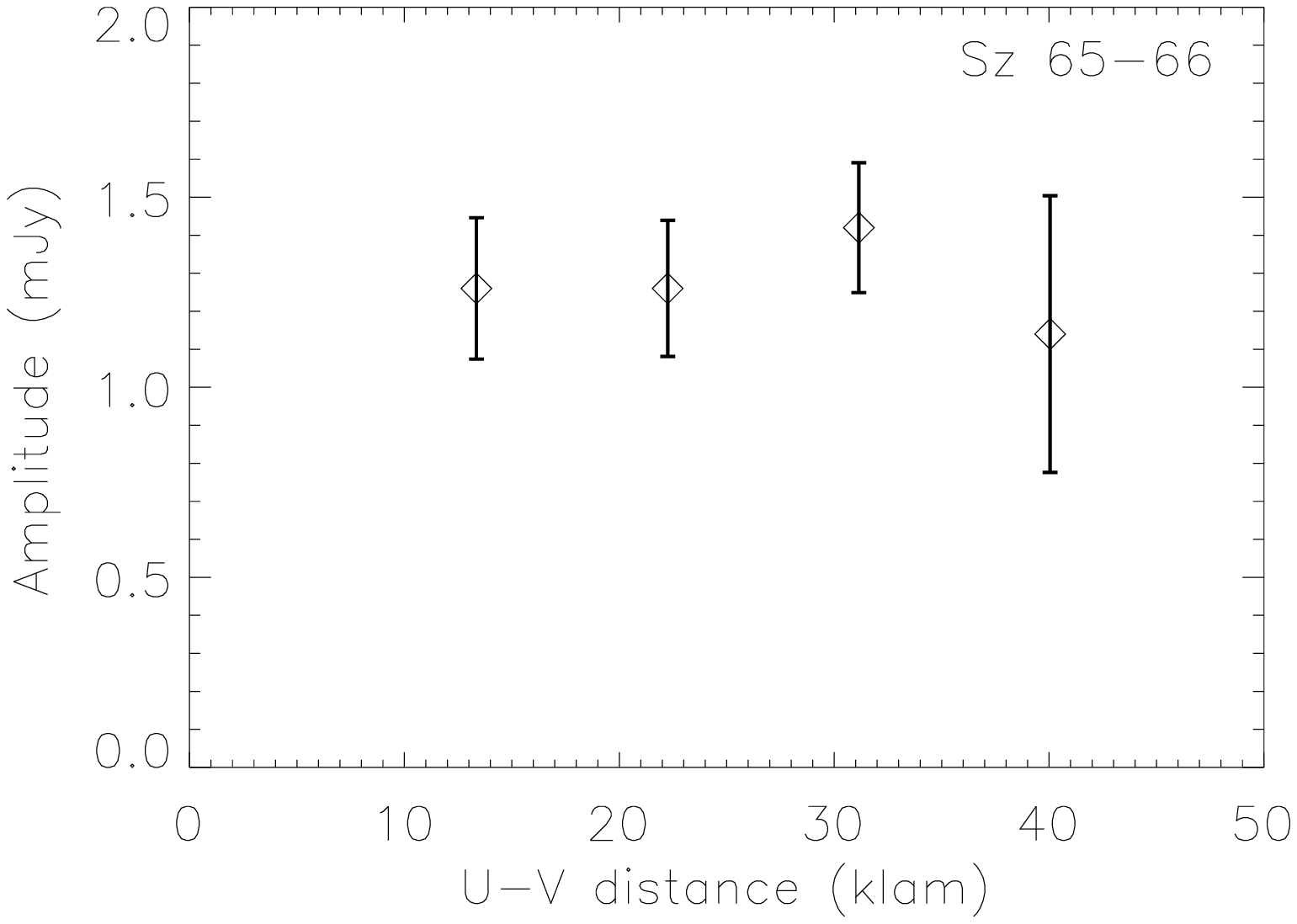}}
	\subfloat{\label{fig:htlup.44030.2}\includegraphics[width=.68\columnwidth]{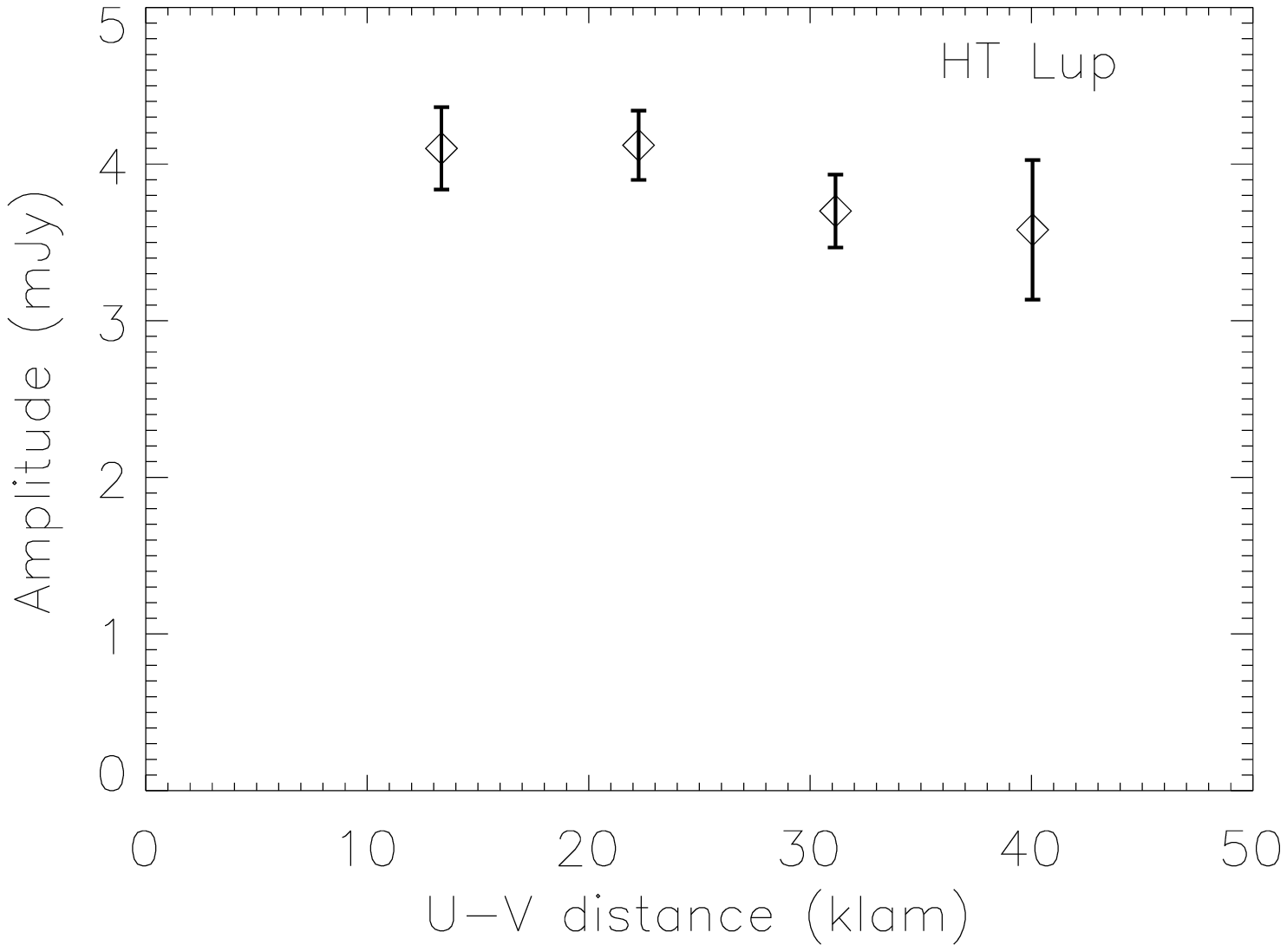}}
	\subfloat{\label{fig:gqlup.44030.2}\includegraphics[width=.68\columnwidth]{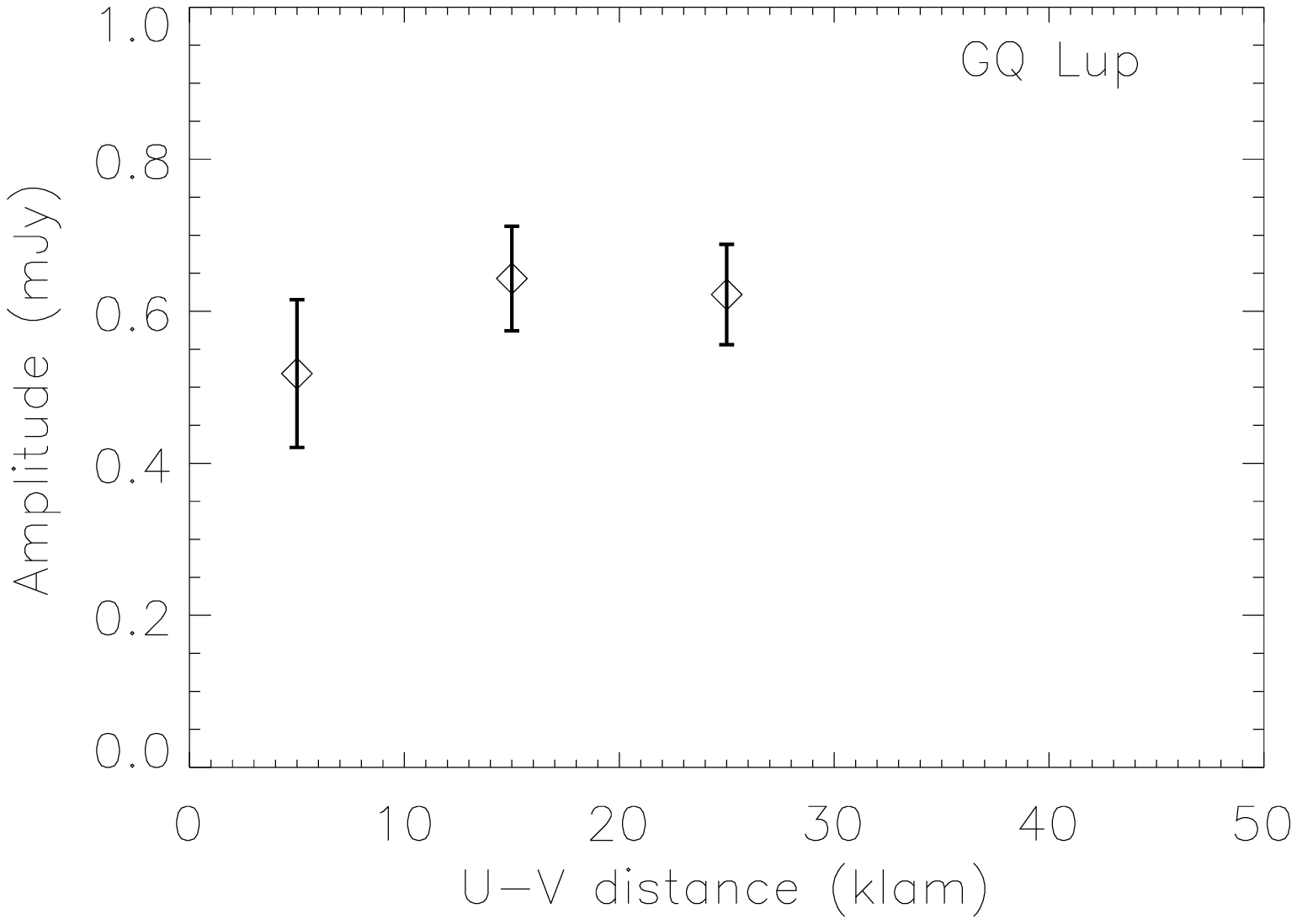}}\\
	\subfloat{\label{fig:gwlup.44030.2}\includegraphics[width=.68\columnwidth]{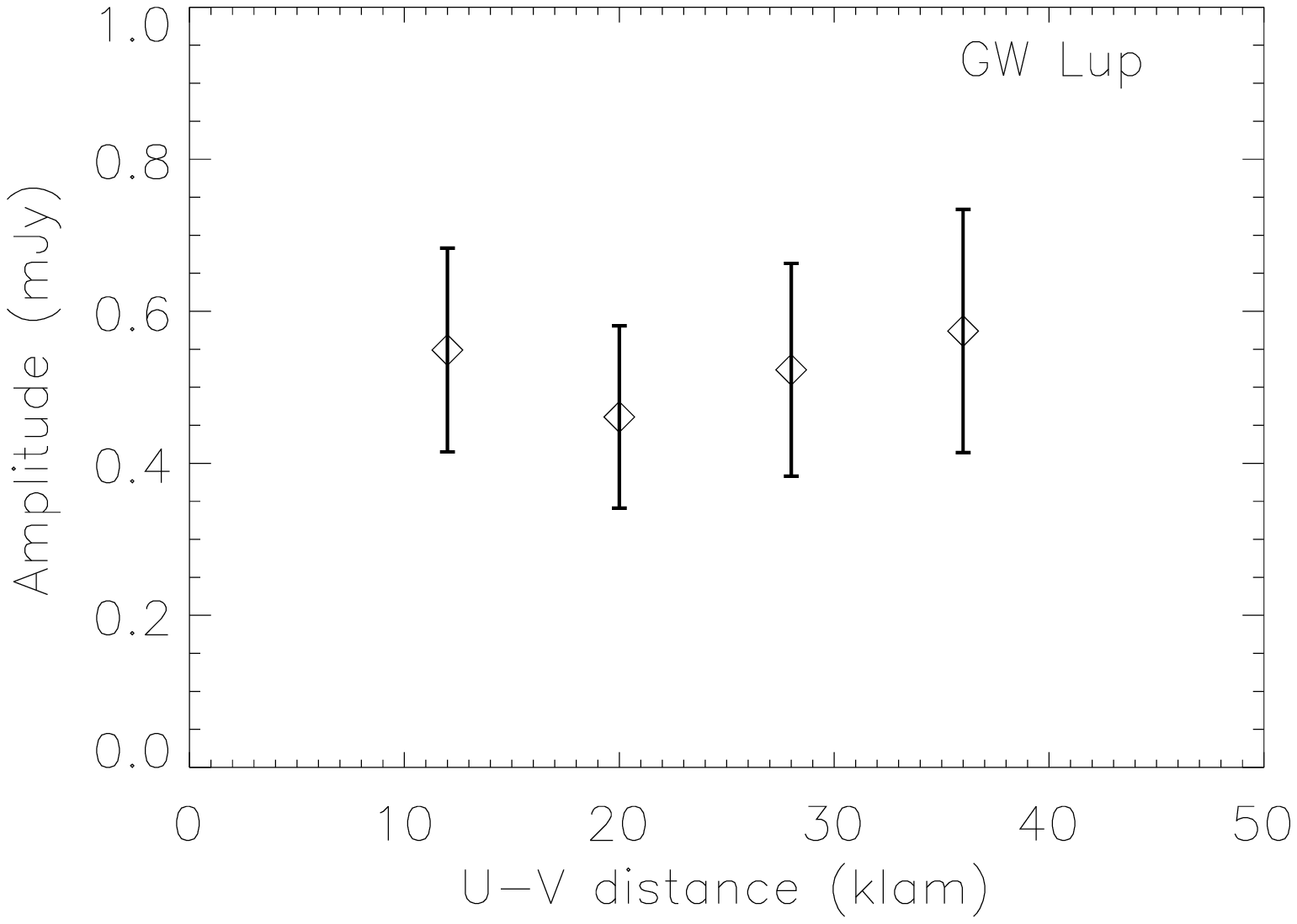}}
	\subfloat{\label{fig:rylup.44030.2}\includegraphics[width=.68\columnwidth]{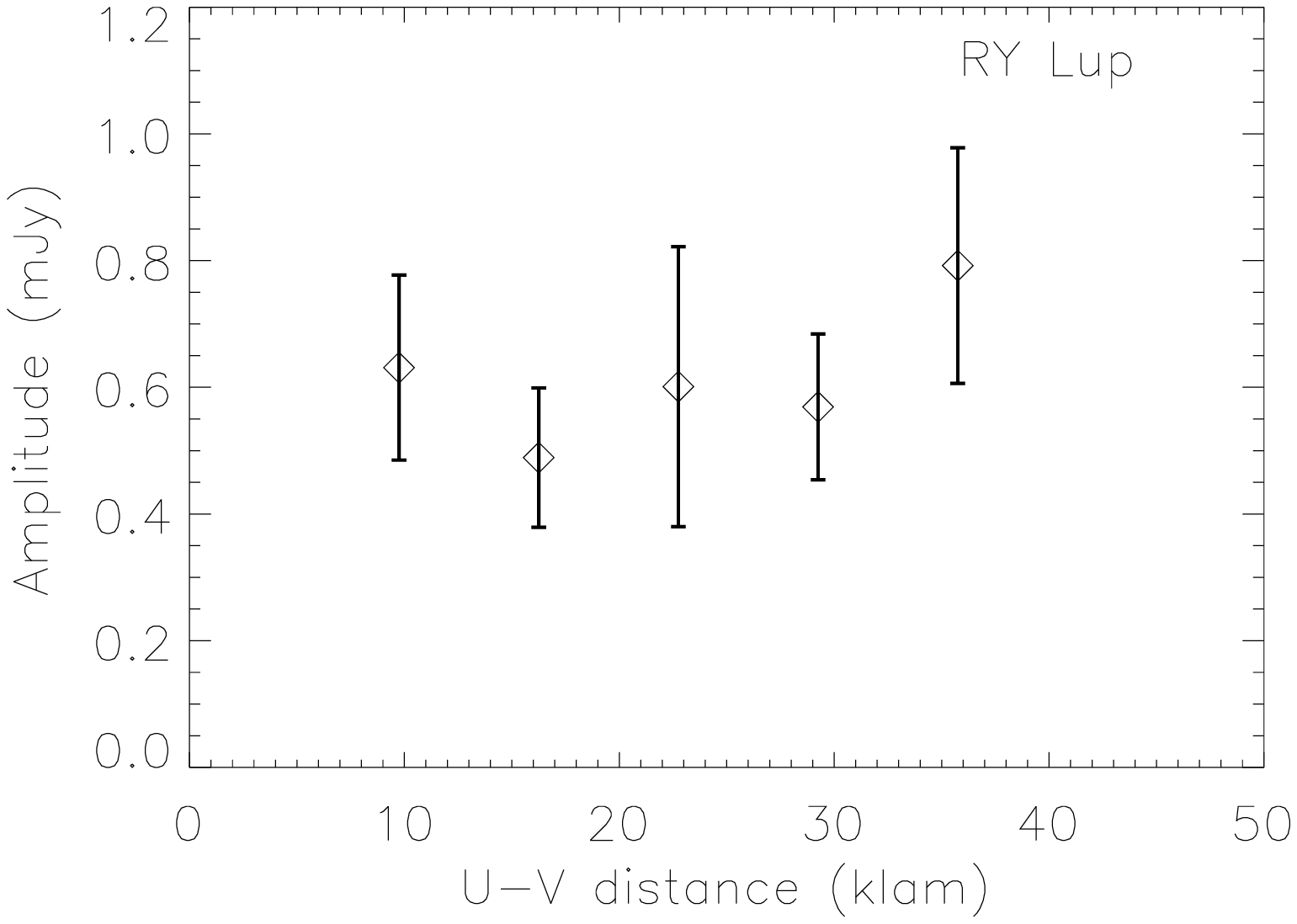}}
	\subfloat{\label{fig:sz111.44030.2}\includegraphics[width=.68\columnwidth]{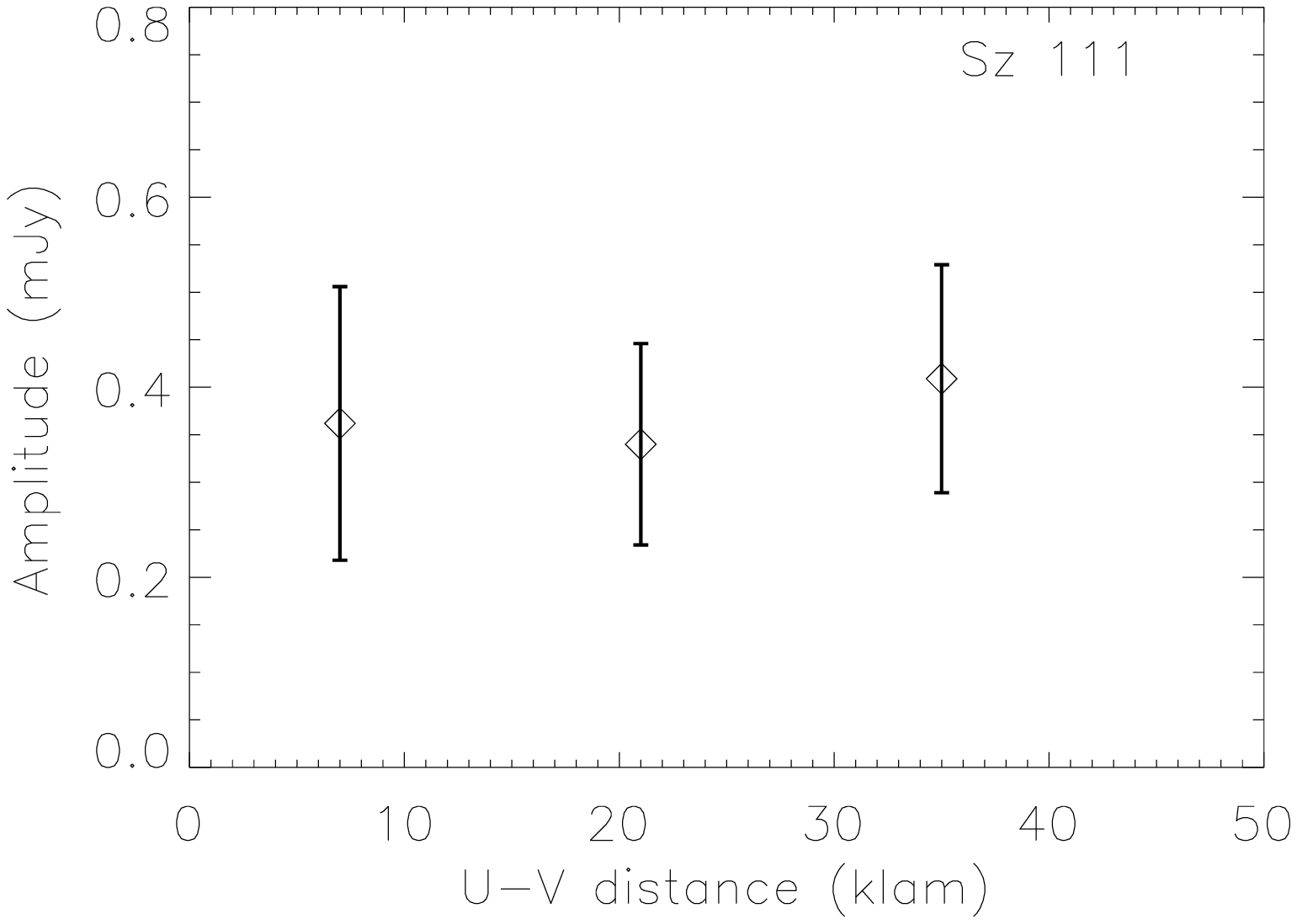}}\\
	\subfloat{\label{fig:hklup.44030.2}\includegraphics[width=.68\columnwidth]{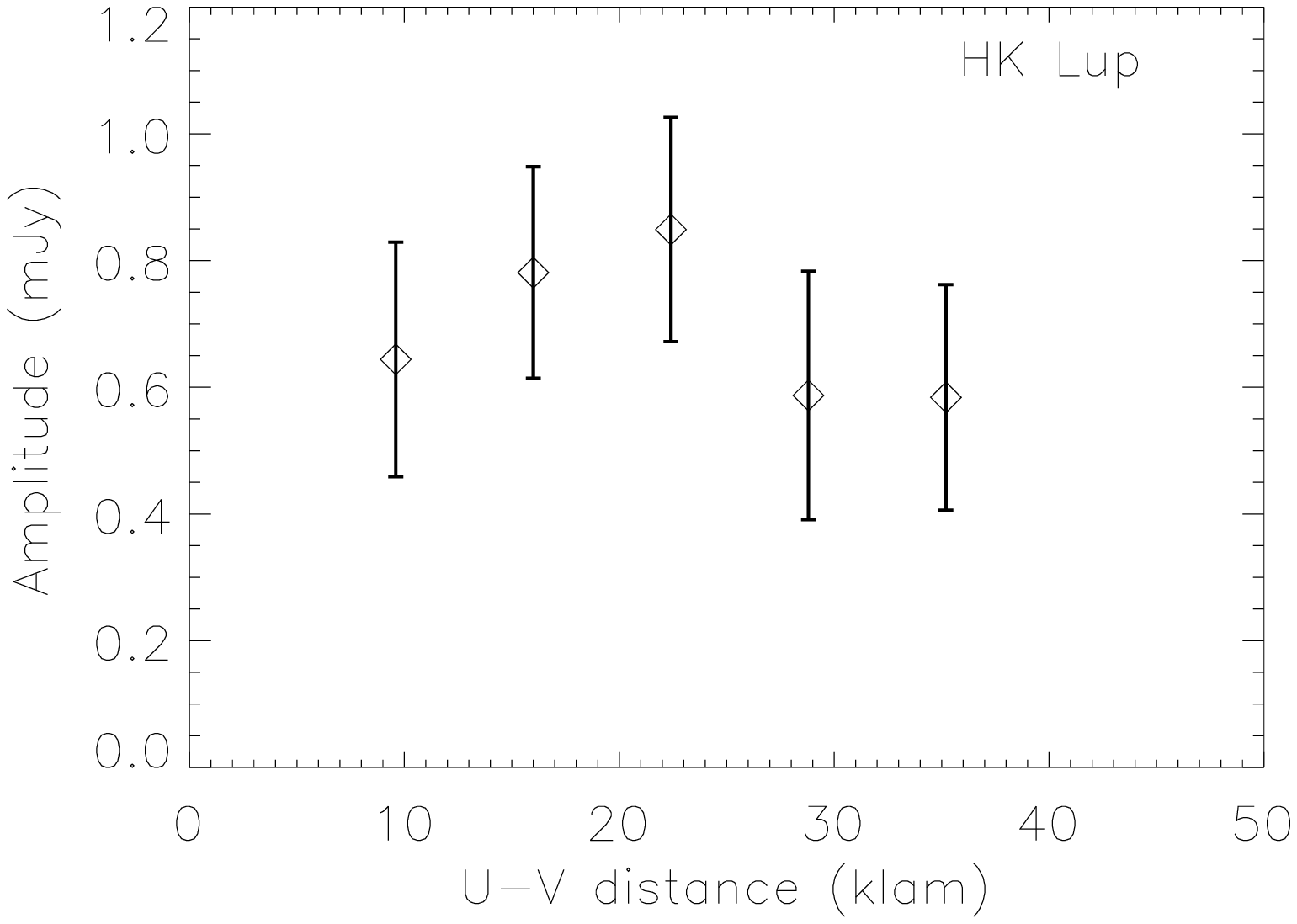}}
	\subfloat{\label{fig:mylup.44030.2}\includegraphics[width=.68\columnwidth]{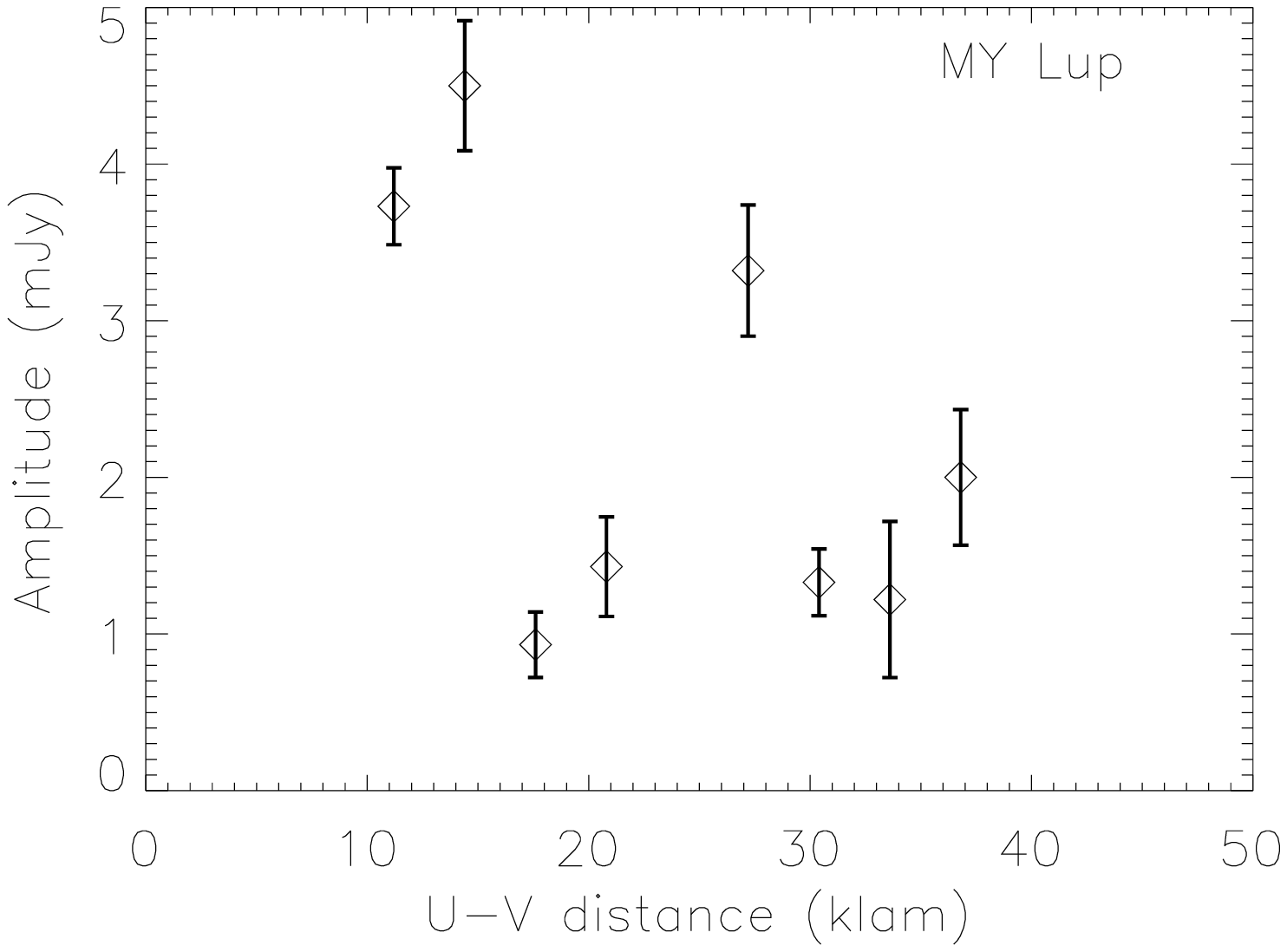}}
	\subfloat{\label{fig:rxj1615.44030.2}\includegraphics[width=.68\columnwidth]{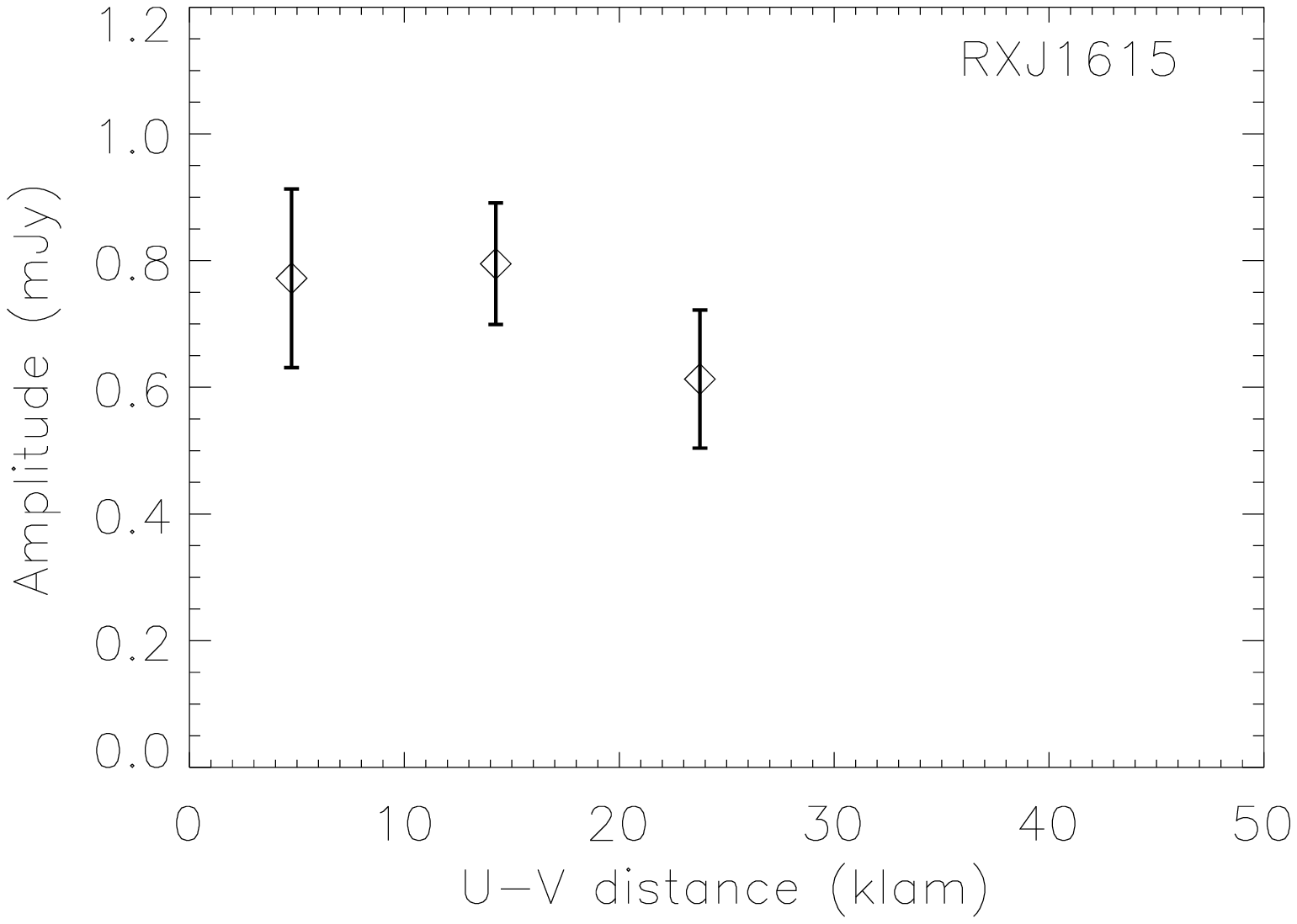}}\\
	\subfloat{\label{fig:exlup.44030.2}\includegraphics[width=.68\columnwidth]{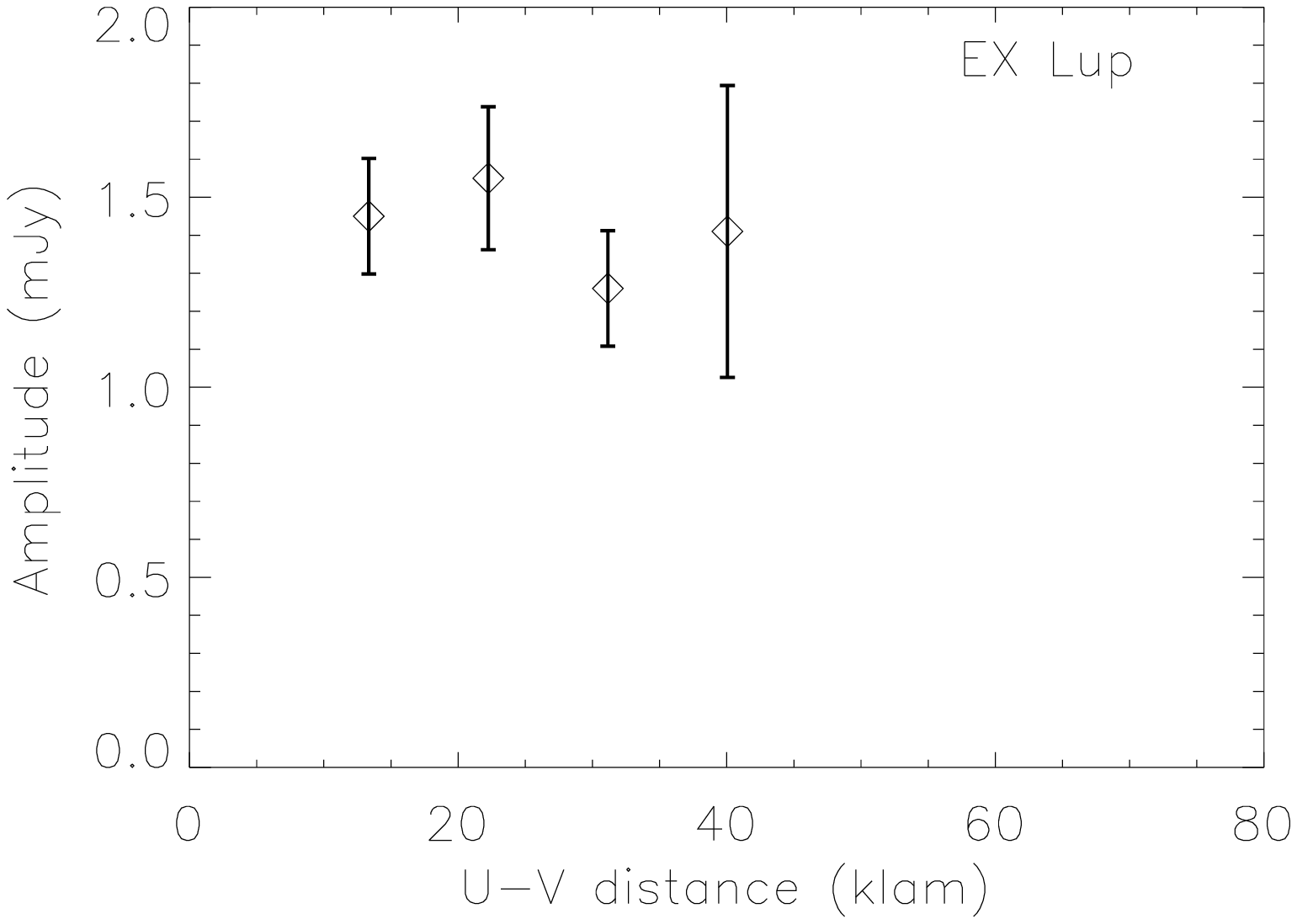}}
	  \caption{Visibility amplitude versus baseline length {{(or u-v distance)}} for Lupus sources at 7~mm, with the  {{1$\sigma$}} statistical error bars for each bin.}
	  \label{fig-uvamp-lup-7mm}
\end{figure*}

\begin{figure*}
	\subfloat{\label{fig-rxj1615map}\includegraphics[width=.73\columnwidth]{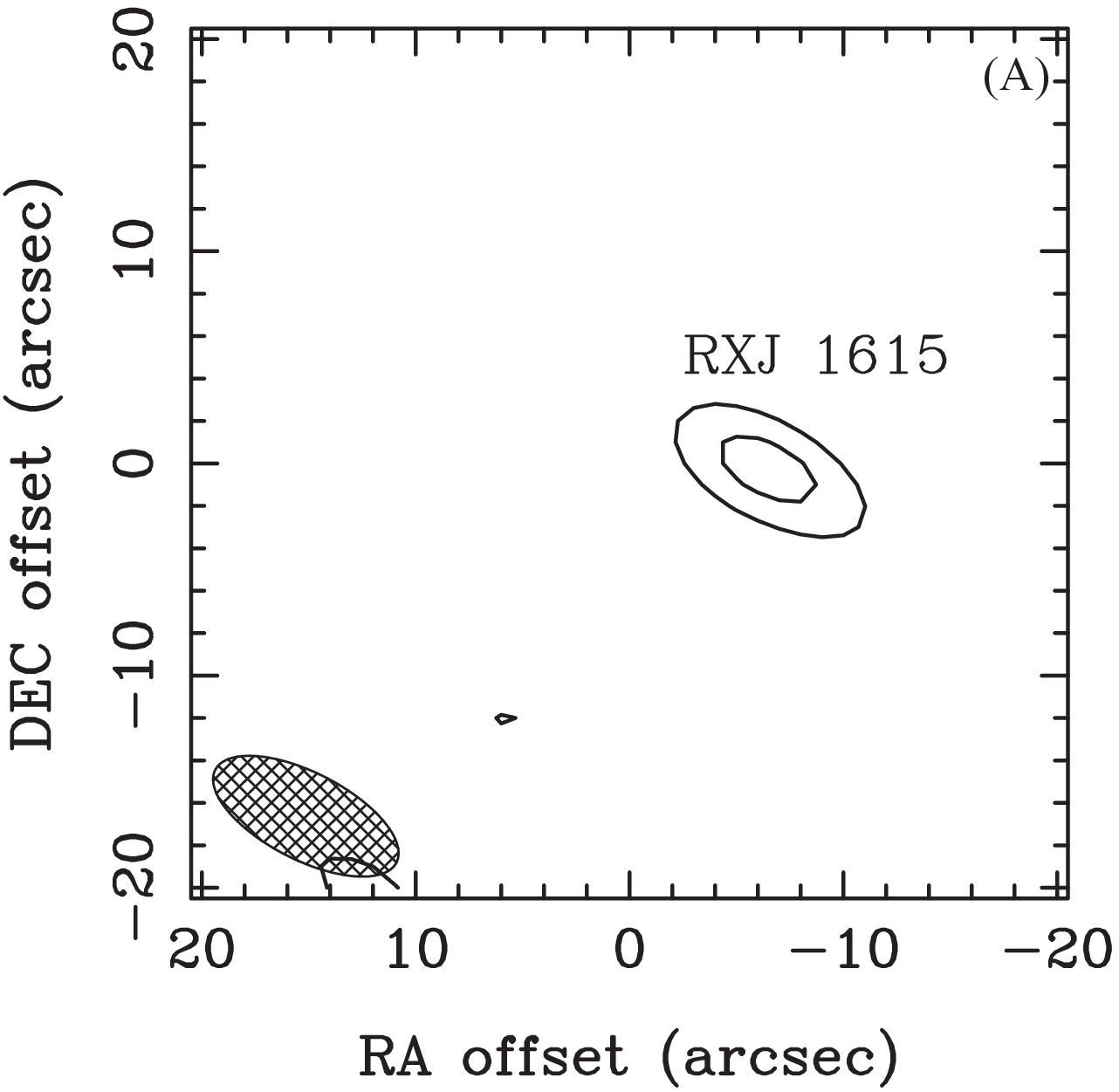}}
	\subfloat{\label{fig-sz6566map}\includegraphics[width=.75\columnwidth]{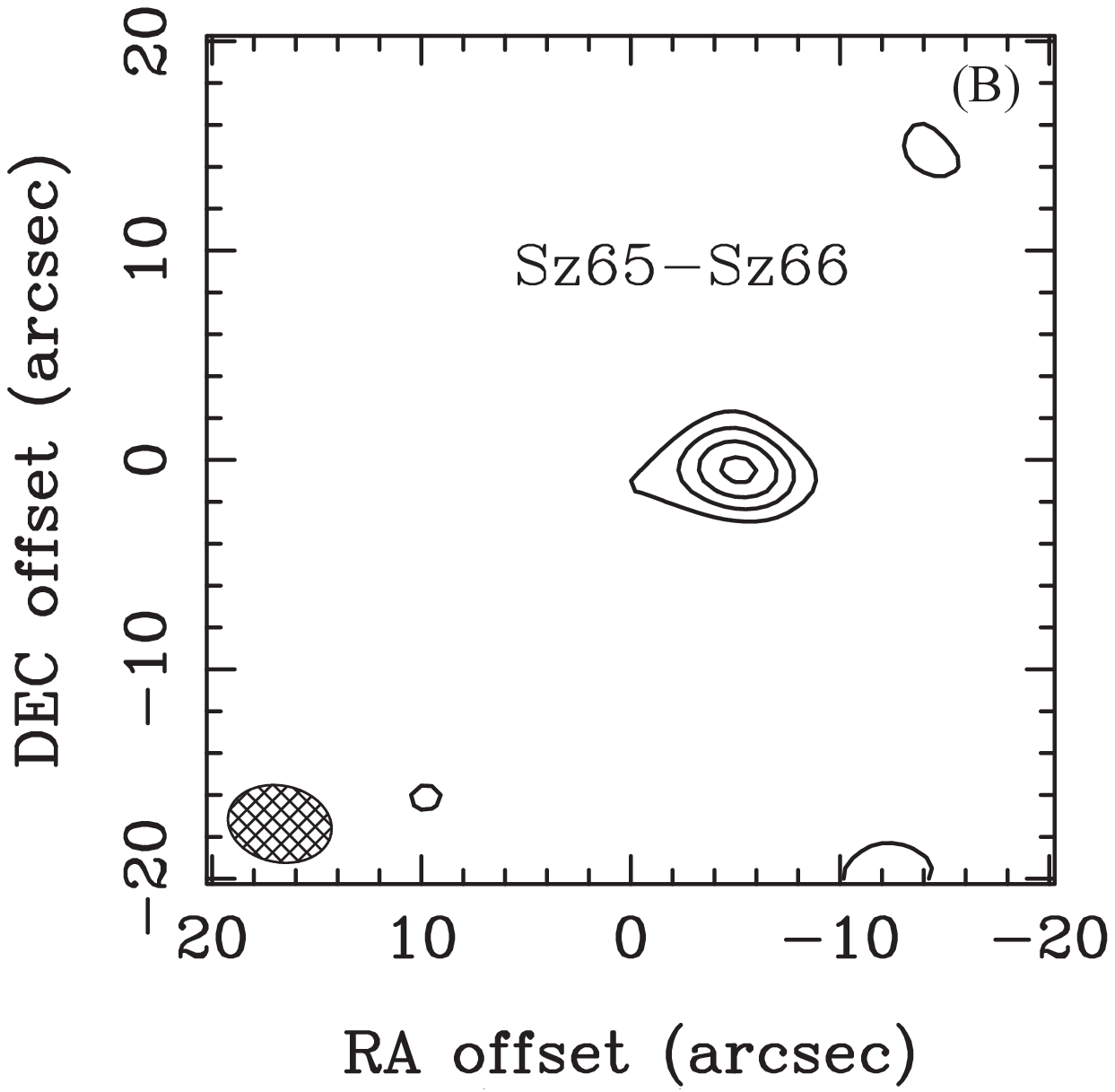}}\\
	\subfloat{\label{fig-mylup}\includegraphics[width=.75\columnwidth]{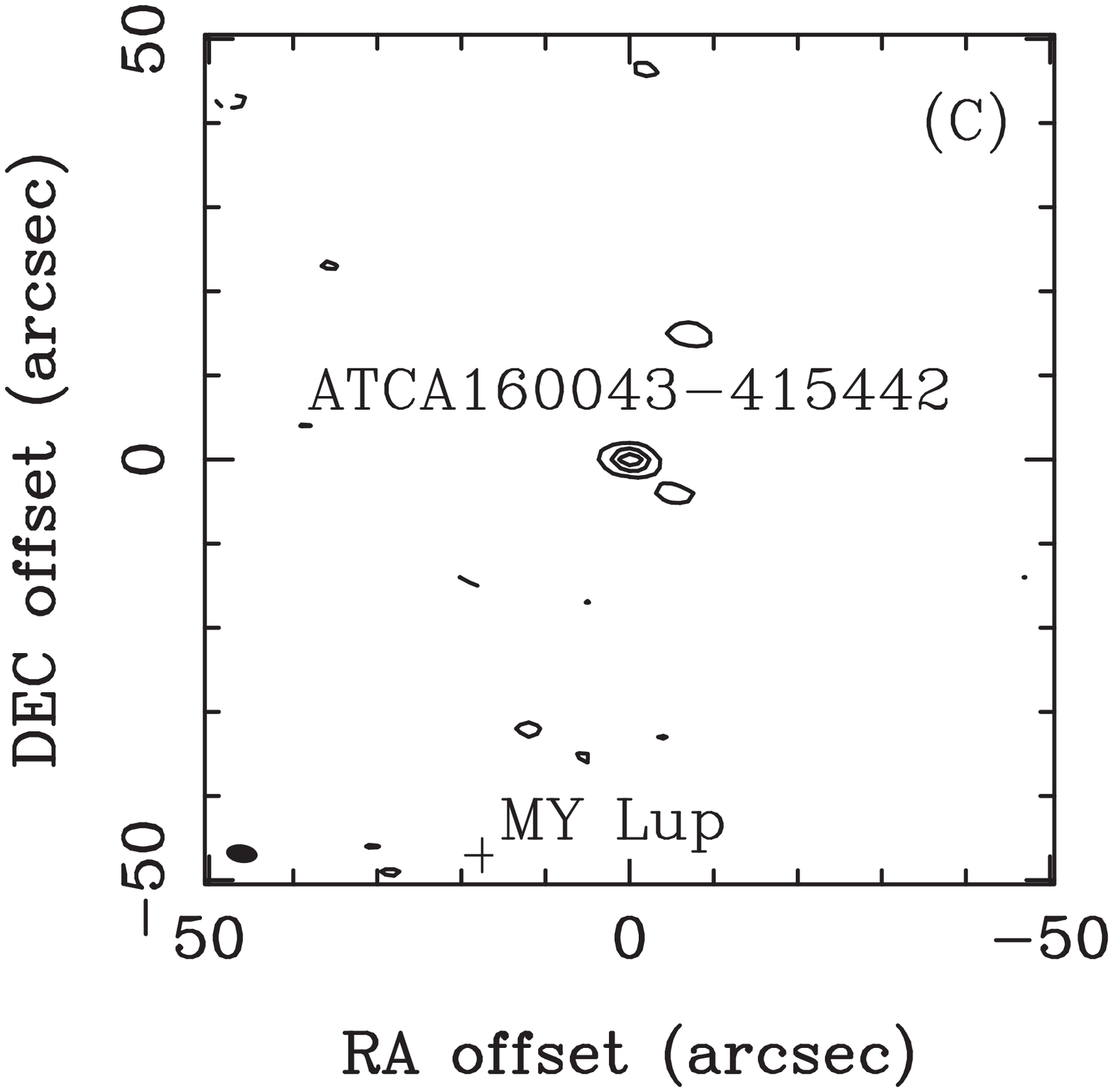}}
	\subfloat{\label{fig-mylup2}\includegraphics[width=.72\columnwidth]{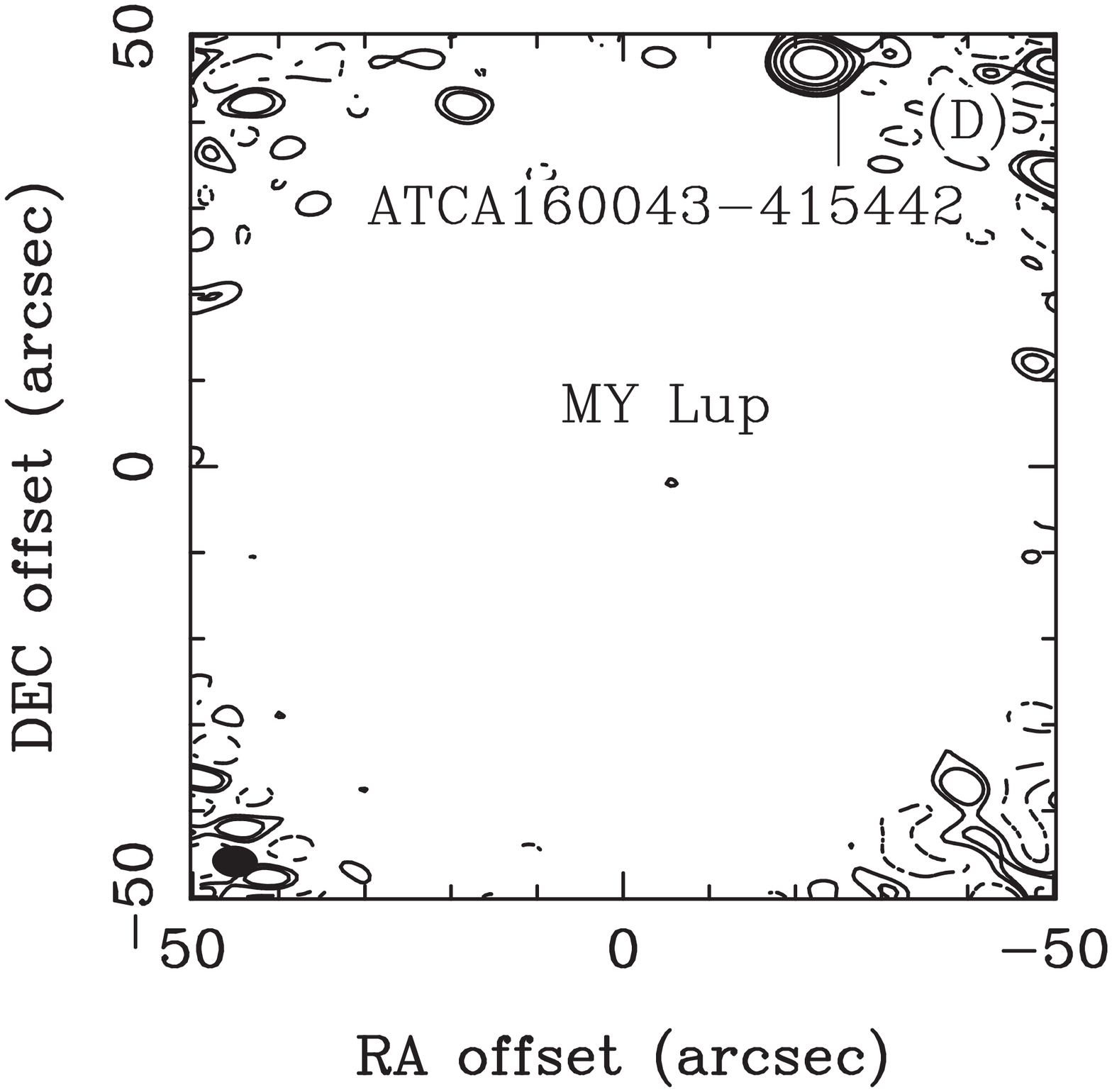}}
	\caption{The cleaned maps using natural weighting for RXJ1615.3-3255 at 7~mm, IK Lup at 7~mm, ATCA160043-415442 at 3~mm and MY Lup at {{7~mm.}} (A) RXJ1615.3-3255 {{at 7~mm}} with contours at -3,3,5 times the image RMS of 0.2 mJy/beam. (B) IK~Lup (Sz~65) at 7~mm with contours at -3,3,6,9,12 times the image RMS of 0.8 mJy/beam. Although the synthesised beam size of $4^{\prime\prime}.9\times3^{\prime\prime}.5$ was sufficient to resolved the binary with a separation of 6$^{\prime\prime}$.4, Sz 66 was not detected. (C) ATCA160043-415442 at 3~mm {{(with a cross in the position of MY Lup)}} with contours at {{-3,3,12,21,31}} times the image RMS of {{0.3}} mJy/beam. (D) {{Beam flux}} corrected clean map of MY Lup at 7~mm with ATCA160043-415442 in the north with contours at -3,3,5,11,22,45 times the RMS of 0.3 mJy/beam.}
\label{fig-maps}
\end{figure*}

\subsection{Source fluxes}
\label{subsec-results-fluxes}
	A summary of all the continuum fluxes obtained in this survey are presented in Table~\ref{tab-results2}. {{The flux values were obtained by combining both sidebands for all available days.}} The Table is supplemented with 1.2~mm {{fluxes}} for Chamaeleon sources from \citet{H1993} with SEST, and for Lupus sources from  \citet{Lommen10} and \citet{Dai10} with the Submillimetre Array (SMA)\footnote{{The Submillimetre Array}
~is a joint project between the Smithsonian Astrophysical Observatory and the Academia Sinica Institute of Astronomy and Astrophysics and funded by the Smithsonian Institution and the Academia Sinica.}, as well as some 3~mm data from \citet{Lommen07,Lommen10} with ATCA {{-- see Table~\ref{tab-sources} for a list of sources observed and wavelengths used in this survey}.} Four of our Chamaeleon sources have 870~$\mu$m LABOCA\footnote{LABOCA: LArge BOlometer CAmera -- service at the 12-m APEX telescope} data  \citep{2011A&A...527A.145B}, however for consistency we will use the SEST {{flux}} data for our analysis of the Chamaeleon sources.

	All seven sources observed at 3~mm in this survey were detected, and are consistent with the previous upper limits reported in \citet{Lommen07,Lommen10}. We present for the first time observations of the sources at 7~mm (for exceptions see Table~\ref{tab-monitoring2}) and beyond, 18/20 sources at 7~mm, 4/6 sources at 15~mm and 1/3 sources at 3+6~cm were detected. A 3$\sigma$ upper limit is provided for all non-detections. 

If the source was considered resolved in Section~\ref{subsec-resolved} or in the literature, the Gaussian fit for the flux is presented in Table~\ref{tab-results2} along with the flux uncertainties in the fits, the RMS values, and the beam size for observations made in this survey. The primary flux calibration uncertainties are not included in Table~\ref{tab-results2}. 
This uncertainty is $\sim$20\% at 1.2~mm for SEST data and $\sim$10\% for SMA data. For the ATCA data this uncertainty is $\sim$30\% at {{3~mm and}} $\sim$10\% at 7--15~mm and 3+6~cm bands.
	
	The beam size of $4^{\prime\prime}.85\times3^{\prime\prime}.53$ at 7~mm was sufficient to resolve the binary pair IK~Lup and Sz~66 with a separation of 6$^{\prime\prime}$.4 \citep{1993A&A...278...81R}. However, Sz~66 was not detected -- see {{Fig.~\ref{fig-sz6566map}}}. Thus the 7~mm fluxes in Tables~\ref{tab-results} are most likely only from IK Lup. Similar results were reported by \citet{Lommen10} for the 1.2 and 3.2~mm observations, with beam sizes of $5^{\prime\prime}.0\times2^{\prime\prime}.1$ and 2$^{\prime\prime}$ respectively. In both cases the beam size was sufficient to resolve the binary, suggesting the 2.2~mJy flux reported by \citet{Lommen10} at 3.2~mm for Sz~66 is likely an over-estimate. If Sz~66 was {{not}} detected at 3.2~mm, a 3$\sigma$ upper limit of 1.2~mJy would be more appropriate. Sz 66 was detected with Spitzer and has a 10~$\mu$m silicate feature -- see {\citet{Lommen10}.}
~However, the lack of cold dust from 1.2 to 7~mm would suggest there is no circumbinary disc, and if Sz 66 has a cold dust disc its mass is too low to have been detected by our observations.

 \begin{landscape}
 \begin{table}
 \caption{Survey results. (1) Source name. (2) 1.2 mm continuum flux with image RMS in parenthesis. (3) 3~mm flux with image RMS in parenthesis. (4) Beam size at 3~mm. (5) 7 mm flux with image RMS in parenthesis. (6) Beam size at 7~mm. {{(7), (8), (9)}} Fluxes at {{15~mm, 3 and 6}}~cm with u-v plane RMS in parenthesis. (10) References: For 1.2~mm continuum: 1--Henning et al. (1993) with SEST, 2--\citet{Lommen10} -- {{with SMA}}, 3--\citet{Dai10} with SMA. For 3~mm continuum with ATCA: 4--\citet{Lommen07}, 5--\citet{Lommen10}, 6--this work.}
 \begin{tabular}{c|ccccc|ccc|c}
	 \hline \hline
	 (1) & (2) & (3) & (4) &(5) & (6) & (7) & (8) & (9) & (10) \\
	 \hline
	Source	&	F$_{\rm 1.2mm}$ (RMS)	& F$_{\rm 3mm}$ (RMS)	&	Beam Size	& F$_{\rm 7mm}$ (RMS) &	Beam Size	& F$_{\rm 15mm}$ (RMS) & F$_{\rm 3cm}$ (RMS) &F$_{\rm 6cm}$ (RMS) & References		\\
		&	mJy		(mJy/beam)		&	mJy				(mJy/beam)		&	arcsec	&	mJy				(mJy/beam)		&	arcsec	&	mJy (mJy/beam) & mJy (mJy/beam) & mJy (mJy/beam) &			\\
	\hline \hline																		
	\multicolumn{10}{c}{Chamaeleon} \\ 																			
	\hline																			
SY Cha	&	$<$172.0	&	$<$4.8				&	---			&	$<$0.9	(0.3)	&	---			&	---			&	---			&	---		&	1,4,6	\\
CR Cha	&	124.9 (24.2)	&	6.2	(1.5)			&	2.5$\times$2.1	&	1.7	(0.1)	&	5.0$\times$4.3	&	0.3	(0.1)		&	---			&	---		&	1,4,6,6	\\
\textbf{CS Cha}	&	128.4 (45.6)	&	9.4	(0.6)$^{G}$*			&	13.2$\times$1.4	&	1.5	(0.1)	&	5.0$\times$4.3	&	$<$0.3	(0.1)		&	---			&	---		&	1,6,6,6	\\
DI Cha	&	38.0 (11.4)	&	2.3	(0.3)*$^{a}$		&	4.6$\times$1.8	&	0.9	(0.1)	&	4.8$\times$4.2	&	$<$0.3	(0.1)		&	---			&	---		&	1,6,6,6	\\
T Cha	&	105.0 (17.7)	&	6.4	(1.0)			&	2.5$\times$2.4	&	3.0	(0.1)	&	5.1$\times$4.1	&	0.3	(0.1)		&	$<$0.3	(0.1)		&	0.3	(0.1)	&	1,4,6,6,6,6	\\
Glass I	&	69.9 (22.4)	&	4.4	(0.1)*			&	2.4$\times$1.8	&	0.7	(0.1)	&	4.9$\times$4.1	&	---			&	---			&	---		&	1,6,6	\\
\textbf{SZ Cha}	&	77.5 (20.3)	&	5.8	(0.5)$^{G}$		&	2.3$\times$2.1	&	0.7	(0.1)	&	5.0$\times$4.1	&	---			&	---			&	---		&	1,5,6	\\
Sz 32	&	93.1 (20.8)	&	3.1	(0.2)*			&	2.3$\times$1.8	&	1.2	(0.1)	&	5.0$\times$4.0	&	0.4	(0.1)		&	$<$0.3 (0.1)	&	$<$0.3	(0.1)	&	1,6,6,6,6,6\\
\textbf{DK Cha}	&	680.0 (22.0)	&	49.8	(1.3)$^{G}$*	&	6.4$\times$1.3	&	6.6	(0.6)	&	47.3$\times$2.5	&	1.6	(0.1)		&	0.8 (0.1)		&	$<$0.3	(0.1)	&	1,6,6,6,6,6\\
	\hline																			
	\multicolumn{10}{c}{Lupus} \\																			
	\hline																			
IK Lup	&	28.0 (2.8)	&	3.4	(0.4)			&	2.1$\times$1.7	&	0.9	(0.1) &	4.5$\times$3.5	&	---	&	---	&	---	&	2,5,6	\\
Sz 66	&	$<$8.0 (2.8)	&	2.2	(0.4)$^{b}$			&	2.1$\times$1.7	&	$<$0.3 (0.1)		&4.5$\times$3.5	&	---	&	---	&	---	&	2,5,6	\\
\textbf{HT Lup}	&	73.0 (4.0)	&	12.0	(1.1)$^{G}$		&	2.4$\times$1.7	&	3.4	(0.1)	&4.4$\times$3.5	&	---	&	---	&	---	&	4,4,6	\\
GQ Lup	&	25.0 (3.0)	&	{{3.6}}	(0.3)			&	{{2.8$\times$2.0}}	&	{{0.6}}	(0.1)	& {{6.7$\times$4.4}}	&	---	&	---	&	---	&	3,6,6	\\
GW Lup	&	64.0 (3.7)	&	8.5	(1.9)			&	5.3$\times$1.7	&	0.6	(0.1)&	4.8$\times$3.5	&	---	&	---	&	---	&	4,4,6	\\
\textbf{RY Lup}	&	89.0 (4.9)$^{G}$	&	2.8	(0.7)$^{G}$		&	2.2$\times$1.9	&	1.0	(0.1)	&4.7$\times$3.5	&	---	&	---	&	---	&	2,5,6	\\
HK Lup	&	101.0 (3.9)$^{G}$	&	7.3	(2.1)			&	---			&	1.0	(0.1)&	5.2$\times$3.5	&	---	&	---	&	---	&	4,4,6	\\
\textbf{Sz 111}	&	52.5 (4.2)$^{G}$	&	5.7	(0.7)$^{G}	$	&	2.1$\times$1.9	&	0.5	(0.1)	&5.2$\times$3.5	&	---	&	---	&	---	&	5,5,6	\\
EX Lup	&	21.3 (4.0)$^{G}$	&	{{2.0}}	(0.3)			&	{{2.9}}$\times$1.9	&	1.4	(0.1)&	4.9$\times$3.6	&	---	&	---	&	---	&	2,6,6	\\
MY Lup	&	66.1 (3.4)$^{G}$	&	8.7	(0.4)			&	2.0$\times$1.7	&	1.1	(0.1)&	4.8$\times$3.4	&	---	&	---	&	---	&	2,5,6	\\
RXJ1615.3-3255	&	169.1 (3.9)$^{G}$	&	6.7	(0.6)			&	2.0$\times$1.7	&	{{0.8}}	(0.2)&	9.5$\times$4.1	&	---	&	---	&	---	&	2,5,6	\\
\hline
 \end{tabular}
    \label{tab-results2}		
       	\begin{tablenotes} 
		\item[1] A 3$\sigma$ upper limit is provided for non-detections. 
		\item[2] Sources in \textbf{boldface} were resolved at 3~mm with ATCA.
	      	\item[3] $^{a}$ Flux at 93~GHz, as 95~GHz was not detected.
		\item[4] $^{b}$ Potentially an over estimate -- see Section~\ref{subsec-results-fluxes}.
		\item[5] $^{*}$ 3~mm continuum fluxes from this work. All 7~mm fluxes are from this work. 
		\item[6] $^{G}$ {{Gaussian flux fit for sources resolved at 1~mm with SMA and 3~mm with ATCA.}}
	\end{tablenotes}	
  \end{table}
\end{landscape}

\subsection{Millimetre spectral slopes}
\label{subsec-results-mmslopes}

The mm spectral slopes $\alpha$ and  $\alpha_{3-7}$ presented in Table~\ref{tab-alphas} were determined using the 1, 3 and 7~mm band fluxes from Table~\ref{tab-results2}. {{The uncertainties in $\alpha$ and $\alpha_{3-7}$ were calculated through propagation of the flux uncertainties presented in Table~\ref{tab-results}.}} Histogram of the distribution of $\alpha$ values is presented in Fig.~\ref{fig-alpha-hist}. The values of $\alpha$ range between 2 to 4. {{The grain size distribution in disc models evolves and can depart substantially from the $n(a) \propto a^{-3.5}$ size distribution of the interstellar medium \citep[e.g,][]{1997Icar..127..290W,2011A&A...525A..11B}.}} {Since fully modelling the grain size distribution {{for all 20 sources in our survey}} is beyond the scope of this paper, for this analysis we will assume the dust emissivity can be represented by a single power law from mm through {{to}} cm wavelengths. Thus if only thermal dust emission is contributing to the flux from 1 to 7~mm the spectral slope would remain constant, whilst an inequality {{between $\alpha$ and  $\alpha_{3-7}$}} would imply a break in the spectral slope, suggestive of the presence of other emission mechanisms such as thermal free-free and non-thermal emissions contributing to the flux at the longer wavelengths. }

{{For this analysis we will consider $\alpha$ and  $\alpha_{3-7}$ to be consistent (no break at 7~mm) when  $\alpha_{3-7}\pm\Delta\alpha_{3-7} = \alpha\pm\Delta\alpha$ and inconsistent (a break at 7~mm) when $\alpha_{3-7}\pm\Delta\alpha_{3-7} < \alpha\pm\Delta\alpha$. The uncertainty $\Delta\alpha$ is 0.4 for all sources and $\Delta\alpha_{3-7}$ is given in Table~\ref{tab-alphas}.}}

The histogram $\alpha$ values show that majority of Lupus sources have a value of $\alpha<3$, while majority of Chamaeleon sources have a value $\alpha>3$ -- see Fig.~\ref{fig-alpha-hist}. From Table~\ref{tab-alphas} we determined that 11 sources do not have a break in their spectral slope at 7~mm, while 7 sources do have a break and two are undetermined.

We plot the flux as a function of wavelength in Figs.~\ref{fig-3pts-sed-cham} and~\ref{fig-3pts-sed-lup} for Chamaeleon and Lupus respectively, to better study the spectral slope. The solid line represents the spectral slope $\alpha$ presented in Table~\ref{tab-alphas} and the dashed-dot lines are the upper and lower limits for $\alpha$ using the flux fit uncertainties and assuming the standard primary flux calibration uncertainties (20\% for 1.2~mm, 30\% for 3~mm, and 10\% for 7, 15~mm and 3, 6~cm). Thus the shaded region represents the flux range expected when thermal dust emission dominants. The dashed line representing free-free emission with $\alpha_{\rm{ff}} = 0.6$ {{\citep{1975A&A....39....1P}}} anchored at the {{15~mm data point is}} included for sources with cm data. Full SEDs can be found in Figs.~\ref{fig-chamSED} and~\ref{fig-lupSED}. 

CR Cha, DI Cha, T Cha, Sz 32, RY Lup, EX Lup and RXJ1615.3-3255 all have 7~mm fluxes in excess of purely thermal dust emission {consistent with Table~\ref{tab-alphas}} -- see Figs.~\ref{fig-3pts-sed-cham} and~\ref{fig-3pts-sed-lup}.
 Note that if we were to use the 870~$\mu$m fluxes obtained with LABOCA by \citet{2011A&A...535A...2B} for the three available Chamaeleon sources instead of the SEST data, the $\alpha$ values would change to 2.3$\pm$0.2 for CS Cha, 2.9$\pm$0.2 for Glass I and 3.1$\pm$0.4 for SZ Cha, which would not change the interpretation of our analysis of these sources.

{{For those sources with observations at 15~mm and longer wavelengths, we can see that all}} six Chamaeleon sources observed {{at 15~mm have excess}} emission above {{that expected from}} thermal dust. {At 3+6~cm, DK Cha, T Cha and Sz 32 the fluxes and upper limits suggest the presence of free-free emission from an ionised wind and/or chromospheric emission, which will be discussed further in the next Section.}

 \begin{table}
 \centering
 \caption{{Spectral slopes {{from 1-3~mm and 3-7~mm}}. (1) Source name. (2) {{Spectral slope from 1-3~mm,}} $\alpha$ (with an uncertainty of 0.4). (3) {{Spectral slope from 3-7~mm,}} $\alpha_{3-7}$. (4) {{Uncertainties in $\alpha_{3-7}$}}. (5) Indication of whether there is a break in the spectral slopes at 7~mm.}}
   \begin{tabular}{ccccc}
  \hline \hline
	Sources &	$\alpha$	&	$\alpha_{3-7}$	& $\Delta\alpha_{3-7}$& Break \\
	\hline \hline																			
	\multicolumn{5}{c}{Chamaeleon} \\ 	
	\hline
	SY Cha*&	$>$3.9	&	$>$2.1	&	0.5	&	--	\\
	CR Cha&	3.4	&	1.7	&	0.4	&	Y	\\
	\textbf{CS Cha}&	2.9	&	2.5	&	0.3	&	N	\\
	DI Cha&	3.2	&	1.2	&	0.3	&	Y	\\
	T Cha&	3.1	&	1.1	&	0.3	&	Y	\\
	Glass I&	3.1	&	2.5	&	0.4	&	N	\\
	\textbf{SZ Cha}&	2.9	&	2.9	&	0.5	&	N	\\
	Sz 32&	3.8	&	1.3	&	0.3	&	Y	\\
	\textbf{DK Cha}&	2.9	&	2.7	&	0.2	&	N	\\
	\hline																			
	\multicolumn{5}{c}{Lupus} \\ 	
	\hline
	IK Lup&	2.4	&	1.7	&	0.4	&	N	\\
	Sz 66*&	$>$1.4	&	$>$3.3	&	0.3	&	--	\\
	\textbf{HT Lup}&	2.0	&	1.7	&	0.4	&	N	\\
	GQ Lup	& {{2.2}}	&	{{2.4}}	&	0.5	&	N	\\
	GW Lup	&2.3	&	3.6	&	0.4	&       N	\\
	\textbf{RY Lup}&	3.9	&	1.4	&	0.4	&	Y	\\
	HK Lup	&2.9	&	2.7	&	0.4	&	N	\\
	\textbf{Sz 111}	&2.5	&	3.3	&	0.5	&	N	\\
	EX Lup&	2.6	&	0.5	&	0.2	&	Y	\\
	MY Lup&	2.3	&	2.8	&	0.5	&	N	\\
	RXJ1615.3-3255	&3.6	&	{{2.8}}	&	0.5	&	Y	\\
	\hline
\end{tabular}
\label{tab-alphas}
	\begin{tablenotes} 
		\item[1] * SY Cha and Sz 66 only have upper limits at 1.2 and 3~mm. 
		\item[2] The sources in \textbf{bold} were resolved at 3~mm.
	\end{tablenotes}
\end{table}

\begin{figure}
   \centering
    \subfloat{\label{fig-alpha-hist13}\includegraphics[width=4.1cm]{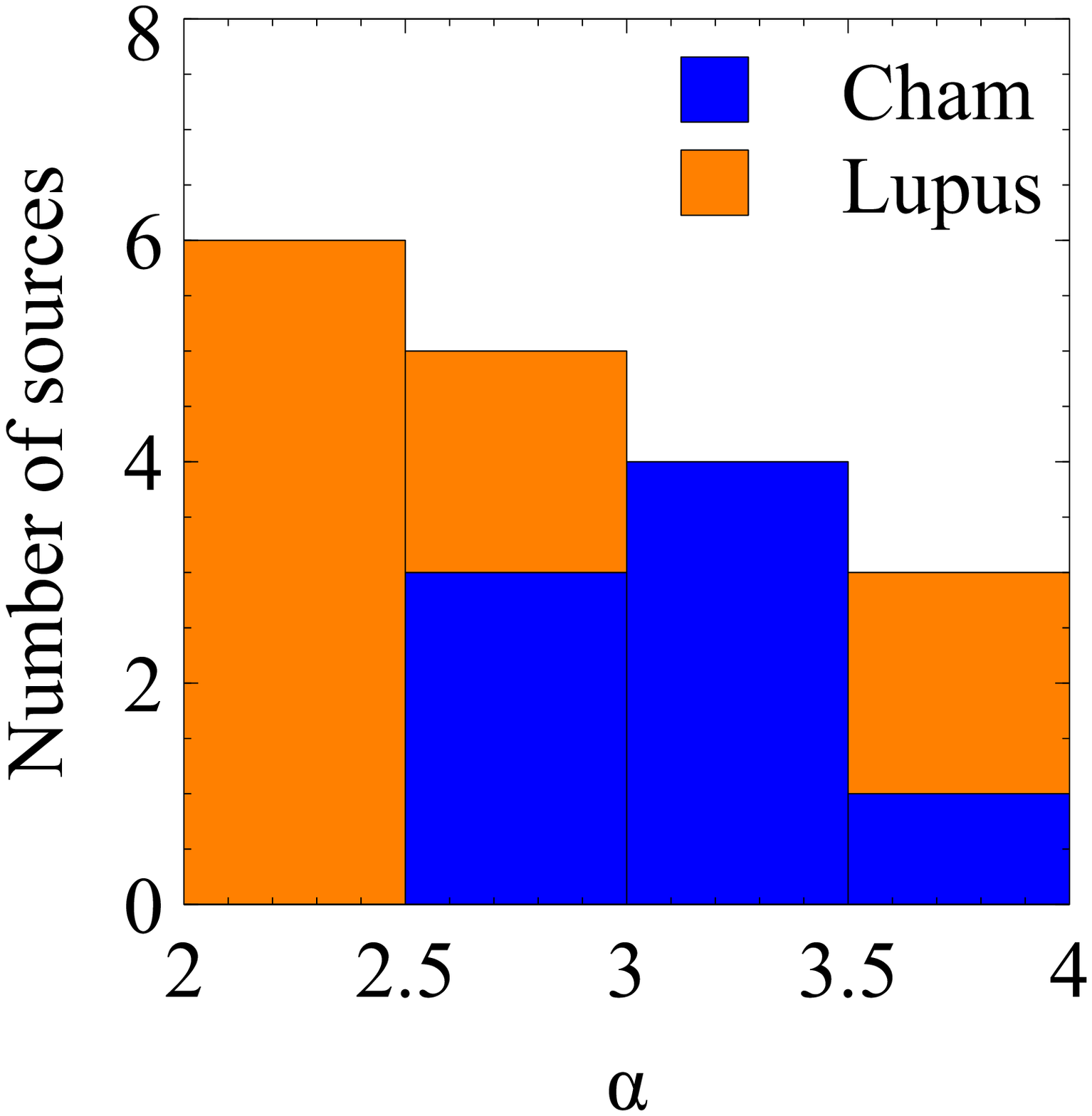}}
    \caption{Histogram of the spectral slopes $\alpha$. SY~Cha and Sz~66 spectral slopes are unconstrained due {{non-detection at 1.2 and 3~mm}}, thus both are excluded from the analyses.}
   \label{fig-alpha-hist}
\end{figure} 

\begin{figure*}
	\subfloat{\label{fig:crcha3}\includegraphics[width=.68\columnwidth]{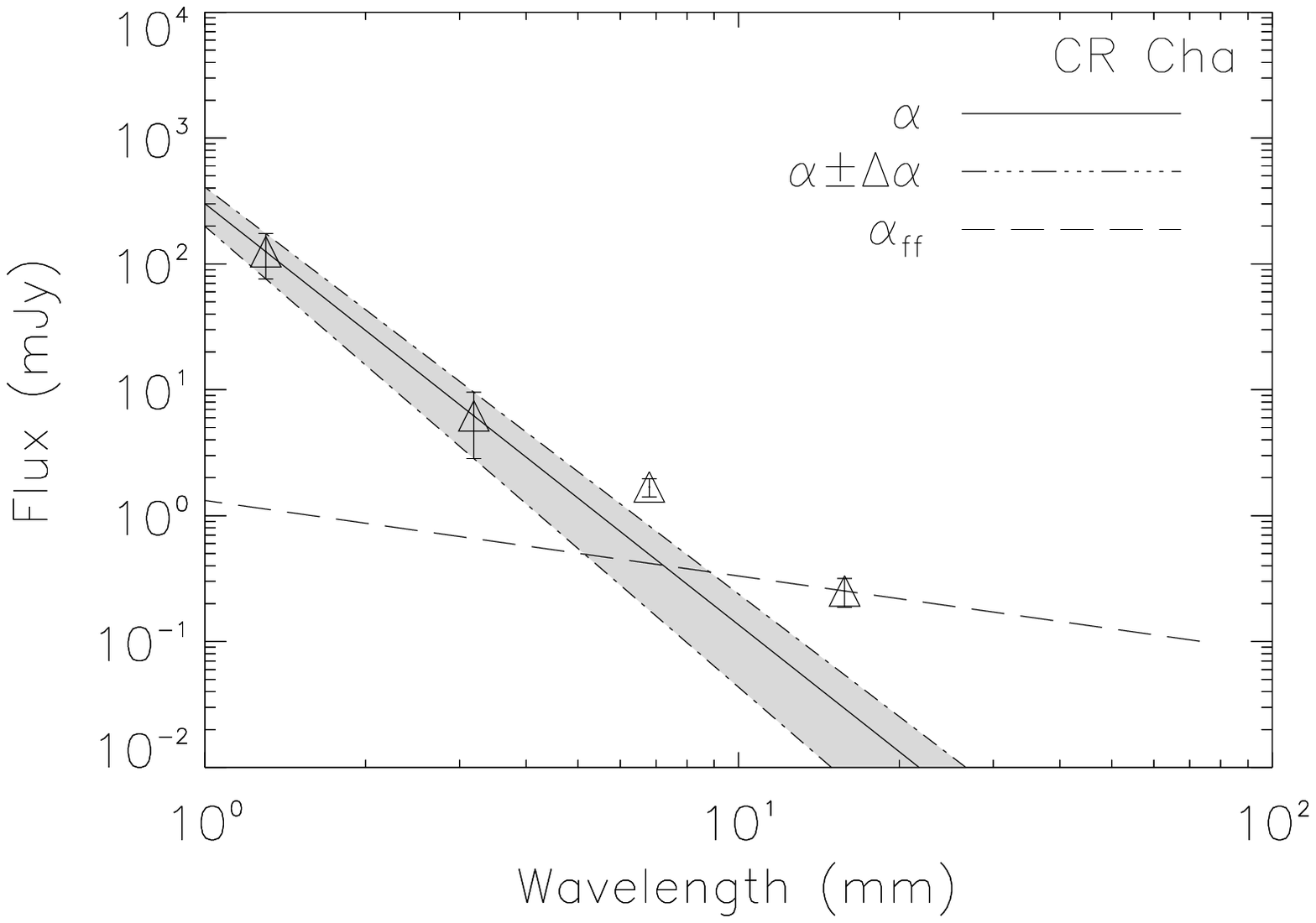}}  
	\subfloat{\label{fig:cscha3}\includegraphics[width=.68\columnwidth]{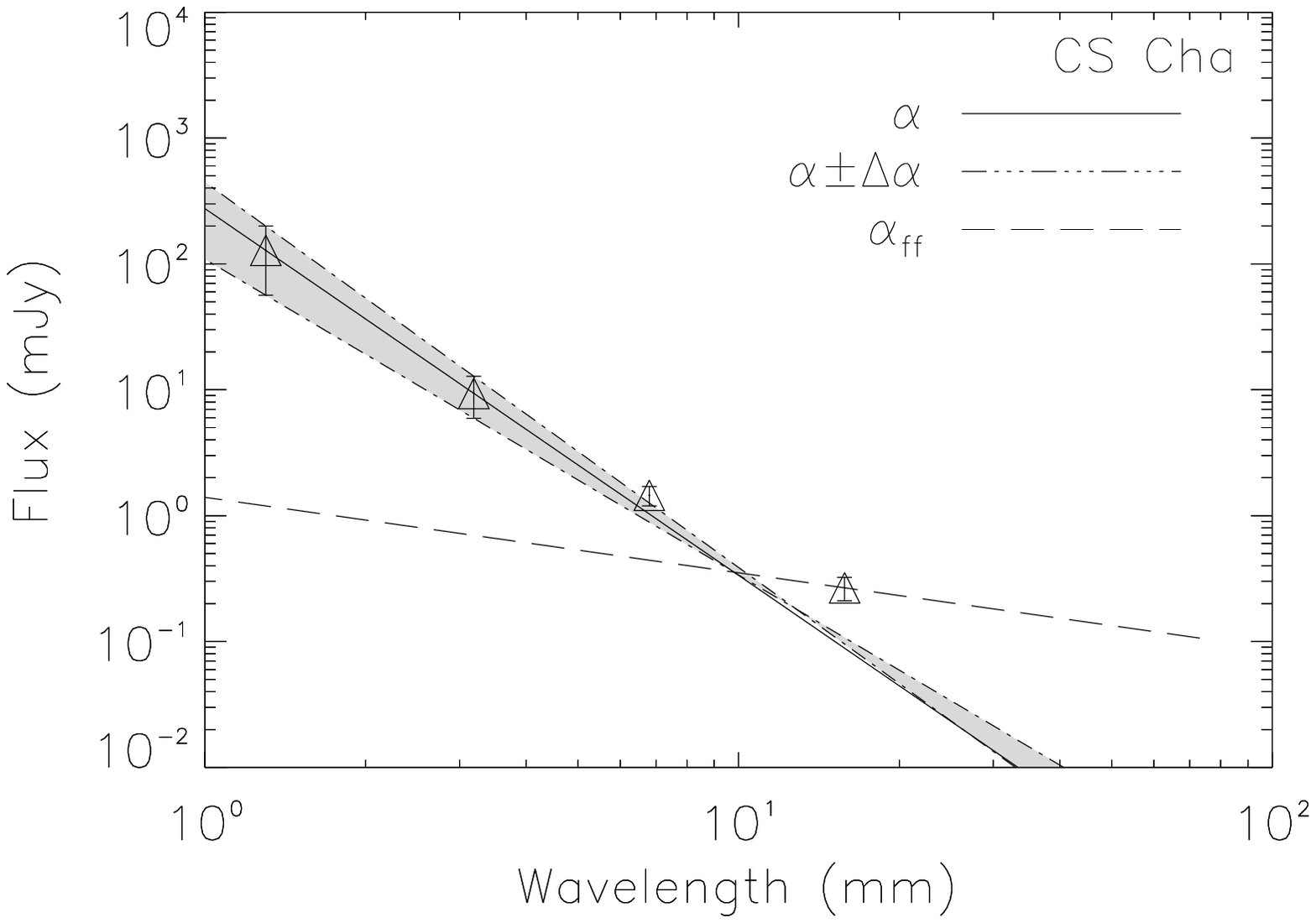}}  \\    
	\subfloat{\label{fig:dicha3}\includegraphics[width=.68\columnwidth]{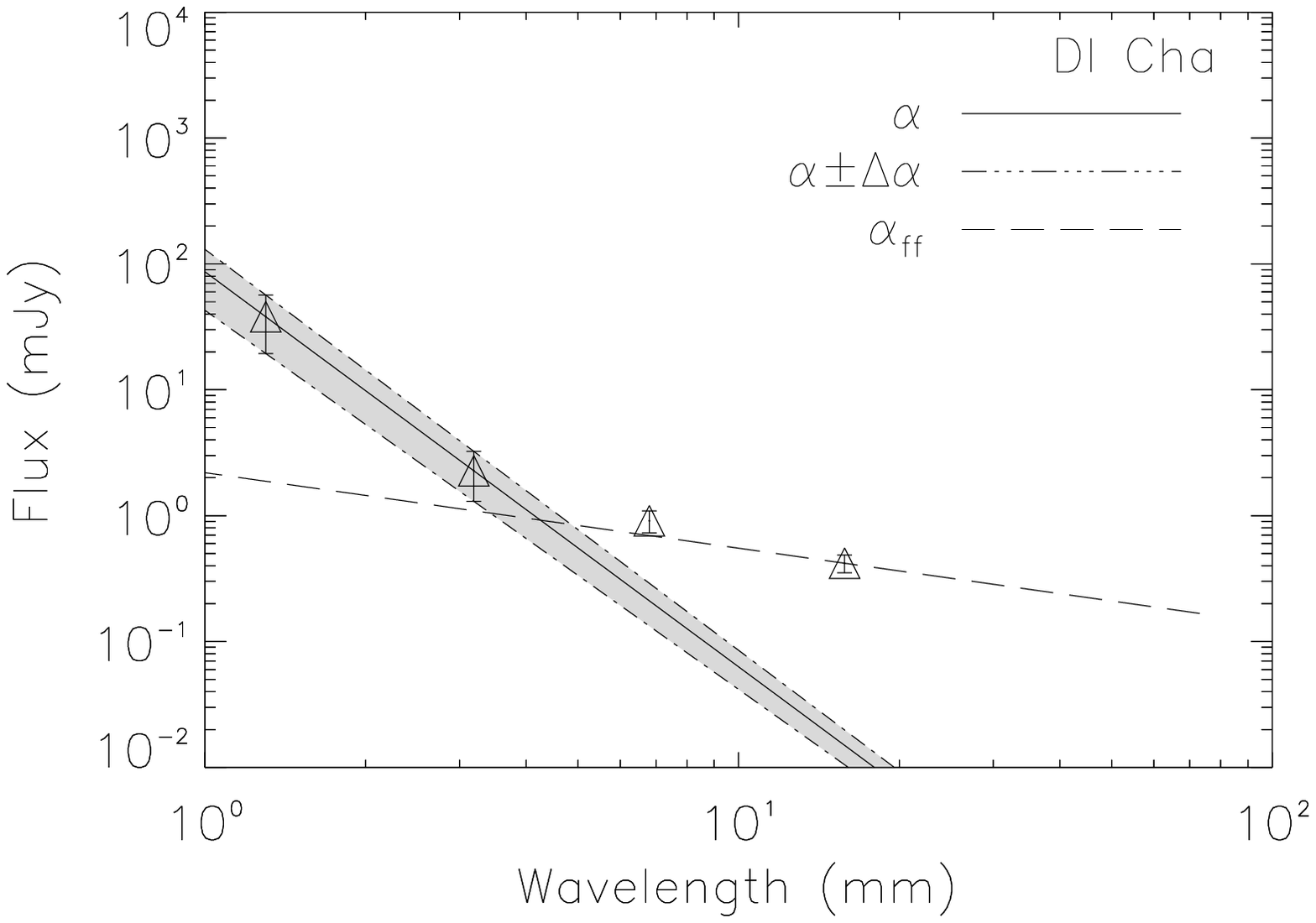}} 
	\subfloat{\label{fig:tcha3}\includegraphics[width=.68\columnwidth]{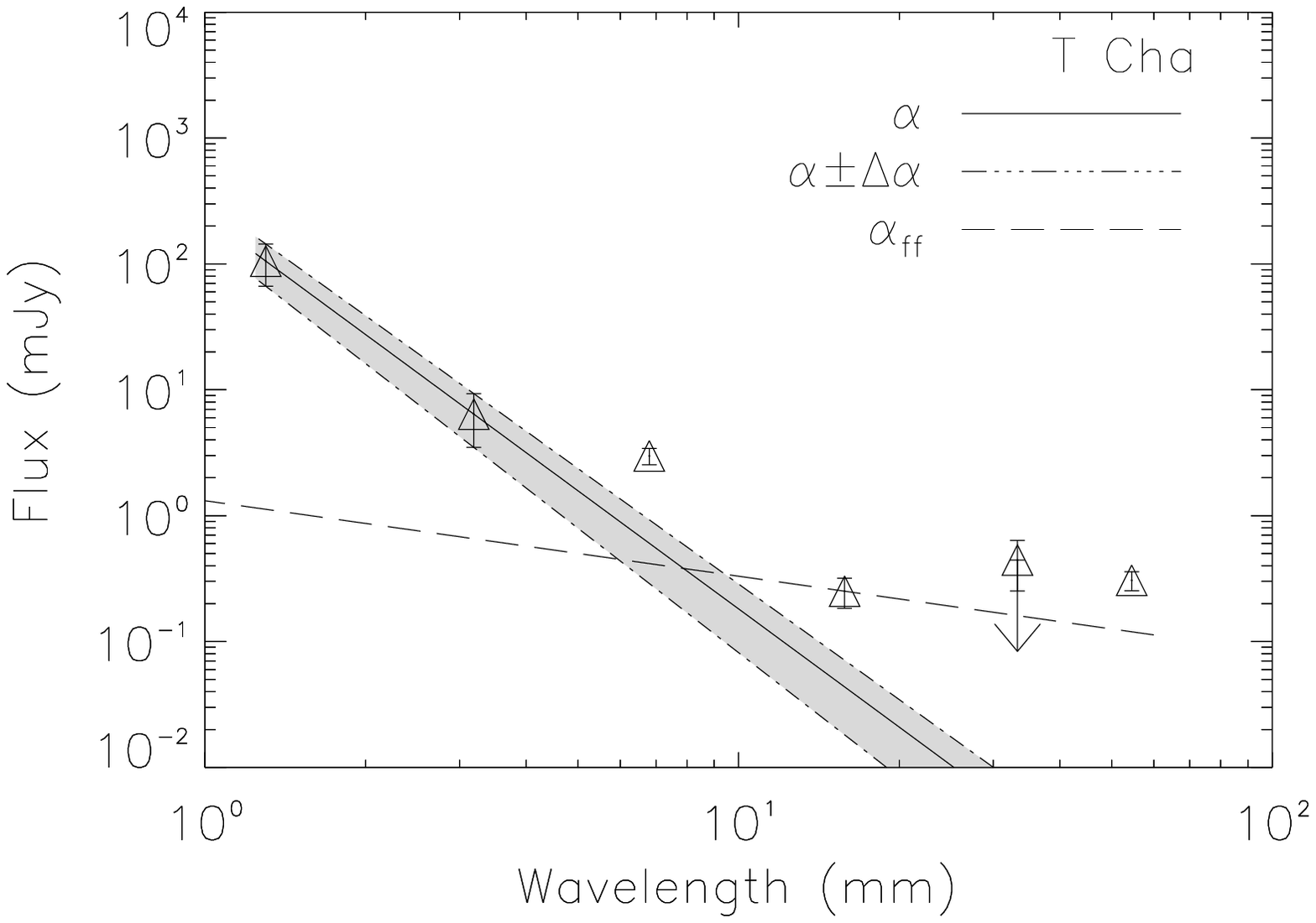}}
	\subfloat{\label{fig:glassi3}\includegraphics[width=.68\columnwidth]{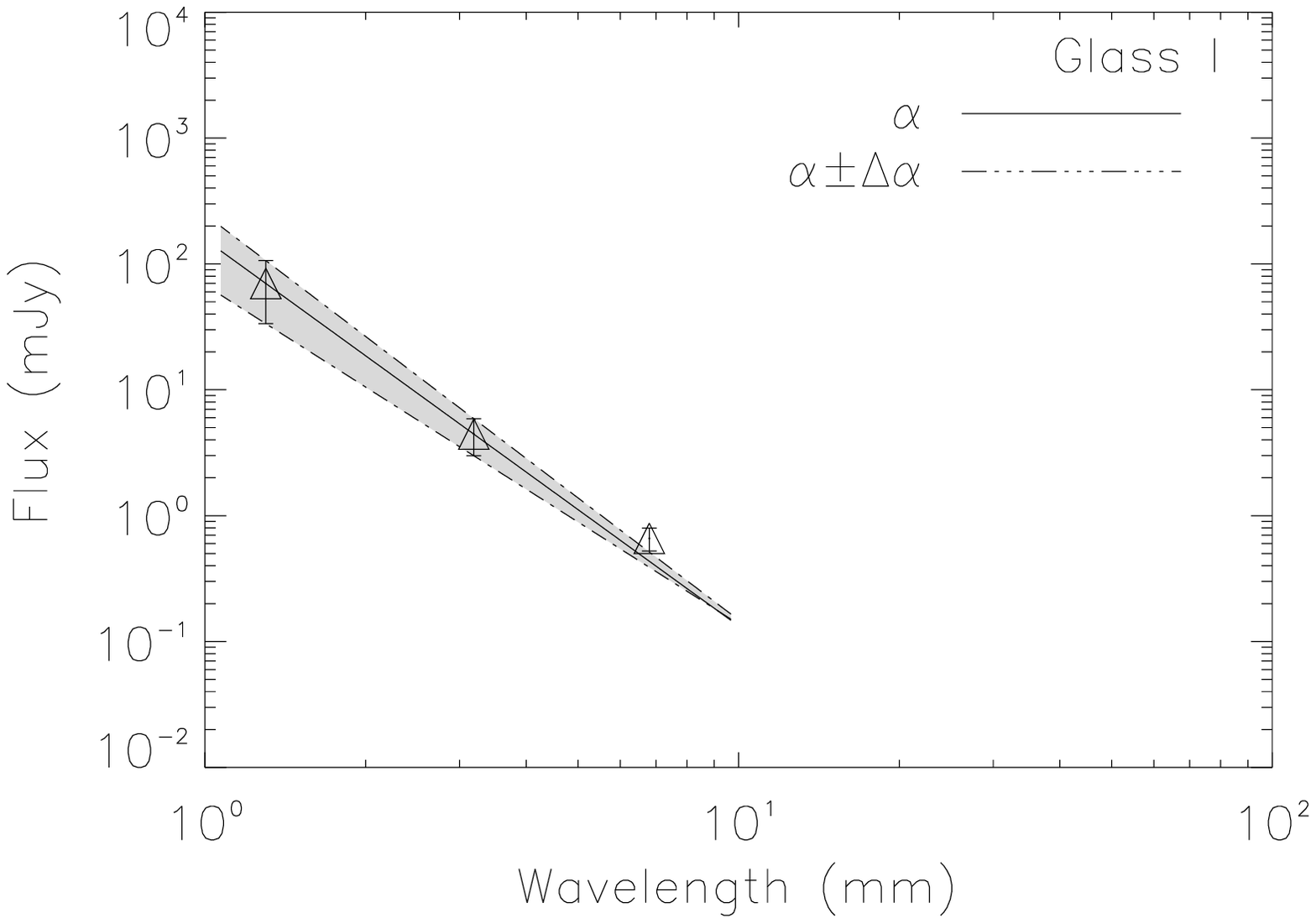}} \\ 
	\subfloat{\label{fig:szcha3}\includegraphics[width=.68\columnwidth]{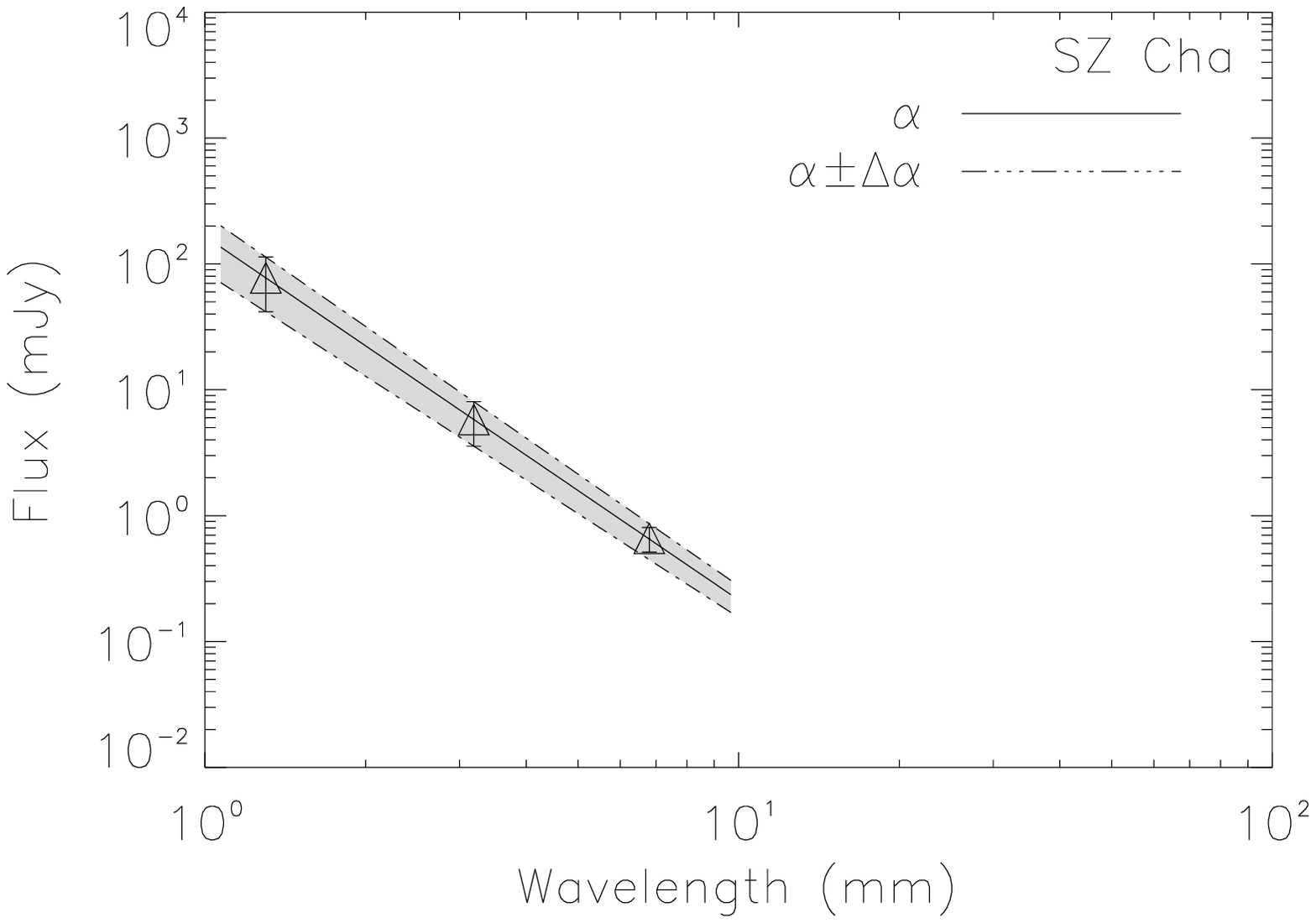}} 
	\subfloat{\label{fig:sz323}\includegraphics[width=.68\columnwidth]{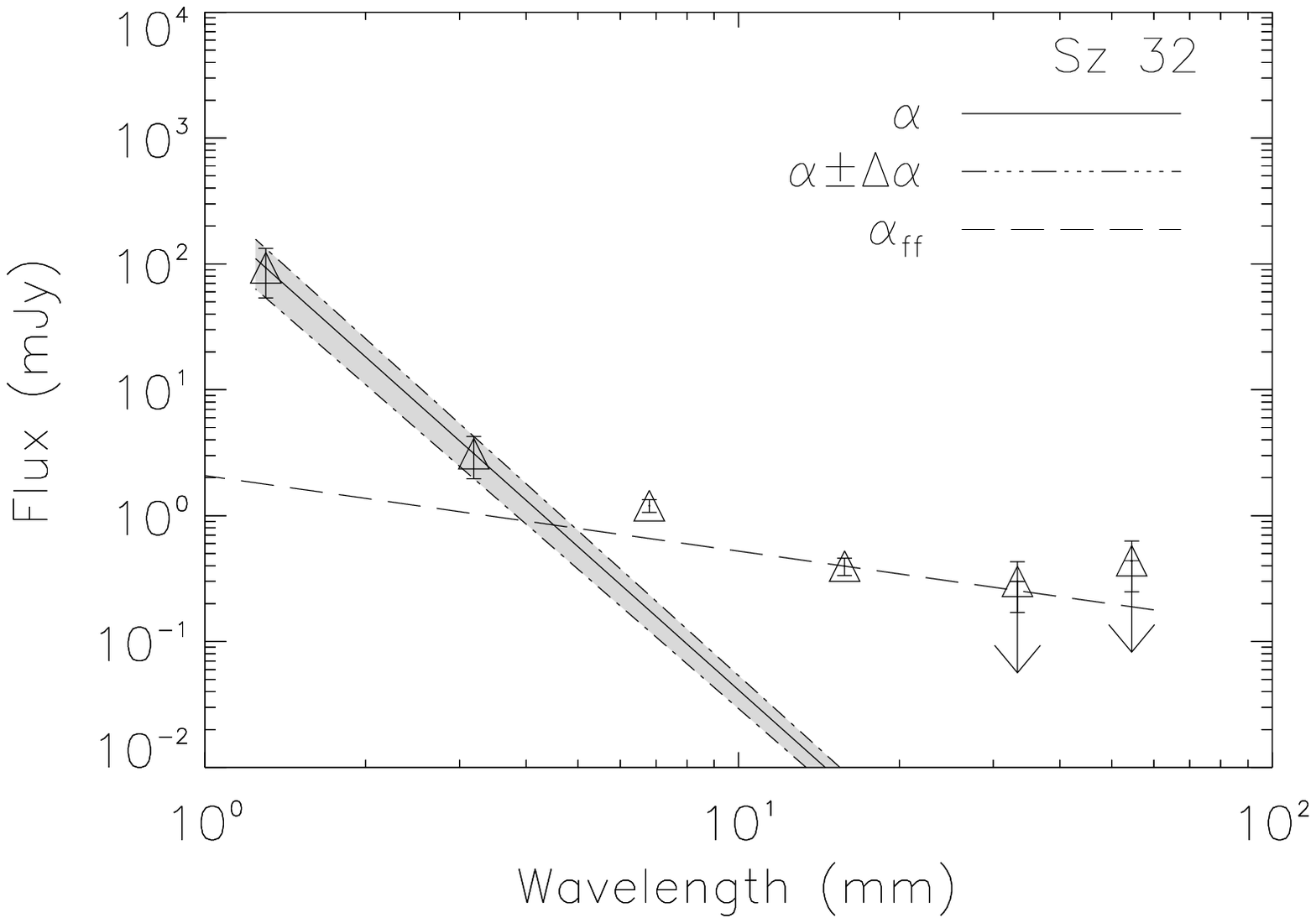}} 
	\subfloat{\label{fig:dkcha3}\includegraphics[width=.68\columnwidth]{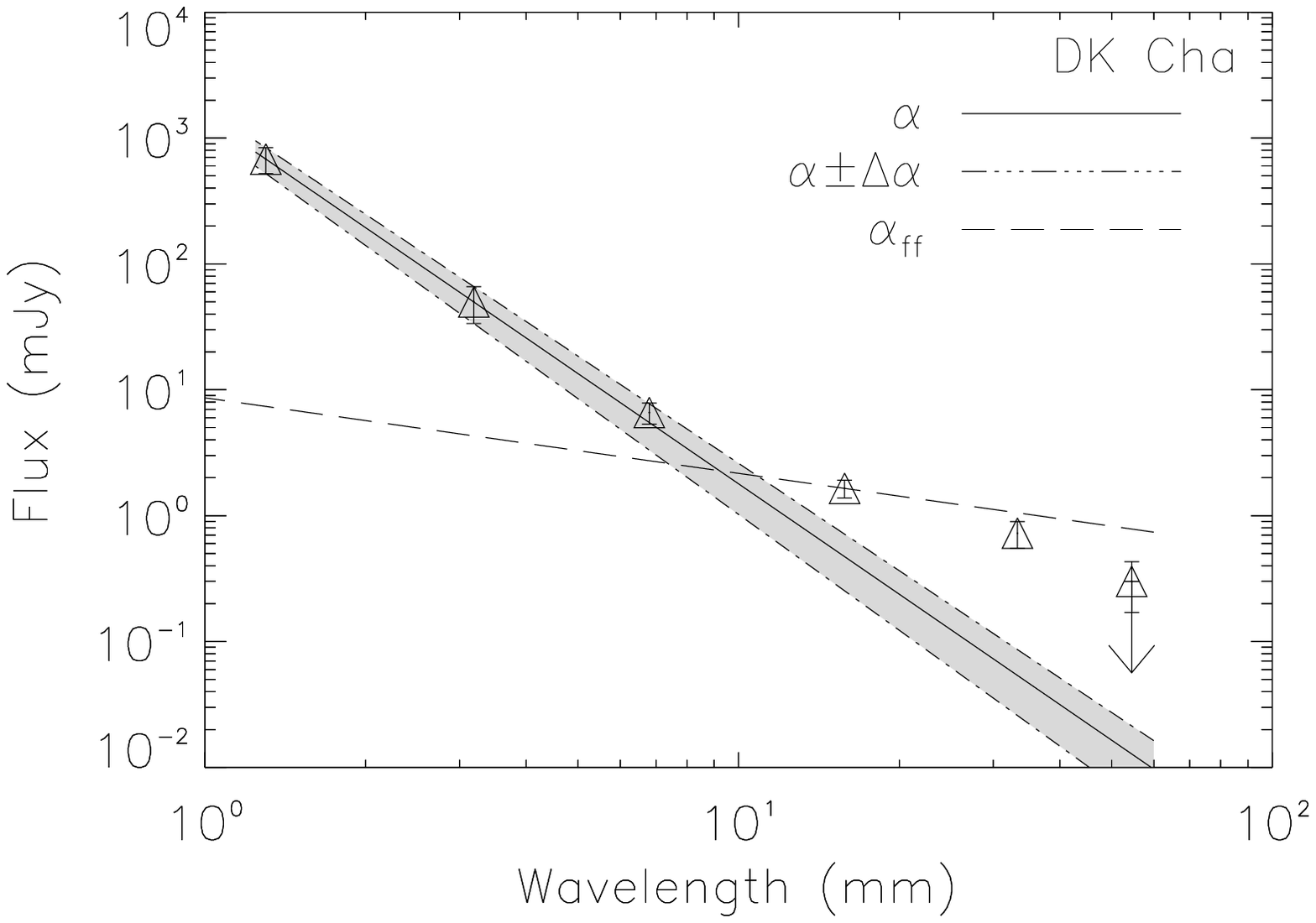}} 
	\caption{Millimetre {{flux}} versus wavelength for Chamaeleon sources. 
	{The solid line represents the spectral slope $\alpha$ and the dashed-dot lines are the upper and lower limits for $\alpha$ using the flux fit uncertainties and assuming the standard primary flux calibration uncertainties (20\% for 1.2~mm, 30\% for 3~mm, and 10\% for {{7 and 15~mm and 3+6~cm)}}. The dashed line represents the free-free emission with $\alpha_{\rm{ff}} = 0.6$ {{anchored at the 15~mm data}} point, is included for sources with cm data. {{Fluxes values from Table~\ref{tab-results2}.}} Note that the error bars include the flux uncertainty from the fitting routine and the primary flux calibration uncertainties.}
~}
	\label{fig-3pts-sed-cham}
\end{figure*}

\begin{figure*}
	\centering
	\subfloat{\includegraphics[width=.68\columnwidth]{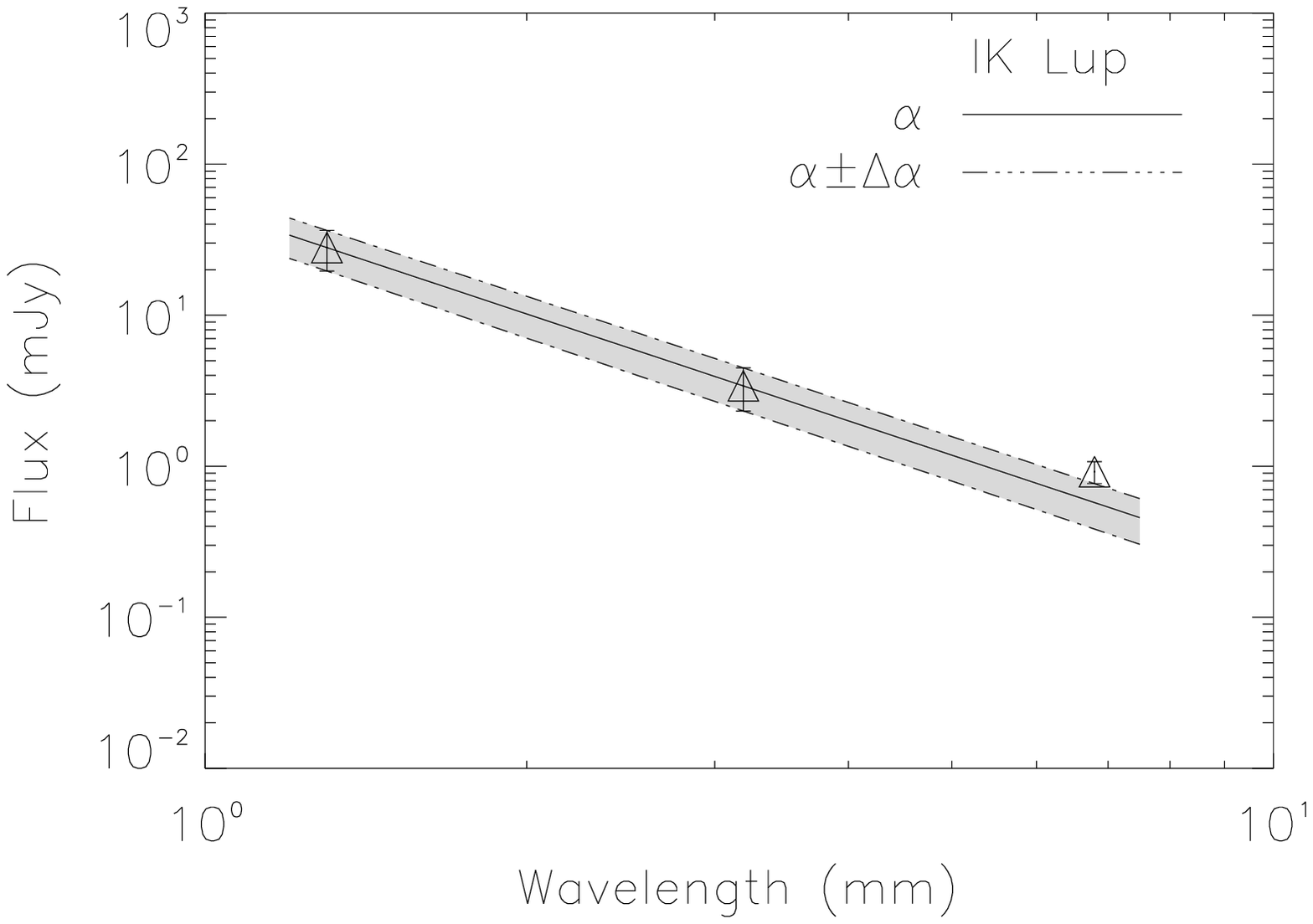} }
	\subfloat{\includegraphics[width=.68\columnwidth]{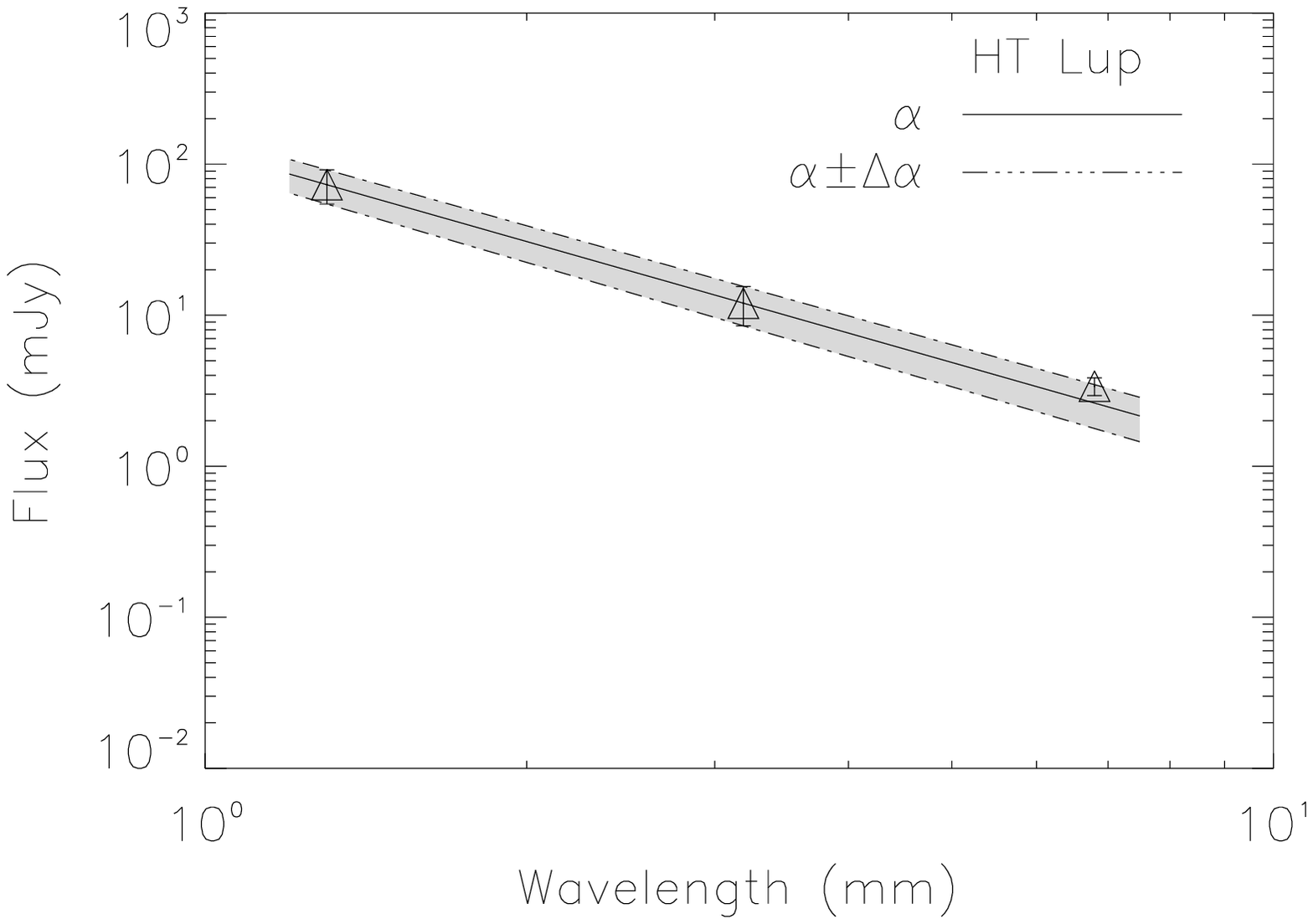} }\\      
	\subfloat{\includegraphics[width=.68\columnwidth]{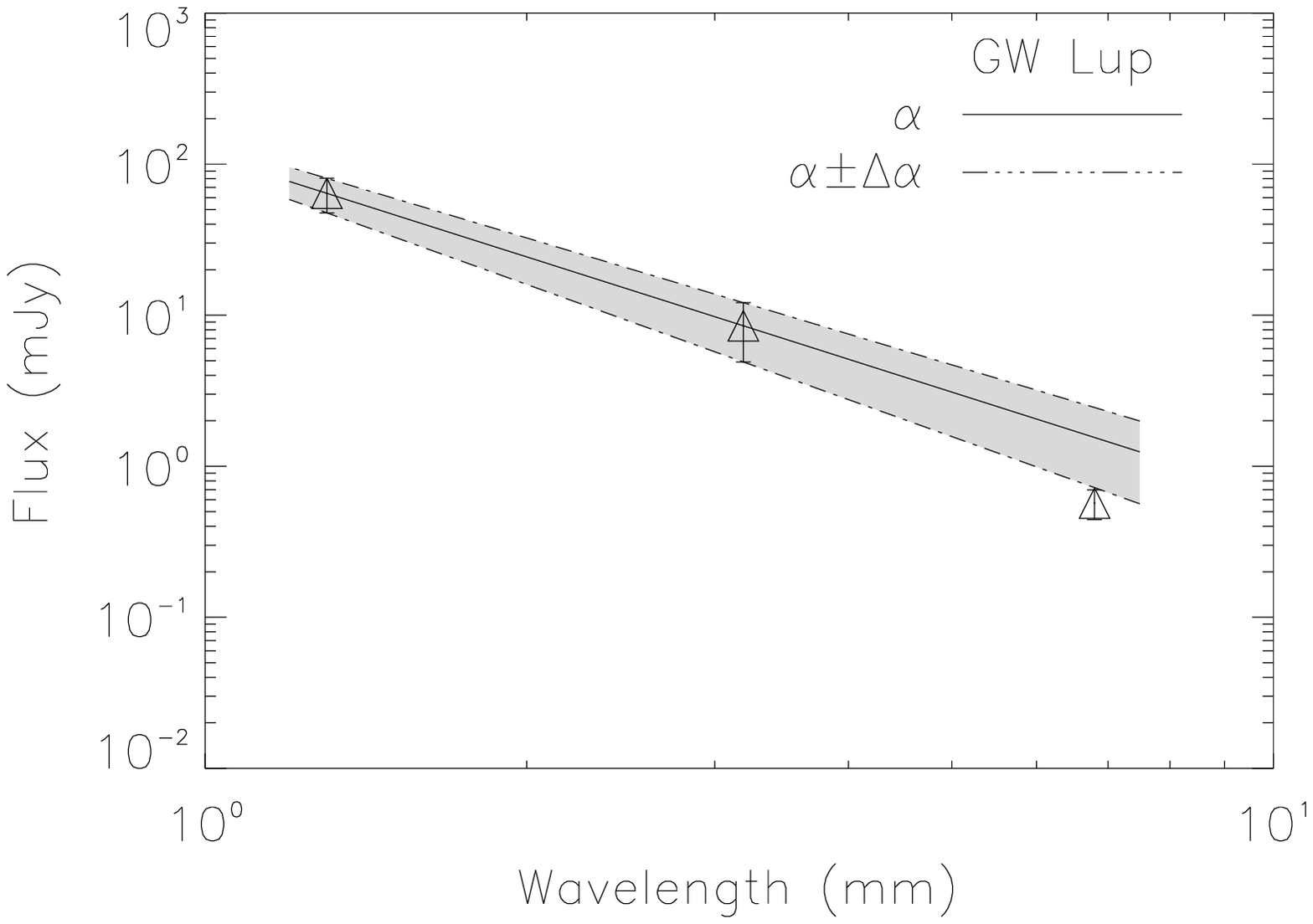} }
	\subfloat{\includegraphics[width=.68\columnwidth]{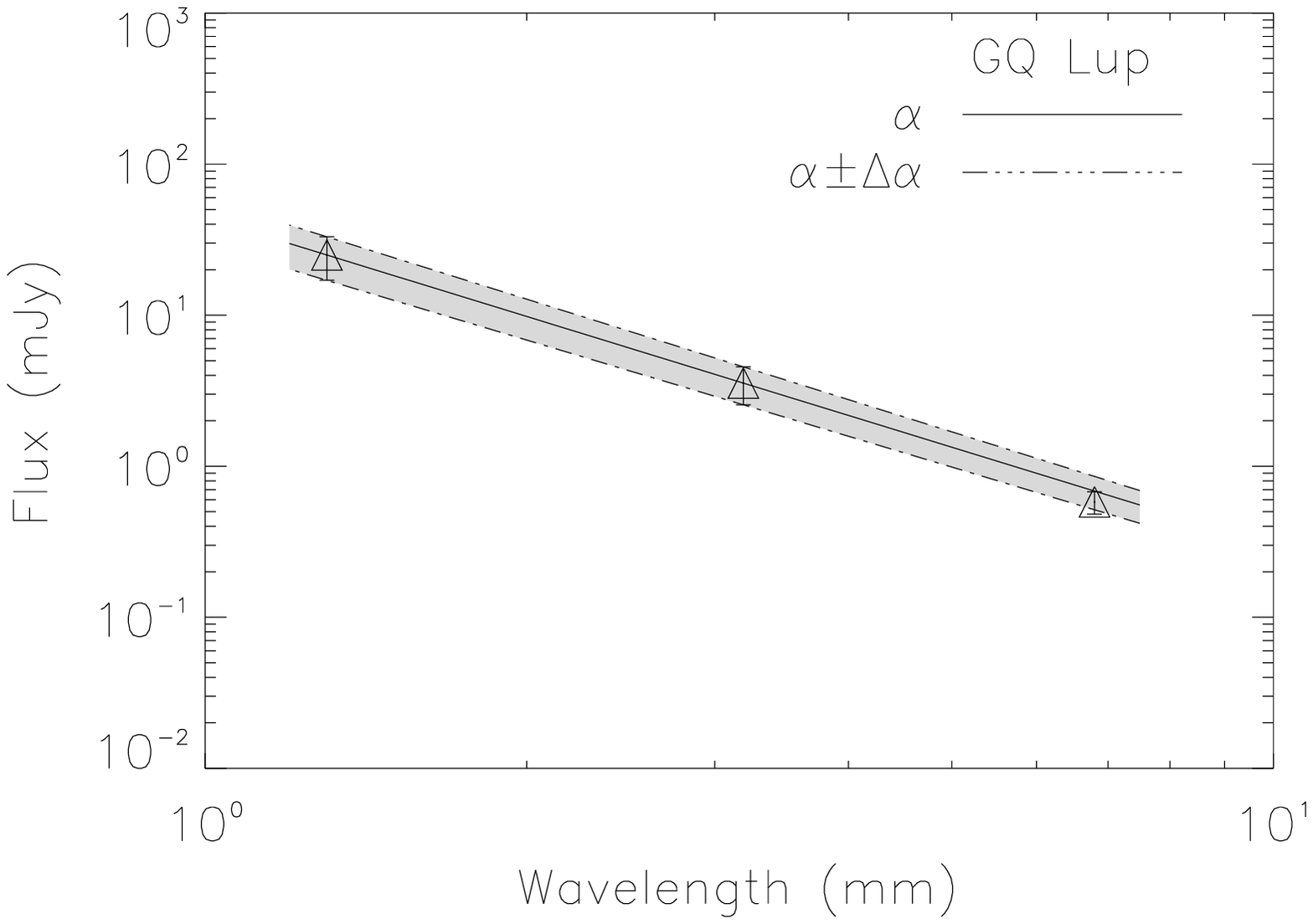}}
	\subfloat{\includegraphics[width=.68\columnwidth]{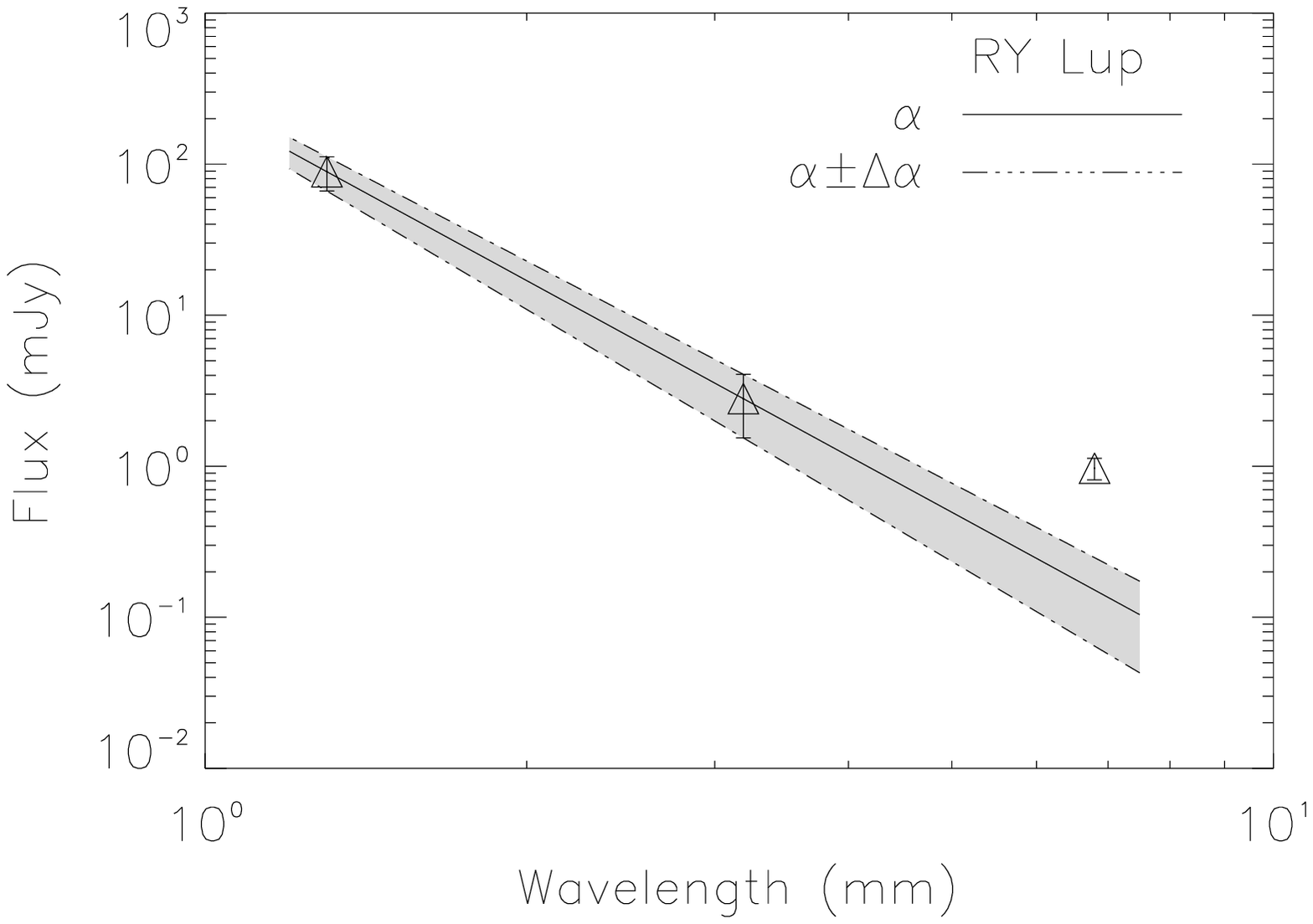} }\\
	\subfloat{\includegraphics[width=.68\columnwidth]{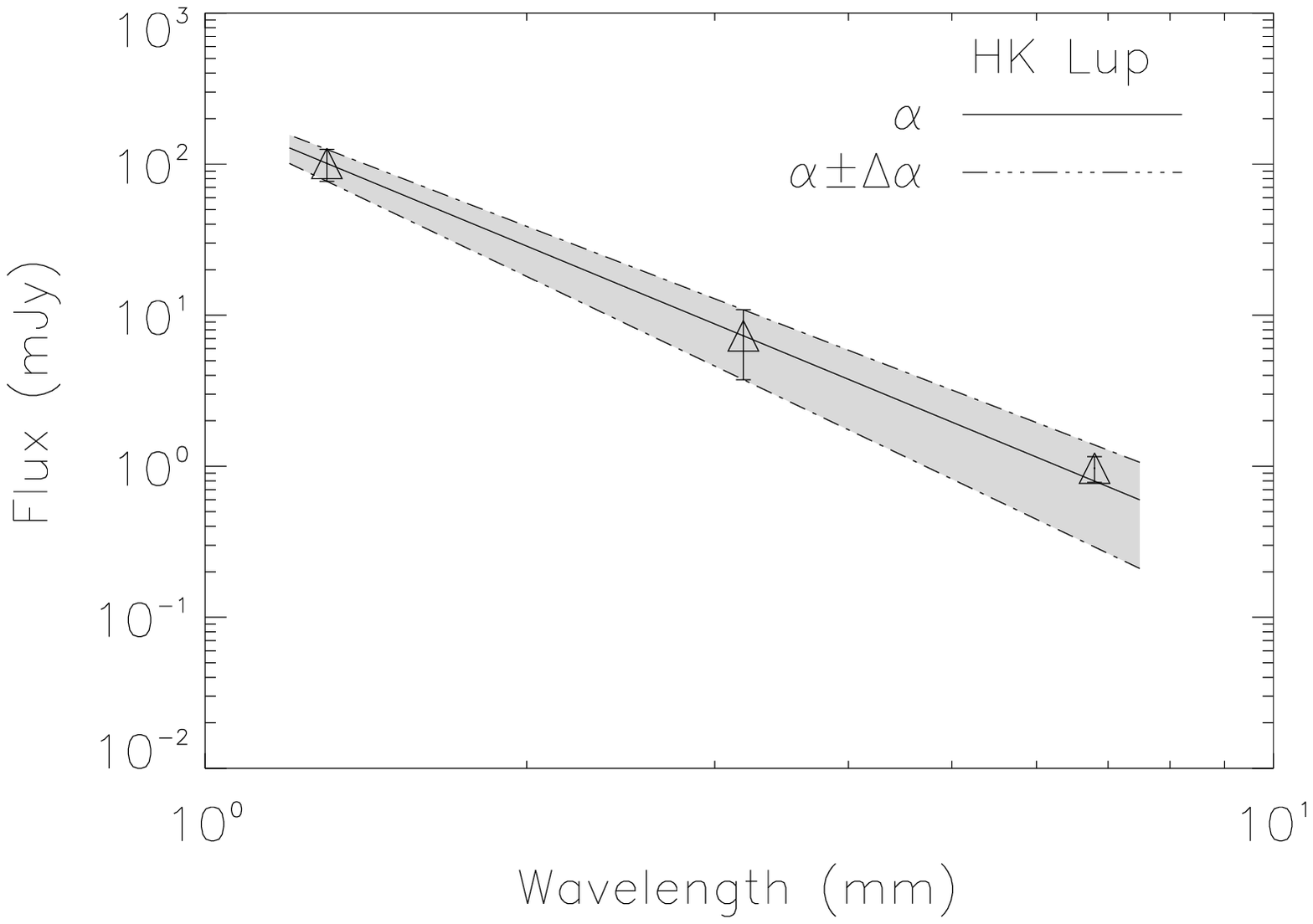}}
	\subfloat{\includegraphics[width=.68\columnwidth]{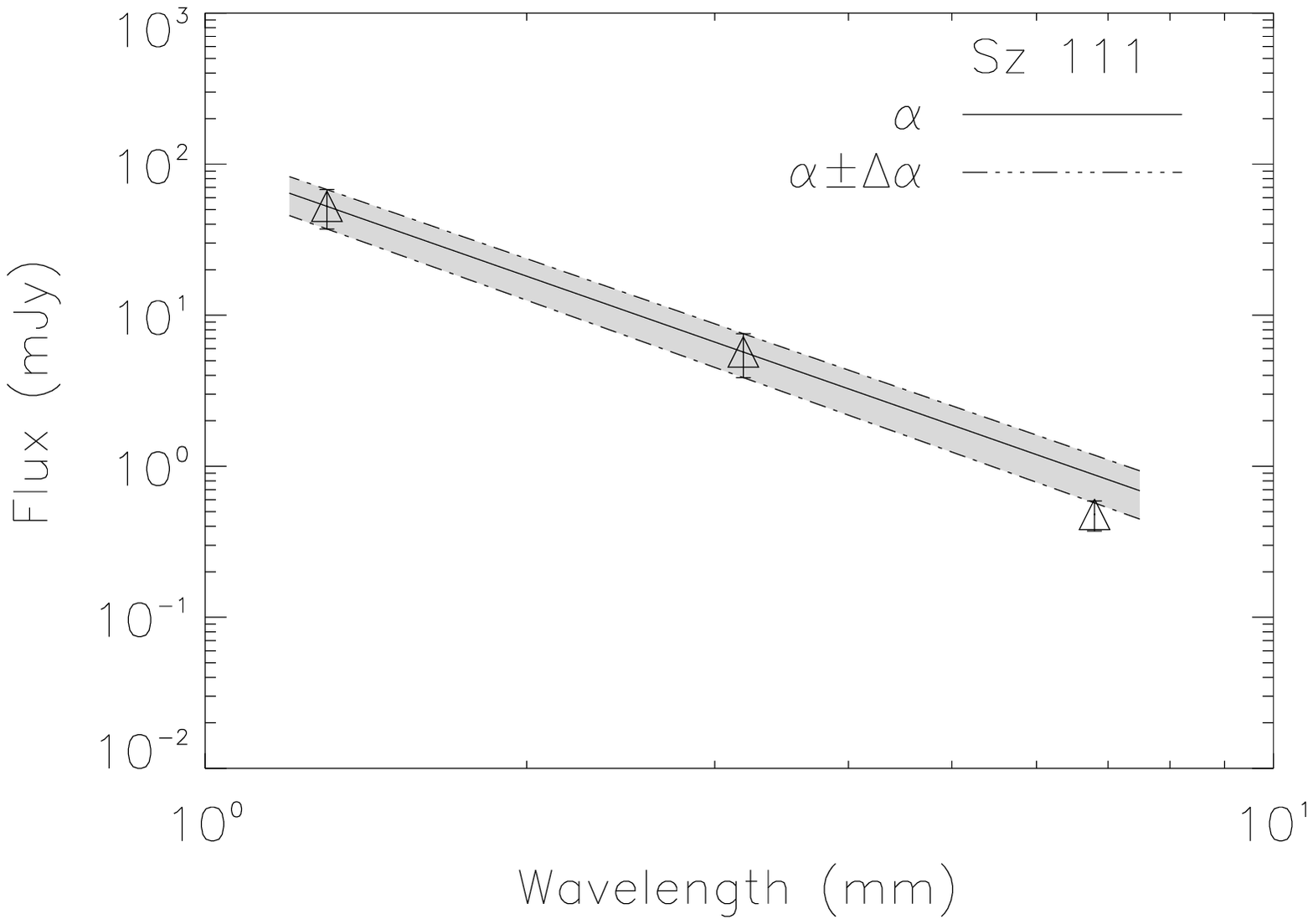} }
	\subfloat{\includegraphics[width=.68\columnwidth]{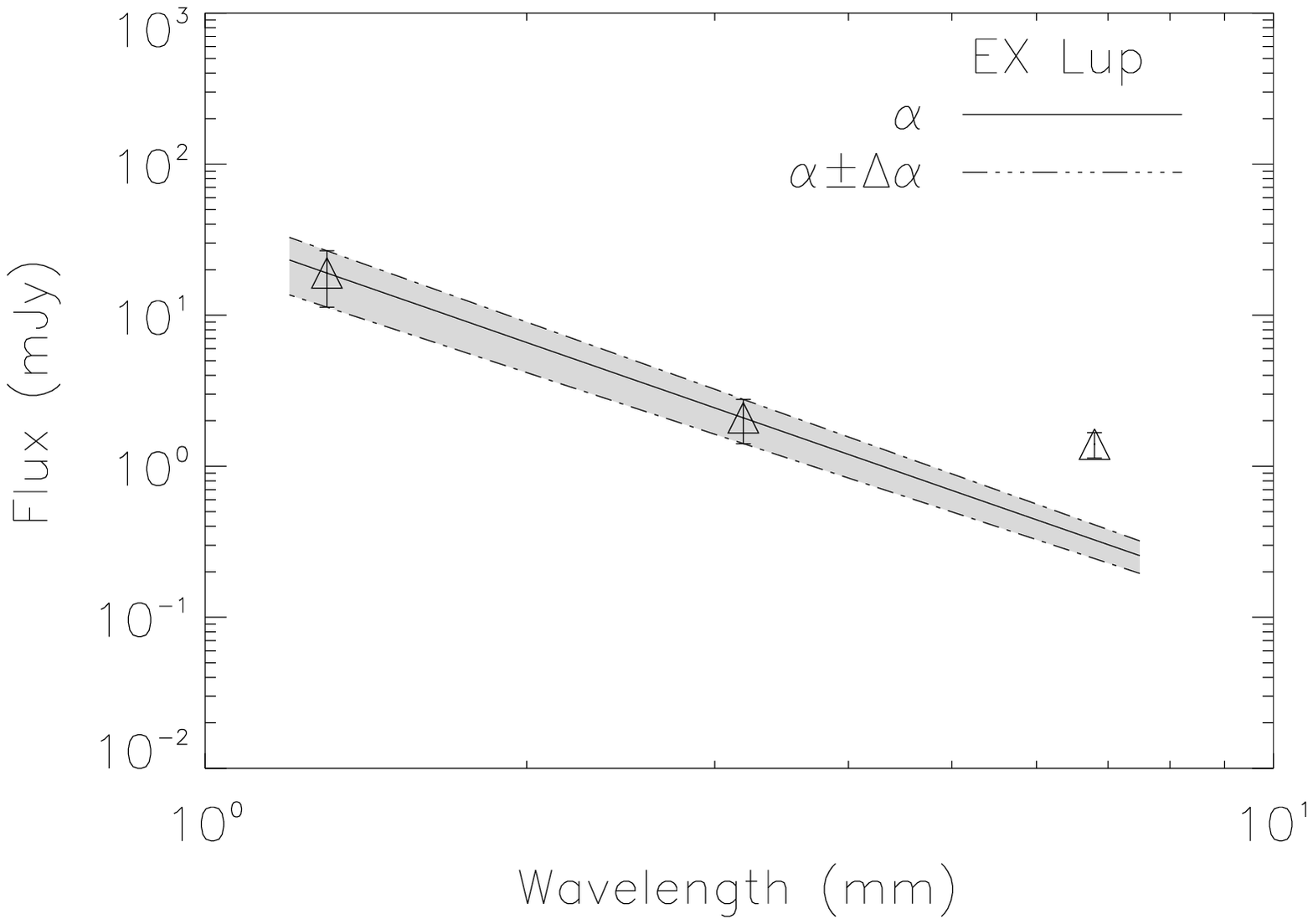} }\\
	\subfloat{\includegraphics[width=.68\columnwidth]{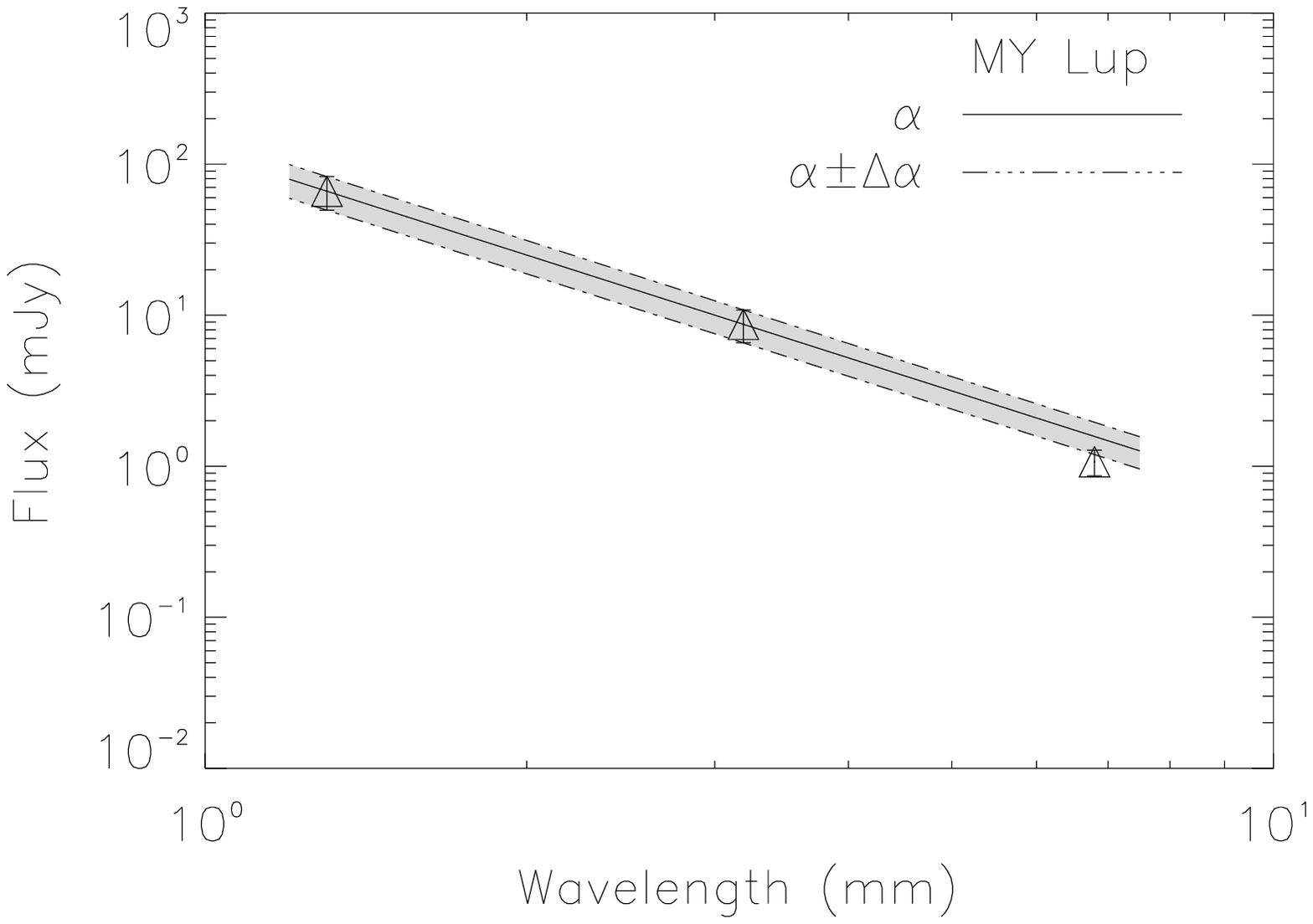} }
	 \subfloat{\includegraphics[width=.68\columnwidth]{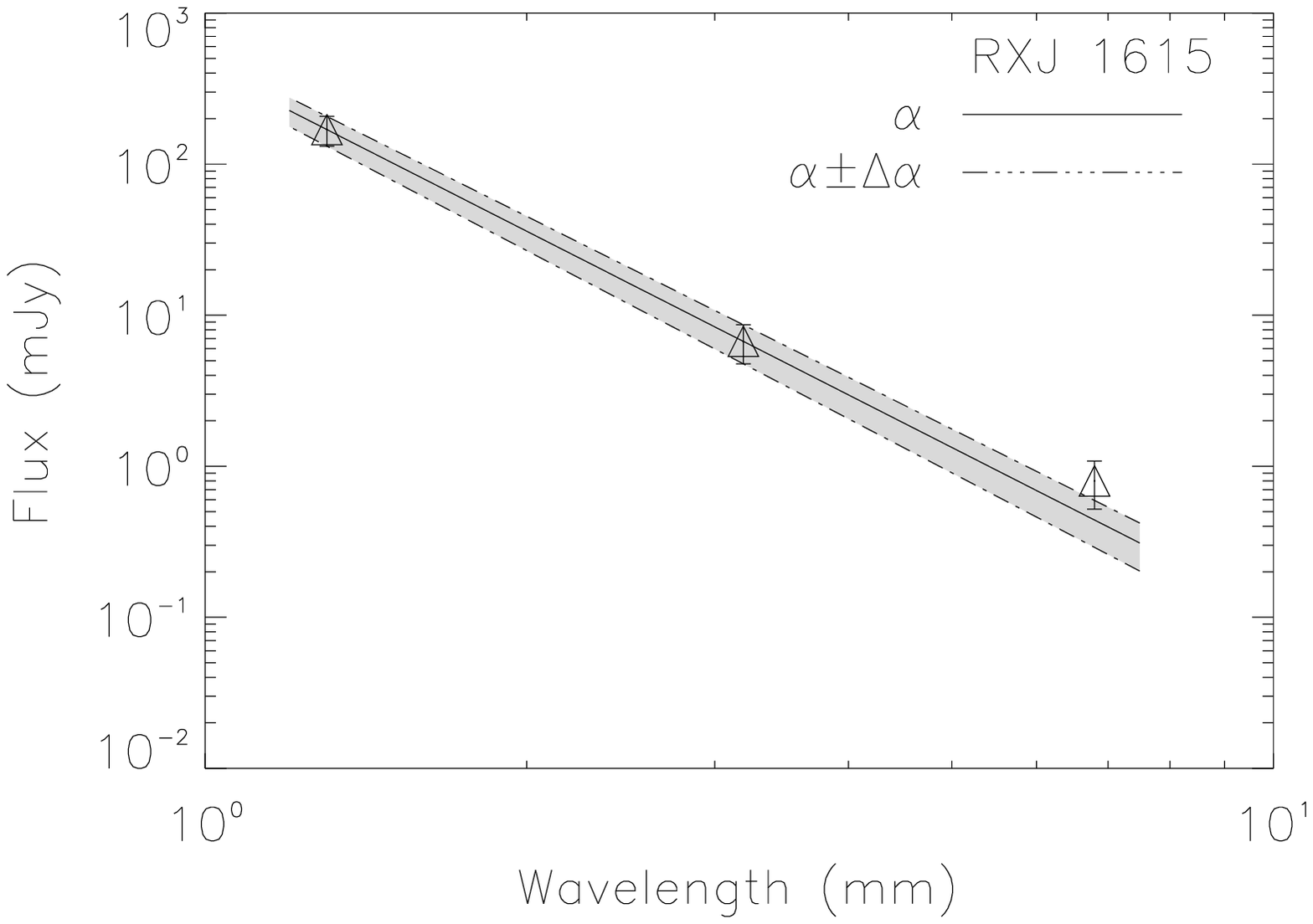}}
	 \caption{Same as Fig.~\ref{fig-3pts-sed-cham}, but for Lupus sources. } 
  \label{fig-3pts-sed-lup}
\end{figure*}	

\subsection{Emission mechanisms at longer wavelengths} 
\label{subsec-discussion-emission}

{{Temporal flux monitoring helps to determine the presence of contributing emission mechanisms. For our sample, a subset of 4 Lupus and 2 Chamaeleon sources which have 7~mm fluxes obtained approximately one year prior to our survey -- see Table~\ref{tab-monitoring2}, allowing us to determine if free-free emission is contributing to the 7~mm fluxes for these sources.}} We also observed {{a sub-sample of}} 6 Chamaeleon sources at 15~mm and 3 {{of these sources}} at 3+6~cm for one epoch {{(see Table~\ref{tab-results2}). We can investigate flux variations in these sources by determining a point source fit in the u-v plane for each scan length at these wavelengths allowing us to determine the contributing emission mechanisms 15~mm wavelength and beyond. The results from the long wavelength temporal monitoring are presented in Figs.~\ref{Fig-15mm-fluxes} and~\ref{Fig-cm-fluxes}, with the {{maximum and minimum}} recorded flux values at 15~mm and 3+6~cm presented in Tables~\ref{tab-15mm} and~\ref{tab-cm} respectively.}}

The literature {{7~mm}} data were obtained from \citet{Lommen09,Lommen10}. {{Their observations}} took place in 2008 with the ATCA using the pre-CABB system. Their {{observing}} frequency pairs at different epochs do not always match our {{central frequencies}}. However small changes in the observing frequency (42.8 to 43.0~GHz and 45.1 to 45.0~GHz) between different epochs would not result in a change in flux by more than the statistical uncertainty and thus should not affect our conclusions. 

For {{our subsample at 7~mm}}, we {{can also investigate}} the potential day-to-day flux variability by obtaining a point source fit flux for the central frequency of the dual sideband for each day the source was observed, resulting in integration times between 10 to 70 minutes depending on the source {{(see Table~\ref{tab-obdetails})}}. In general, sources observed on consecutive days do not have the same integration times, and since Chamaeleon never sets at ATCA, more time was spent on the Chamaeleon than {{the}} Lupus sources.

For this analysis we define ``short temporal monitoring" as scans separated by less than a day, and ``long temporal monitoring" as scans separated by a day or more. 
Thermal dust emission is considered dominant when no intra-epoch variability is observed. Thermal free-free emission is likely present when the flux varies during long temporal monitoring by a factor of 20--40\%. Non-thermal emission will be considered present when the flux varies by a factor of 2 or more {{on timescales of minutes to hours}}. For this analysis {{we take}} into account the flux fit uncertainty and the primary flux calibrator uncertainty for CABB data. {We only claim variability if the same behaviour is observed in both frequency sidebands.}

\subsubsection{Lupus sources}
\label{subsubsec-discussion-emission-lupus}

Four Lupus sources (RY Lup, MY Lup, RXJ1615.3-3255 and Sz 111) were observed three times at 7~mm, once in 2008 and twice in 2009 on consecutive days -- fluxes are presented in {{Table~\ref{tab-monitoring2}.}} {{RY Lup, MY Lup and Sz 111}} show no evidence of flux variability over daily or yearly timescales, indicating that thermal dust emission is dominant, {{while \rm{RXJ1615.3-3255} is consistent with a presence of thermal free-free emission.}} {{The $\alpha$ values for all sources}} are consistent with the upper limits calculated by \citet{Lommen10}.

\subsubsection{Chamaeleon sources}
\label{subsubsec-discussion-emission-Cham}

{{Temporal flux monitoring at 7~mm is available for CS Cha and Sz 32 -- see Table~\ref{tab-monitoring2}. The 15~mm temporal monitoring for 6 Chamaeleon sources are presented in Fig.~\ref{Fig-15mm-fluxes} with the maximum and minimum recorded flux presented in Table~\ref{tab-15mm}. The 3+6~cm temporal monitoring for 3 of those 6 Chamaeleon sources are shown in Fig.~\ref{Fig-cm-fluxes} with the recorded fluxes in Table~\ref{tab-cm}.}}

The temporal monitoring {{of CR Cha and DI Cha}} suggest that the excess emission {{seen in Fig.~\ref{fig-3pts-sed-cham}}} at 15~mm is not from a fast varying chromospheric {{emission but}} most likely from thermal free-free {{emission, consistent}} with the spectral slope of {{0.6 seen}} in Fig.~\ref{fig-3pts-sed-cham}. 

CS Cha was observed at 7~mm three times in 2008 {{with extended}} array configurations (1.5 and 6~km), and twice in 2009 {{on consecutive}} days with a compact array {{configuration.}} There is no evidence of flux variability {{at 6.7~mm}} on consecutive days in 2009 (see Table~\ref{tab-monitoring2}), {{while the three fluxes obtained in 2008 are inconsistent with each other and with the fluxes obtained in 2009.}} {{The 2008 and 2009 changes are consistent with the}} fact that CS Cha was resolved with the extended arrays in 2008 and unresolved with the compact array in {{2009, which}} {{likely caused}} an {{underestimate}} of the flux {{when CS Cha}} was resolved. The {{variability in the}} 7~mm flux observed in {{2009}} is consistent with thermal free-free emission, {{and the likely cause of the excess emission observed in Fig.~\ref{fig-3pts-sed-cham}.}}

{The temporal monitoring of DK Cha at 15~mm is consistent with the presence of some excess emission from thermal free-free emission -- see Fig.~\ref{Fig-15mm-fluxes} and Table~\ref{tab-15mm} -- and contamination from chromospheric emission was detected at 3+6~cm -- {{see Fig.~\ref{Fig-cm-fluxes} and}} Table~\ref{tab-cm}.} {{These results are consistent with the excess emission at 15~mm and 3+6~cm seen in Fig.~\ref{fig-3pts-sed-cham}.}}

Sz 32 was observed once in 2008 and twice in 2009 on consecutive days at 7~mm. The level of flux variability on consecutive days and between the 2008 and 2009 {{(see Table~\ref{tab-monitoring2})}} indicate thermal free-free emission is {{contributing to}}  the 7~mm flux.

T Cha and Sz 32 have excess emission at 15~mm likely due to contamination from thermal free-free emission from an ionised wind, and excess emission at 6~cm from chromospheric emission --- see Tables~\ref{tab-15mm},~\ref{tab-cm}, and Figs.~\ref{Fig-15mm-fluxes} and~\ref{Fig-cm-fluxes}. These results are consistent with the excess emission seen in Fig.~\ref{fig-3pts-sed-cham}. This is also consistent with Sz 32 being the suggested source that drives the short East-West jet, HH 914 {{seen at a distance of}} 0.4$^{\prime\prime}$ from Sz 32 \citep{2006ApJ...643..985W}.

\begin{figure*}
	\includegraphics[width=.65\columnwidth]{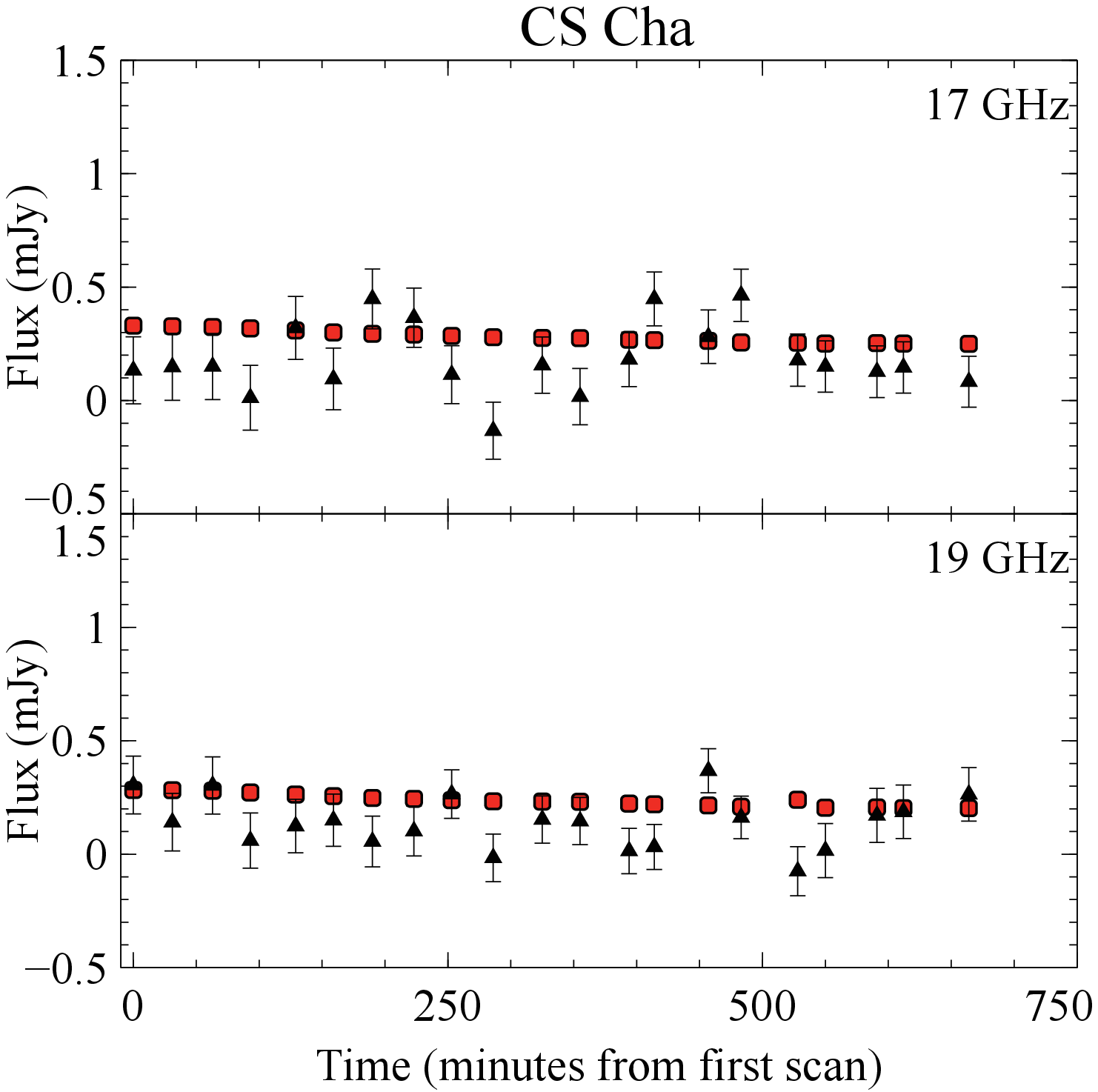}
	\includegraphics[width=.65\columnwidth]{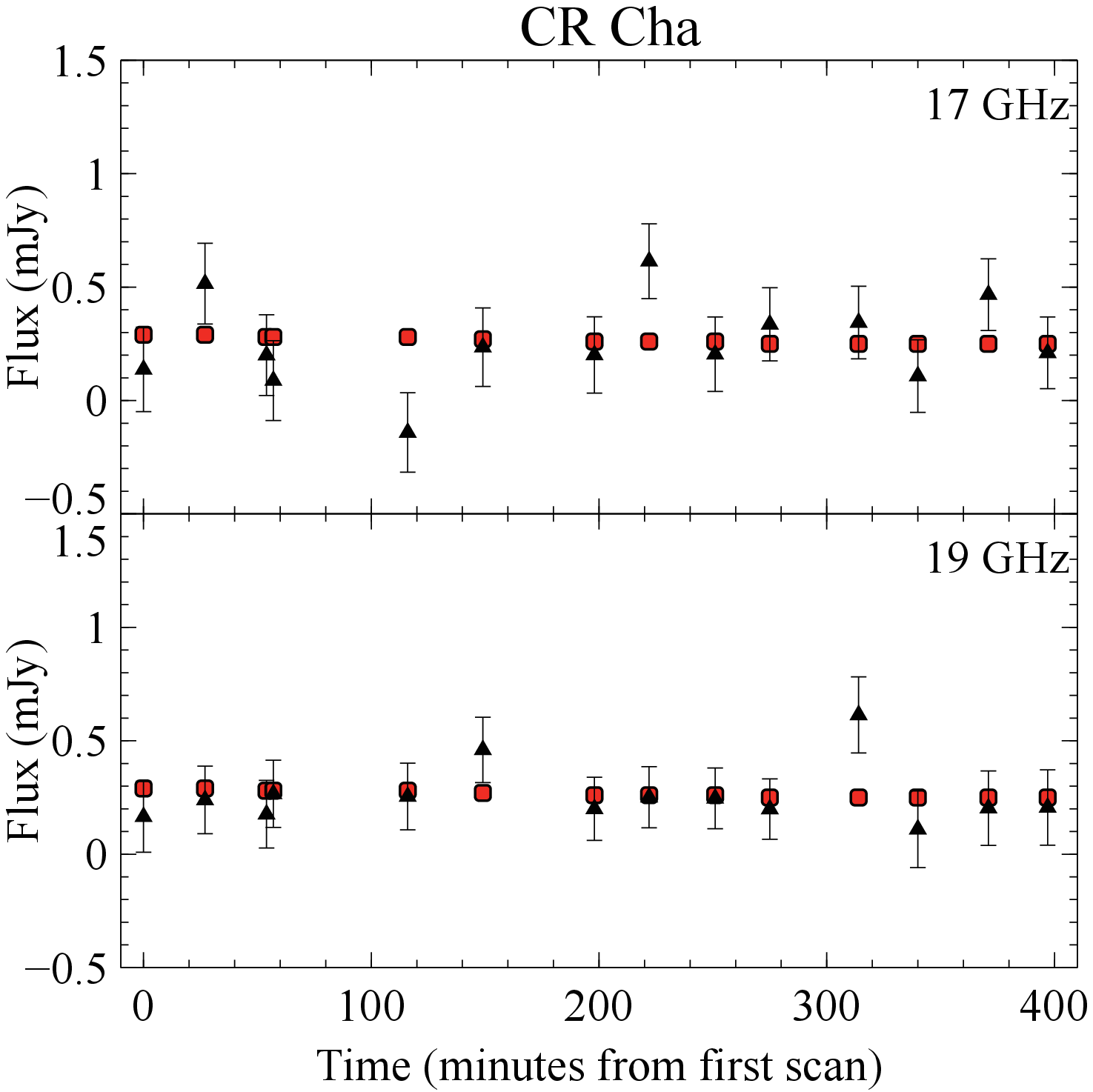}
	\includegraphics[width=.65\columnwidth]{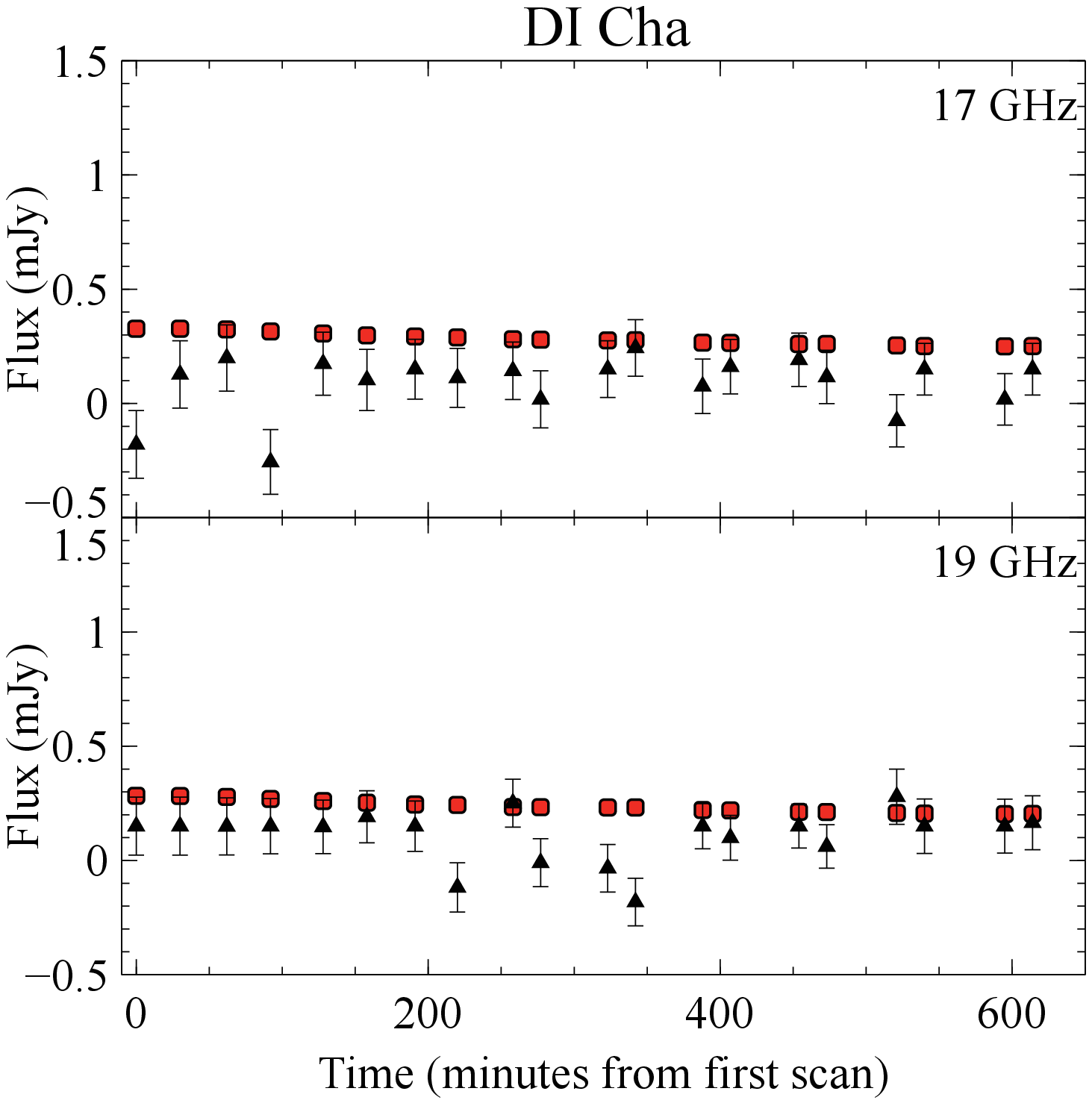}\\
	\includegraphics[width=.65\columnwidth]{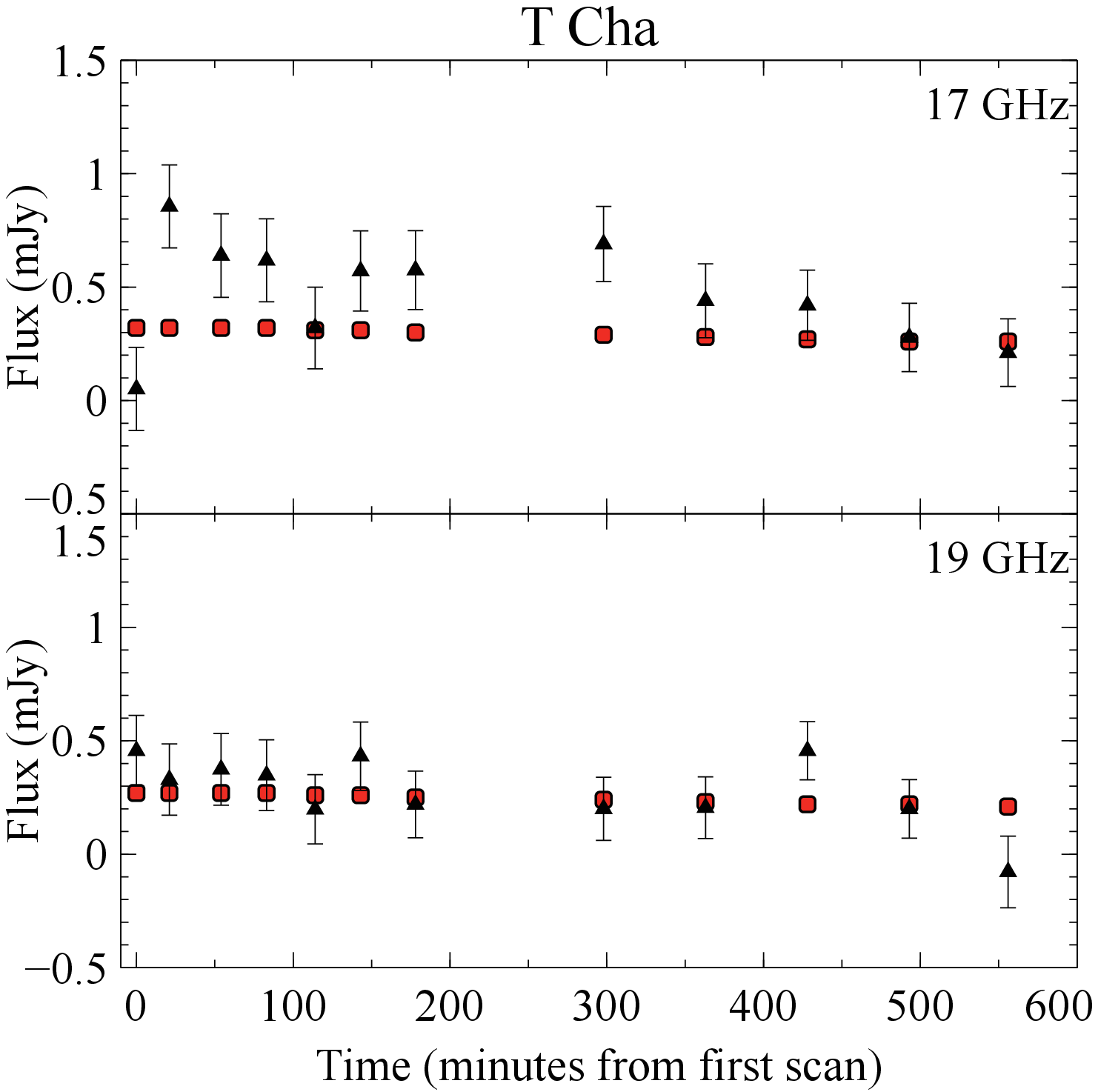}
	\includegraphics[width=.65\columnwidth]{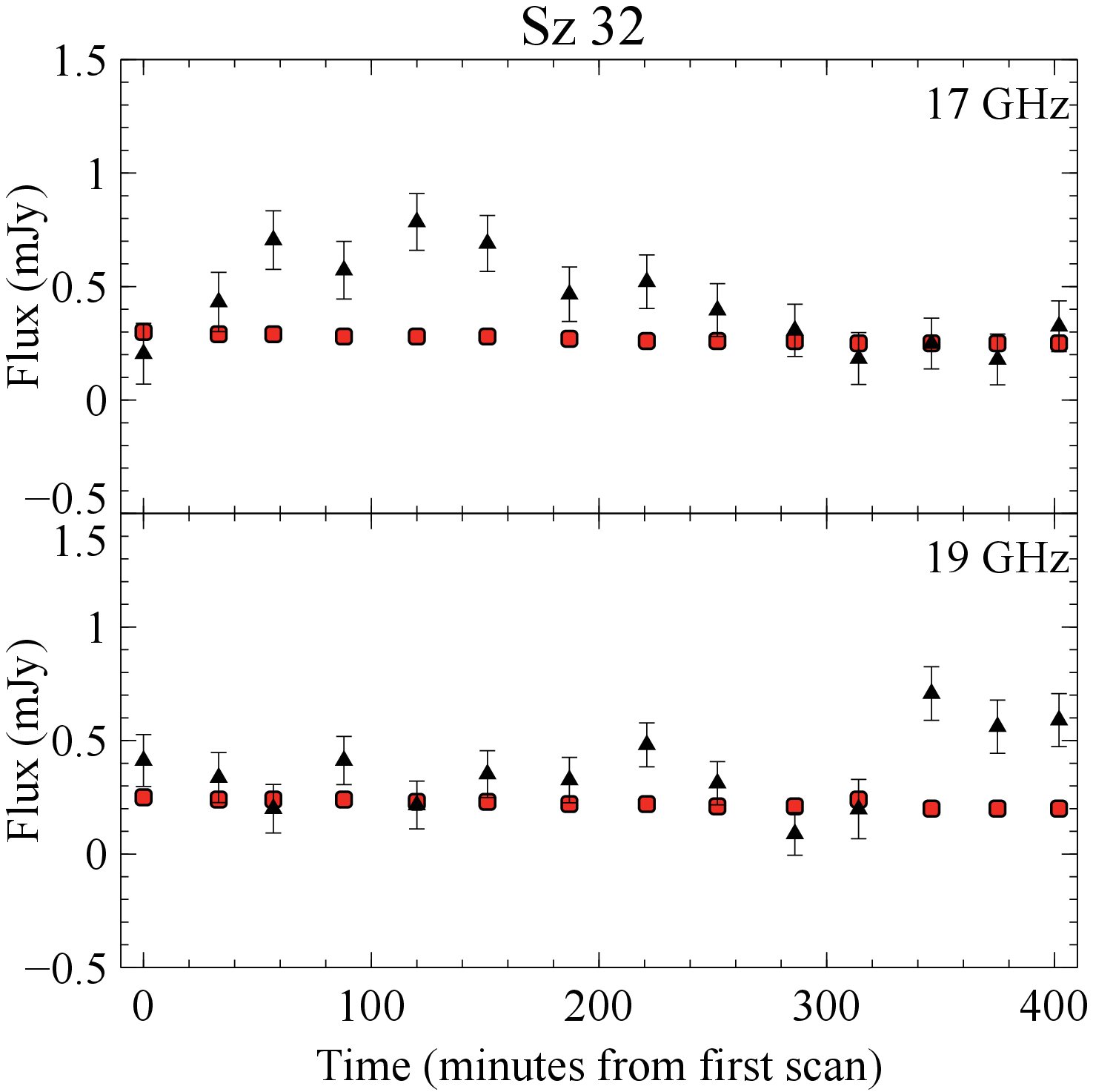}
	\includegraphics[width=.65\columnwidth]{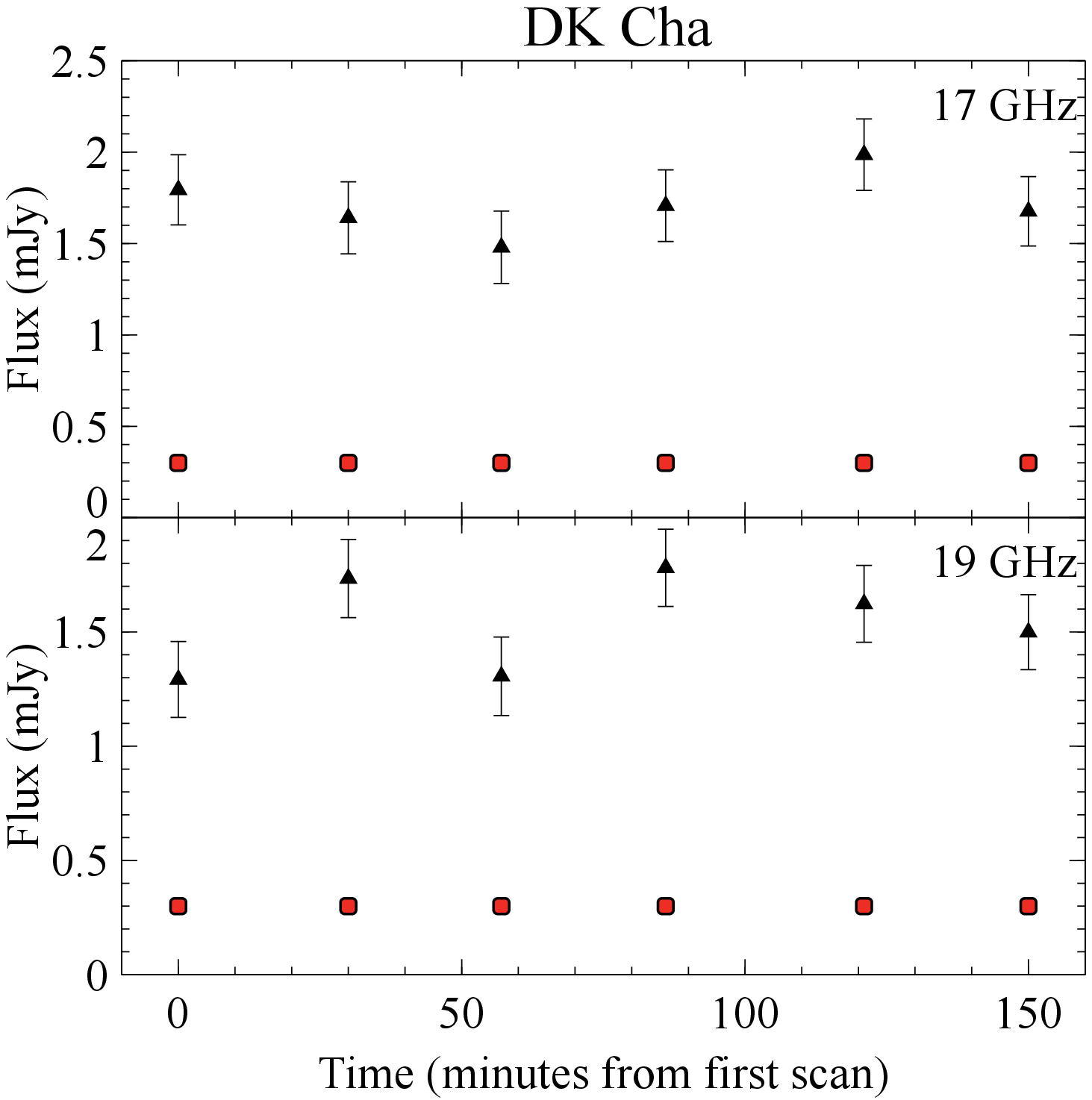}
\caption{Temporal monitoring of 15~mm band continuum flux at 17 and 19~GHz {{of 6 Chamaeleon sources}}. The scan lengths were between 3 and 5 minutes {{with total integration times given in Table~\ref{tab-obdetails}.}} The fluxes are represented by triangles with corresponding uncertainties, and the red squares represent 3$\sigma$ detection values. {{Note that}} when a source is not detected (fluxes below the red squares), \texttt{UVFIT} can sometimes estimates a negative {{flux. These}} are included for completeness, however the value itself is incorrect.}
\label{Fig-15mm-fluxes}
\end{figure*}

\begin{figure}
\centering
	\includegraphics[width=.65\columnwidth]{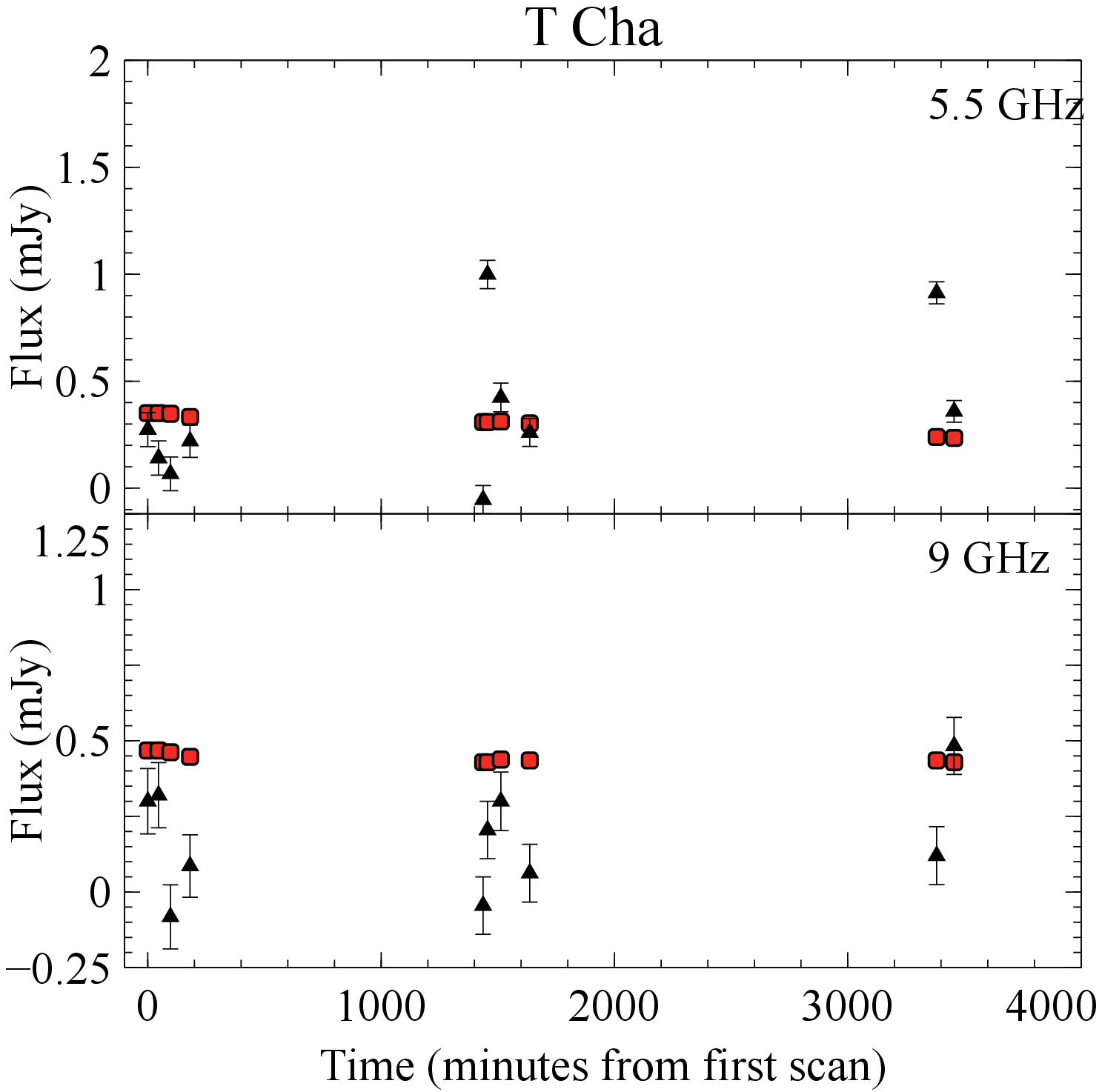}\\
	\includegraphics[width=.65\columnwidth]{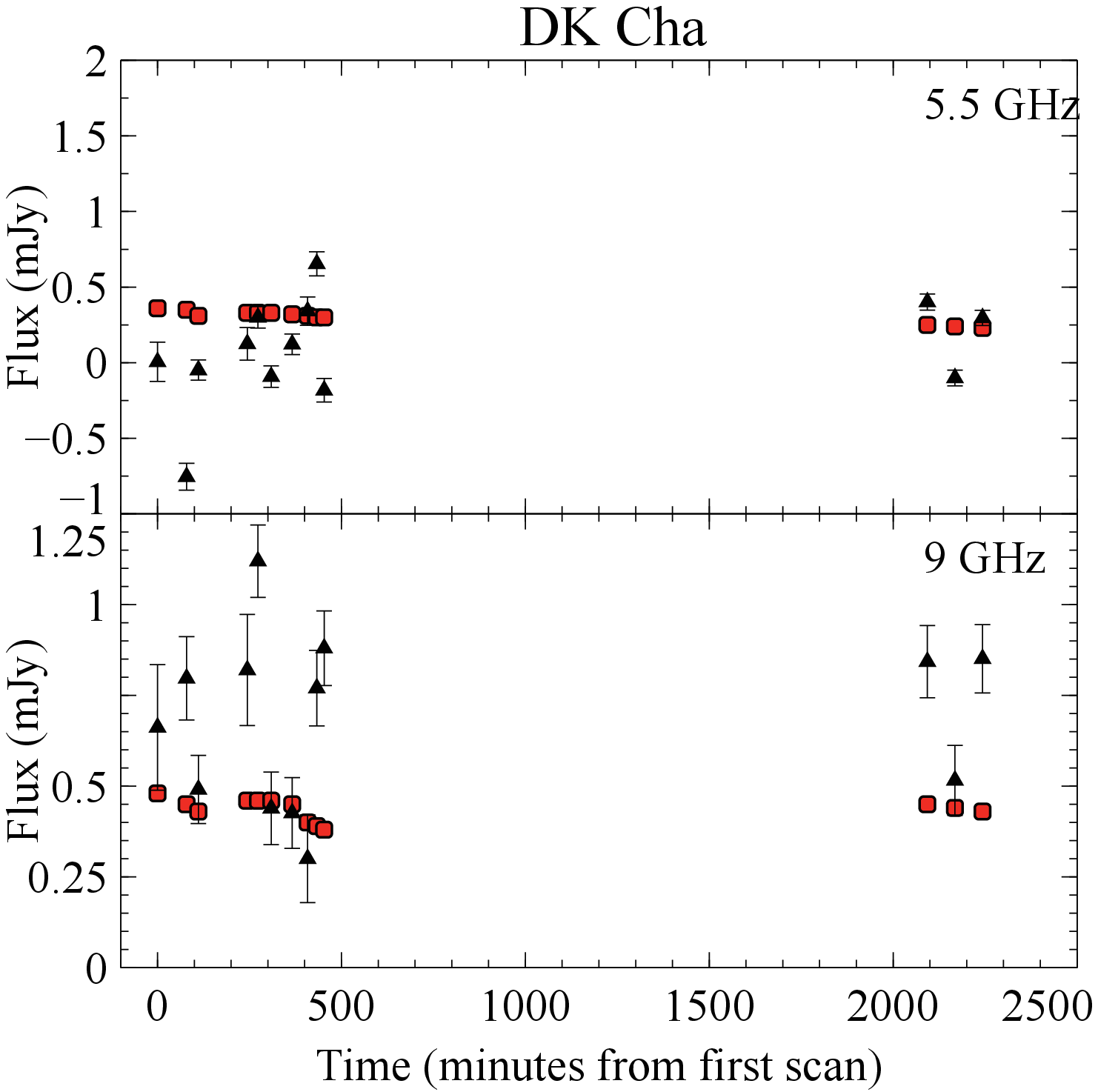}\\
	\includegraphics[width=.65\columnwidth]{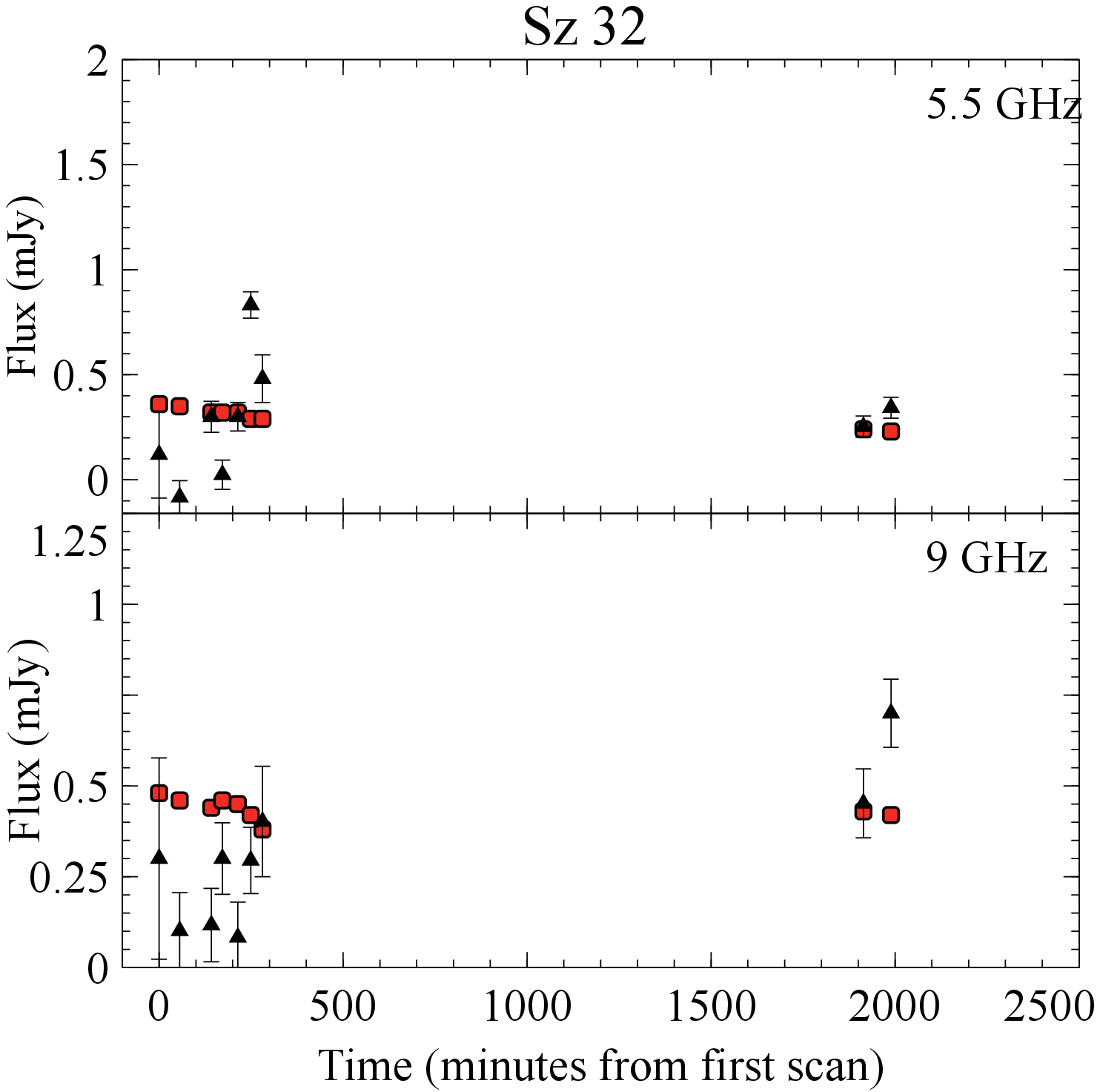}
\caption{Temporal monitoring of 3+6~cm band continuum flux at 9 and 5.5~GHz {{of 3 Chamaeleon sources}}. The scan lengths were 15 minutes {{with total integration times given in Table~\ref{tab-obdetails}.}} The fluxes are represented by triangles with corresponding uncertainties, and the red squares represent 3$\sigma$ detection values. {{Note that}} when a source is not detected (fluxes below the red squares), \texttt{UVFIT} can sometimes estimates a negative {{flux. These}} are included for completeness, however the value itself is incorrect.}
\label{Fig-cm-fluxes}
\end{figure}

\subsection{Additional Sources}
\label{subsec-results-othersources}
WW Cha was detected in the field of view of the Sz 32 observations at 7, 15~mm and 3+6~cm. The continuum fluxes obtained are consistent with \citet{Lommen09}. 

A second source was detected at 7~mm {{to the northwest}} of MY Lup -- {{see Fig.~\ref{fig-mylup2} for beam flux corrected clean map}}. This is an unknown source which we identify by ATCA160043-415442 (instrument~$+$~J2000 location), first observed in May 2009 at RA 16:00:43, Dec~-41:54:42 with an estimated Gaussian {{flux}} of $25.0\pm5.5$ mJy after a beam flux correction. Since the source was at the edge of the beam, observations were conducted at 3~mm to confirm the detection and investigate the spectral slope. In August 2010, the source was detected in the same location at 3~mm with a point {{source fit}} of ${{9.0\pm0.4}}$~mJy and RMS of {{0.3}} Jy/beam -- {{see Fig.~\ref{fig-mylup}}}. There are no known 1.2~mm or cm band detections in the literature. The mm spectral slope from 3 to 7~mm, $\alpha_{3-7} ={{-1.4}}\pm0.8$, {{suggests}} non-thermal emission and the source is likely a background radio galaxy.

A range of extra-galactic sources were detected during the 3+6~cm observations, including: \rm{SUMS~J115749-792819}, \rm{SUMS J11558-792403}, \rm{SUMS J1201169-790726}, \rm{G 2MASX J11574287-7923537}, \rm{PMN J110-7618}, \rm{PMN J1113-7625}, \rm{SUMSS J110856-763624}, \rm{CXO J110841.4-763451}, \rm{SUMS J125216-770903} and \rm{SUMS J125225-770345}.

\section{Discussion}
\label{sec-discussion}

In this section we aim to identify the protoplanetary discs with signs of mm and cm-sized grains. Thus far we have determined the sources with extended emission (i.e. {{which we assume to be from}} resolved discs) in the 3~mm band, breaking the degeneracy between radially extended optically thin discs and compact optically thick discs with shallow spectral slopes. 

Analysis of the continuum fluxes in the mm and cm bands and the resulting spectral slopes allow us to identify sources with a break in spectral slope at 7~mm, indicating {{contributions to the emission at}} 7~mm and beyond from other emission mechanisms {{other than thermal dust}}. Here {{we calculate}} the dust opacity index at 3~mm {{in order}} to estimate the maximum grain sizes at this wavelength, and then estimate the dust disc masses at {3~mm. Finally, we explore a previously claimed tentative correlation}
~between grain growth signatures in the IR and mm regimes using this data set.

\subsection{Dust opacity index at 3~mm}
\label{subsec-discussion-index}

The 1 and 3~mm band emission are in the Rayleigh-Jeans limit and is expected to be mostly optically thin. The spectral slope $\alpha$ can be used to estimate the dust opacity index $\beta$. Assuming the emission is optically thin, the dust opacity index can be written as $\beta \sim \alpha-2$ and an opacity index ${{\beta \leq 1}}$ indicates grain growth up to mm sizes \citep{draine06}.
Table~\ref{tab-betas} presents the $\beta$ values determined {{assuming optically thin emission}} for all the sources in our sample. We found that 2 of the 6 resolved sources and 5 of the 12 unresolved sources have $\beta < 1$, suggestive of grains up to mm sizes -- see columns two and three of Table~\ref{tab-betas}. {{Due to the uncertainties in the $\beta$ values (determined through the propagation of the $\alpha$ uncertainties) the symbol \textbf{``?"} is used when $0.7 < \beta < 1.4$ in Table~\ref{tab-betas}.}}

\begin{table*}
 \centering
 \caption{Dust opacity indices and dust disc masses. {(1) Source name. (2) Dust opacity index (uncertainty is 0.4). (3) Large grains? Y when maximum grain {{size up}} to mm-sizes (assumed when $\beta \leq 1$), N when $\beta > 1$, \textbf{?} when $0.7 < \beta < 1.4$ {{due to the uncertainty in $\beta$ values}}. (4) Temperature power-law exponent, $q$, using the IRAS 60 and 100~$\mu$m fluxes. (5) Product of dust disc mass and dust opacity at 3~mm.  (7) The dust opacity evaluated using Eq.~\ref{eqn-opacity}. (8) The dust disc mass evaluated using the dust opacity $\kappa_{3\rm{mm}}$. {{(9) The total disc mass (gas+dust) assuming a gas-to-dust ratio of 100.}}}
 ~}
  \begin{tabular}{cccccccc}
  \hline 
  \hline
Sources		&		$\beta$		&	Large		&	 $q_{\rm{IRAS}}$ 	& $M_{\rm{dust}}\times \kappa_{3\rm{mm}}$ &		$\kappa_{3\rm{mm}}$		&		M$_{\rm{dust}}$ & M$_{\rm{total}}$\\			
&	  $\beta \approx \alpha-2$  	&	grains?   	&	$q = 2/(3-\alpha_{\rm{IR}})$& $10^{-4}$M$_{\odot} \times \rm{cm}^2\rm{g}^{-1} $ &	cm$^2$g$^{-1}$		&		$10^{-4}$~M$_{\odot}$ &		($\rm{g}/\rm{d}=100$)	\\
	\hline 
	\hline																		
	\multicolumn{8}{c}{Chamaeleon} \\ 	
	\hline	
	SY Cha*	&	---	&	---	&	0.3	&---	&	---	&	---	&	---		\\
	CR Cha	&	1.3	&	?	&	0.3	&1.2	&	0.4	&	3.0	&	3.0E-02		\\
	\textbf{CS Cha}	&	0.9	&	?	& 0.4	&1.9	&	1.2	&	1.6	&	1.6E-02		\\
	DI Cha	&	1.1	&	?	&	---	&0.5	&	0.7	&	0.7	&	6.7E-03		\\
	T Cha	&	1.1	&	?	&	0.4	&0.5	&	0.7	&	0.7	&	7.0E-03		\\
	Glass I	&	1.2	&	?	&	---	&0.8	&	0.6	&	1.3	&	1.3E-02		\\
	\textbf{SZ Cha}	&	0.9	&	?	& ---	&0.4	&	1.2	&	0.9	&	9.4E-03		\\
	Sz 32	&	1.8	&	N	&	---	&0.6	&	0.1	&	4.3	&	4.3E-02		\\															
	\textbf{DK Cha}	&	0.9	&	? &0.7	&12.2	&	1.2	&	10.6	&	1.1E-01		\\
	\hline
	\multicolumn{8}{c}{Lupus} \\ 	
	\hline																									
	IK Lup	&	0.4	&	Y	&	---	&	0.6	&	---	&	0.1	&	1.4E-03		\\
	Sz 66*	&	---	&	---	&	---	&	---	&	---	&	---	&	---		\\
	\textbf{HT Lup}	&	0.0	&	Y	&	2.9	&	2.1	&	9.7	&	0.2	&	2.1E-03		\\
	GQ Lup	&	0.2	&	Y	&	2.1	&	0.5	&	5.9	&	0.1	&	8.7E-04		\\
	GW Lup	&	0.3	&	Y	&	0.2	&	1.5	&	5.6	&	0.3	&	2.7E-03		\\
	\textbf{RY Lup}	&	1.9	&	N	&	0.7	&	0.9	&	0.1	&	6.9	&	6.9E-02		\\
	HK Lup	&	0.9	&	?	&	0.4	&	2.3	&	1.1	&	2.0	&	2.0E-02		\\
	\textbf{Sz 111}	&	0.5	&	Y	&	0.2	&	1.8	&	3.3	&	0.5	&	5.4E-03		\\
	EX Lup	&	0.6	&	Y	&	0.2	&	0.6	&	2.3	&	0.2	&	2.4E-03		\\
	MY Lup	&	0.3	&	N	&	0.4	&	1.8	&	5.4	&	0.3	&	3.4E-03		\\
	RXJ1615.3-3255	&	1.6	&	N	&	0.3	&	1.8	&	0.2	&	7.7	&	7.7E-02		\\
	\hline
\end{tabular}
\label{tab-betas}
	\begin{tablenotes} 
		\item[1] * SY Cha and Sz 66 only have upper limits at 1.2 and 3~mm.
		\item[2] \textbf{Boldface} sources were resolved at 3~mm.
		\item[3] {Sources with ``---" for $q_{\rm{IRAS}}$ have no available IRAS 60 and/or 100~$\mu$m fluxes.}
	\end{tablenotes}	
\end{table*}

This approximation of $\beta$ does not account for the contribution {{from}} optically thick emission. Taking this into account, \citet{Beckwith90} showed that 
\begin{eqnarray}
	 \label{eqn-beta2} \beta &\approx& (\alpha-2)(1+\Delta),
 \end{eqnarray} 
where $\Delta$ is the ratio of optically thin to thick emission given by 
\begin{eqnarray}
 \label{eqn-deltapprox} \Delta &\approx& -p \times [(2-q)\rm~ln{(1-p/2)\bar{\tau}}]^{-1},
  \end{eqnarray} 
where \textit{q} and \textit{p} are the power-law exponents of the disc temperature and surface density profiles respectively, and $\bar{\tau}$ is the average disc opacity at the specific frequency. The logarithmic dependence on $\Delta$ is only valid when $2-p-q=0$ -- see \citet{Beckwith90} for details. 

We determined the temperature profile exponent \textit{q} for our sources using IRAS 60 and 100~$\mu$m fluxes obtained from the Infrared Science Archive\footnote{Housed at the Infrared Science Archive (IRSA) at the Infrared Processing and Analysis Centre (IPAC), California Institute of Technology.} using $q = 2/(3-\alpha_{\rm{IR}})$\footnote{Where $\alpha_{\rm{IR}} =~$d(log F$_{\nu}$)/d(log~$\nu)$ is the IR spectral slope for $\lambda \leq 100~\mu$m \citep{Beckwith90, Rod06}}, and found $0.2 < q < 0.7$ with an uncertainty of 0.4, with the exception of HT Lup and GQ Lup with $q > 2$ --- see Table~\ref{tab-betas}. 

For a flat spectrum disc a value of $q = 0.5$ and $p = 1.5$ is expected \citep{Beckwith90}, and high angular resolution observations suggest $p\leq1.5$ \citep{2000ApJ...534L.101W,2003A&A...403..323T}. If we adopt $p = 1.5$, a $\bar{\tau}=0.02$ at 3.3~mm \citep{Lommen07} and use the $q$ values in Column 4 of Table~\ref{tab-betas}, with uncertainties such that the $q$ values are consistent with 0.5, $\Delta\sim0.2$ (with exception of HT Lup and GQ Lup where $\Delta$ approximation in Eq.~\ref{eqn-deltapprox} is invalid).

There are three sources (HT Lup, HK Lup, and MY Lup) for which Spitzer 24, 60, 70, and 100~$\mu$m fluxes exist. Using the MIPS\footnote{Multiband Imaging Photometer for Spitzer is operated by the Jet Propulsion Laboratory, California Institute of Technology, under contract with the National Aeronautics and Space Administration.} fluxes at 24 and 70~$\mu$m obtained by \citet{2008ApJS..177..551M}, we re-calculated $q$ and found values of 0.4, 0.9, and 0.6 for HT Lup, HK Lup, and MY Lup respectively. The significant decrease of $q$ from 2.9 to 0.4 for HT Lup is well above the 0.4 uncertainty level, while the HK Lup and MY Lup increases are within the uncertainty. {The increase in $q$ for HK Lup and MY Lup leads to no significant change in the $\Delta$ and $\beta$ values, while the decrease in HT Lup to $q=0.4$ allows for the use of the $\Delta$ approximation (obtaining a $\Delta = 0.2$).}

Since knowledge of the inner and outer radius of the disc is needed to determine an exact value for $\Delta$ (see eq. 20 of \citealt{Beckwith90}) and taking $\Delta=0.2$ changes $\beta$ by 0.3 or less which does not change the interpretation of our analysis (with exception of CR Cha which increases to 1.6 and thus would be considered to have ISM sized grains); for further analysis we will use the values for $\beta$ calculated in Table~\ref{tab-betas}.

\subsection{Dust disc masses}
\label{subsec-discussion-mass}

The dust disc mass can be determined via
\begin{eqnarray}
	\label{eqn-dustmass} M_{\rm{dust}} &=& \frac{F_{\upsilon}D^{2}}{\kappa_{\upsilon}B_{\upsilon}(T_{\rm{dust}})},
\end{eqnarray}
using the distances $D$ presented in Table~\ref{tab-sources}, the 3~mm fluxes $F_{\upsilon}$ from Table~\ref{tab-results2}, the brightness, $B_{\upsilon}(T_{\rm{dust}})$, for a dust temperature $T_{\rm{dust}}$ {{(assumed to be 25~K)}} given by the Planck function, and the dust opacity, $\kappa_{\upsilon}$ at $\upsilon = 94$~GHz, evaluated using the $\beta$ values in Table~\ref{tab-betas} via the \citet{Beckwith90} equation
\begin{eqnarray}
	\label{eqn-opacity} \kappa_{\upsilon} &=& 10(\upsilon/10^{12}\rm~Hz)^{\beta} \mbox{cm}^{2} \mbox{g}^{-1}.
\end{eqnarray}
The resulting $\kappa_{3\rm{mm}}$ and $M_{\rm{dust}}$ values are presented in Table~\ref{tab-betas}. Note that $\kappa_{3\rm{mm}}$ varies from the canonical 0.9~cm$^{2}$g$^{-1}$ used in the past works \citep[e.g.,][]{Beckwith90,Andrews07,Ricci10b}. 

{{The dust}} {disc masses {{which range from $10^{-3}$--$10^{-5}~$M$_\odot$}} are similar to those found in other star forming regions  \citep[e.g.,][]{Ricci10a}. 
Assuming a gas-to-dust ratio of 100, we find eight sources have a {{total disc mass (gas+dust)}} greater than 0.01~M$_{\odot}$, the minimum mass solar nebula according to \citet{1977MNRAS.180...57W}{{,}} and six have {{a total disc mass (gas+dust)}} greater than 0.02~M$_{\odot}$, the minimum mass solar nebula according to \citet{1981PThPS..70...35H}.

Note {{that}} there is a lot of uncertainty in the dust mass calculations. One of the main sources of uncertainty comes from the dust opacity, which is dependent on the chemical composition, size, and shape of the grain \citep[e.g.,][]{draine06}. A second source of uncertainty comes from the assumed dust temperature, where a change of $\pm5$~K can cause a $\sim$20\% change in the dust mass. The uncertainty in the primary flux calibration and the uncertainty in the source distances of $\pm50$~pc \citep{2007ApJS..173..104L, 1994AJ....108.1071H, 2008hsf2.book..295C} also contributes to the total uncertainty of the dust masses. That said, the method employed here is {{that}} generally used in the literature \citep[e.g.][]{Andrews05,Lommen09,Lommen10,Ricci10a} and {one can expect the disc dust masses to be constrained within a factor of 2 to 10 due to these systematic uncertainties.}
  
 \subsection{Millimetre-sized grains?}
Our  3 and 7~mm fluxes presented in Table~\ref{tab-alphas} (and in Figs.~\ref{fig-3pts-sed-cham} and~\ref{fig-3pts-sed-lup}) indicate that 11 sources do not have a break in their spectral slopes at 7~mm, suggesting that thermal dust emission is dominant up to 7~mm. From the derived dust opacity indices (assuming $\beta \sim \alpha-2$) in Table~\ref{tab-betas} --- {{where the second column presents the $\beta$ values and third column indicates}} the existence of grains up to mm sizes --- we find that 6/11 sources, all in Lupus, have $\beta < 1$, suggesting grain growth up to mm sizes, while 3 sources have ISM sized grains with $\beta >1$. The majority of the Chamaeleon sources (7 sources) have $0.7 < \beta < 1.4$, making it difficult to analyse the results given that the uncertainty in $\beta$ is 0.4. 

Assuming the emission from 1-7~mm is optically thin, our results would {{suggest dust}} grains in Lupus sources are generally larger than {{in}} Chamaeleon sources, and the lack of excess emission observed at 7~mm for Lupus sources {{could}} suggest less {{chromospheric emission}} and potentially more evolved systems than Chamaeleon sources and hence potentially older. However, compared to published ages of the different star forming regions, it is unclear if this is indeed the case. Lupus 3 and Chamaeleon I have similar ages (3--6~Myr), while Lupus 1 is thought to be younger than Lupus 3 \citep{1994AJ....108.1071H,2007ApJS..173..104L,2008ApJS..177..551M}. Individual sources with signs of large grains have ages ranging from $< 1$--14~Myrs \citep{1994AJ....108.1071H,1996MNRAS.280.1071L}. The lack of a break at 7~mm, is an indication of no excess emission which could be a signature of age. However, no distinction is found in the ages of individual sources with a break (3--10~Myrs) and sources without a break (0.9--14.1~Myrs) \citep{1994AJ....108.1071H,1996MNRAS.280.1071L,2009MNRAS.398..189H,2009A&A...501.1013S}.

	The temporal monitoring results for both mm and cm wavelengths presented in Section~\ref{subsec-discussion-emission} provide further clues, allowing us to rule out other sources of emission besides thermal dust at 7~mm and beyond. Note of the 7 sources with $\beta < 1$ only Sz 111 and  MY Lup were monitored for flux variability at 7~mm. We found no flux variability at 7~mm for MY Lup, Sz 111, and conclude that thermal dust emission dominates in these sources and hence they have grains up to $\sim$1~cm in size. 
	
	Our results also suggest all six sources observed at 15~mm have excess emission above thermal dust, with temporal monitoring suggesting the emission comes from thermal free-free emission. At 3+6~cm all three sources were found to have some excess emission. Thus it is difficult to obtain grain size information for these source beyond 7~mm.

\subsection{Correlating grain growth signatures}
\label{subsec-discussion-IR}
	Thus far we have determined signatures of mm grain growth in {{our survey sample}} by evaluating $\beta \sim \alpha -2$, and found 35\% of the discs have grains up to mm sizes. Another grain growth signature is the 10~$\mu$m silicate feature, which probes {{the warm inner ($\sim$1--5~AU) and upper layers of typical T~Tauri stars discs}}, while the mm band probes the {{the cooler outer disc regions ({{$>10$~AU}}) and mid-plane.}}

	 \citet{Lommen07,Lommen10} suggested that a tentative correlation exists between the strength and shape of the 10~$\mu$m feature and mm spectral slope for a sample of discs in Chamaeleon, Lupus and Taurus. However, \citet{Ricci10b} found no such correlation in their Taurus sample. As noted by \citet{Lommen10}, a correlation between these two regions of the disc is unexpected, and if confirmed, it could imply that grain growth in the inner upper layers of the disc and in the mid-plane occur simultaneously. Here we use our sample to further investigate this potential correlation.
	 
	 {{For this analysis, Spitzer Infrared Spectrograph (IRS)\footnote{Spitzer Space Telescope is operated by the Jet Propulsion Laboratory, California Institute of Technology under a contract with NASA. For more information on IRS see \citet{2004ApJS..154...18H} } spectrum data from 5.3--16~$\mu$m was obtained from the Spitzer Heritage Archive\footnote{Housed at the Infrared Science Archive (IRSA) at the Infrared Processing and Analysis Centre (IPAC), California Institute of Technology} for all our sources and an additional seven $\rho$ Ophiucus sources (to coincide with the \citealt{Ricci10b} sample), {{as well as}} 11 Taurus-Auriga and two additional Chamaeleon and Lupus sources (to coincide with {{the}} \citealt{Lommen10} sample) -- see Table~\ref{tab-10micronresults}. All the infrared data presented in Table~\ref{tab-10micronresults} were processed with the S18.18 pipeline.}}
	
	Following {{\citet{2005ApJ...622..404K} and \citet{Furlan06}}}, we define the shape of the 10~$\mu$m silicate feature as the flux ratio at $11.3 ~\mu$m and $9.8~\mu$m, $\frac{\rm F_{11.3 \mu \rm m}}{\rm F_{9.8  \mu \rm m}}$, where F$_{11.3  \mu \rm m}$ and F$_{9.8  \mu \rm m}$ are the integrated flux over a 0.4~$\mu$m band centred at 11.3 and 9.8~$\mu$m, and define the strength of the 10~$\mu$m silicate feature as:
	\begin{equation}
	\rm strength = \frac{(F_{10\mu\rm m} - F_{\rm cont})}{ F_{\rm cont}},
	\label{eq-strength}
	\end{equation}
	where F$_{\rm 10\mu\rm m}$ is the integrated flux from 8--12.4~$\mu$m and F$_{\rm cont}$ is a third-order polynomial fit to the continuum from 5--7.5 $\mu$m and 12.5--16~$\mu$m. 
		
	{{For each source presented in Table~\ref{tab-10micronresults}, the strength and shape of the 10~$\mu$m silicate feature were calculated by}} using the three {{IRS}} spectrum tables containing the wavelength ranges 5--8.6~$\mu$m, 9--19~$\mu$m or 13--21~$\mu$m, and 7--14~$\mu$m (corresponds to modules SL, SH, LL and SL respectively), with the exception of Taurus-Auriga sources which used two tables (9--19~$\mu$m and 7--14~$\mu$m). Note that the method of finding the 10~$\mu$m strength and shape is sensitive to the wavelength range chosen. Here we used the ranges from 5--7.5~$\mu$m, 12.5--16~$\mu$m, 8--12.4~$\mu$m to determine the 10~$\mu$m strength, and 0.4~$\mu$m bandwidth to determine the 10~$\mu$m shape, which are consistent with the ranges used by \citet{Furlan06} -- {see Figure~\ref{fig-fits} for the third-order polynomial fits to the infrared continuum.}
	
	 No IRS spectrum was available for SY Cha, DK Cha and Sz 111, and only the 9--19~$\mu$m table was available for GQ Lup and RU Lup and hence these sources are not listed in Table~\ref{tab-10micronresults}. T Cha and RXJ1615.3-3255 are isolated sources and were not included in the statistics of the Chamaeleon and Lupus sample. SZ Cha has a strong Polycyclic Aromatic Hydrocarbons emission band at 11.3~$\mu$m and {{thus the bandwidth ranged of 11.28-11.31~$\mu$m was used to determine F$_{11.3  \mu \rm m}$.}} 

	The {{1--3~mm}} spectral slopes $\alpha$ are from Table~\ref{tab-alphas} for Chamaeleon and Lupus, from \citet{Ricci10b} for the $\rho$ Ophiucus sources and from \citet{Rod06,2007ApJ...659..705A} for Taurus-Auriga -- see Table~\ref{tab-10micronresults}.

\begin{table}
	 \centering
	 \caption{Strength and shape of the 10~$\mu$m silicate feature for all Chamaeleon and Lupus sources from this work, and supplementary $\rho$ Ophiucus and Taurus-Auriga sources. (1) Source name. (2) 10~$\mu$m strength values. (3) 10~$\mu$m shape. (4) 1--3~mm spectral slope.}
	 \begin{tabular}{ccccc}
	 \hline \hline											
		Source	&	Strength	&	Shape	&	$\alpha$	\\
		\hline
		\multicolumn{4}{c}{Chamaeleon}\\
		\hline
	CR Cha	&	1.07	&	0.71	&	3.4	\\
	CS Cha	&	0.40	&	0.58	&	2.9	\\
	DI Cha	&	0.42	&	0.84	&	3.2	\\
	Glass I	&	1.19	&	0.72	&	3.1	\\
	SZ Cha*	&	0.60	&	0.80	&	2.9	\\
	Sz 32	&	0.18	&	1.05	&	3.8	\\
	WW Cha$^a$	&	0.77	&	0.78	&	2.8	\\
	\hline
	\multicolumn{4}{c}{Lupus}\\
	\hline
	IK Lup	&	0.13	&	0.83	&	2.4	\\
	HT Lup	&	0.31	&	0.85	&	2.0	\\
	GW Lup	&	0.28	&	0.92	&	2.3	\\
	RY Lup	&	1.24	&	0.60	&	3.9	\\
	HK Lup	&	0.67	&	0.73	&	2.9	\\
	EX Lup	&	0.54	&	0.74	&	2.5	\\
	MY Lup	&	0.55	&	0.81	&	2.3	\\
	IM Lup$^a$	&	0.51	&	0.85	&	3.6	\\
	\hline
	\multicolumn{4}{c}{Isolated}\\
	\hline
	T Cha	&	-0.03	&	0.77	&	3.1	\\
	RXJ1615.3-3255	&	0.93	&	1.02	&	3.6	\\
	\hline
	\multicolumn{4}{c}{$\rho$ Ophiucus}\\
	\hline
	SR 4	&	0.33	&	1.38	&	2.5$^b$		\\
	IRS 49	&	0.51	&	0.98	&	1.8$^b$	\\
	WSB 52	&	0.06	&	1.01	&	1.8$^b$	\\
	WSB 60	&	0.28	&	0.99	&	1.9$^b$	\\
	DoAr 44	&	1.39	&	0.65	&	2.2$^b$	\\
	RNO 90	&	0.39	&	0.79	&	2.3$^b$	\\
	Wa Oph 6	&	0.18	&	0.90	&	2.4$^b$	\\	
	\hline
	\multicolumn{4}{c}{Taurus-Auriga }\\
	\hline
	DG Tau	&	-0.03	&	0.67	&	2.1$^c$	\\
	DO Tau	&	0.18	&	1.24	&	2.3$^c$	\\
	AA Tau	&	0.32	&	0.54	&	3.2$^d$	\\
	CI Tau	&	0.54	&	0.58	&	2.2$^d$	\\
	DL Tau	&	0.10	&	0.61	&	2.0$^d$	\\
	DM Tau	&	0.41	&	0.80	&	2.9$^d$	\\
	DN Tau	&	0.14	&	0.62	&	2.3$^d$	\\
	DR Tau	&	0.22	&	0.54	&	2.2$^d$	\\
	FT Tau	&	0.20	&	0.61	&	1.8$^d$	\\
	GM Aur	&	0.84	&	0.48	&	3.2$^d$	\\
	AS 205	&	0.50	&	0.54	&	0.7$^d$	\\
	\hline
	 \end{tabular}	
	 \label{tab-10micronresults}	
	 	\begin{tablenotes} 
		\item[1] * {{SZ Cha has a}} strong Polycyclic Aromatic Hydrocarbons emission band at 11.3~$\mu$m.
		\item[2] $^a$ Additional Chamaeleon and Lupus sources from \citet{Lommen10}. 
		\item[2] $^b$ $\alpha$ obtained from \citet{Ricci10b}.
		\item[3] $^c$ $\alpha$ obtained from \citet{Rod06}.
		\item[4] $^d$ $\alpha$ obtained from \citet{2007ApJ...659..705A}.
	\end{tablenotes}
	  \end{table}

\begin{figure}
\centering
	    \includegraphics[width=5cm]{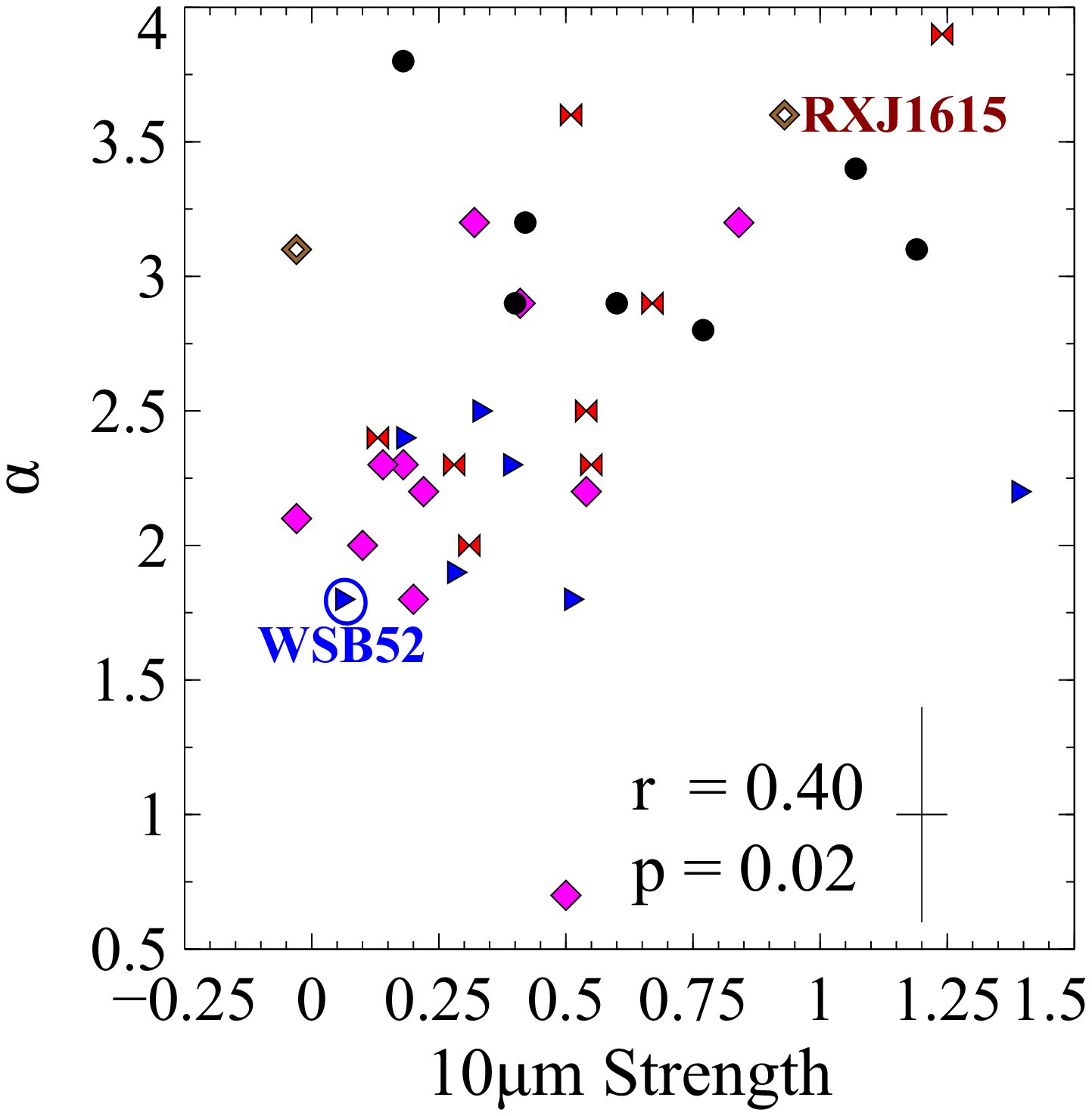}\\
	    \includegraphics[width=5cm]{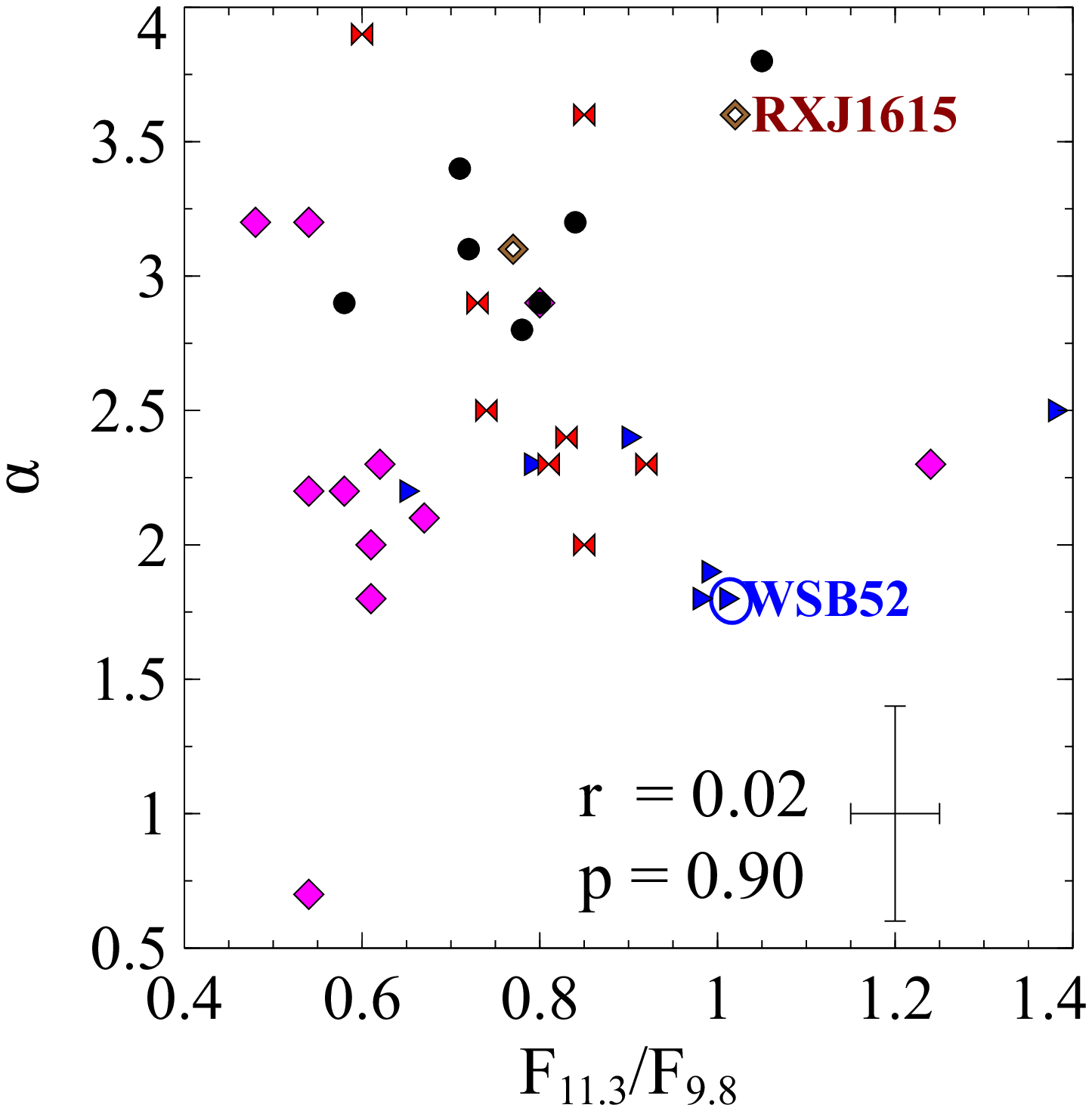}
	    \caption{The 1--3~mm spectral slope $\alpha$ as a function of the 10~$\mu$m strength (top) and shape (bottom) using data from Table~\ref{tab-10micronresults}. Black circles are Cham sources, red {{ties}} are Lupus sources, {{blue triangles}} are $\rho$ Ophiucus, {{open diamonds}} are the isolated sources and pink diamonds are Taurus-Auriga sources. {{The obtained Spearman rank correlation coefficient $r$ and the corresponding p-value (probability the obtained r value happen by chance) are given for the 10~$\mu$m strength and $\alpha$ and the 10~$\mu$m shape and $\alpha$.}}}
	    \label{fig-10micron}
\end{figure}

The millimetre spectral slope {{$\alpha$}} as function of 10~$\mu$m strength and shape where plotted in Fig.~\ref{fig-10micron}.
{{To test these two correlations we used the statistical package \texttt{R} \citep{R2011} to determine the Spearman rank correlation coefficient, where the rank coefficient, $r$, is a measure of the how two values are related ($r=0$ weak correlation, $r=\pm1$ strong correlation), p-value is a measure of the probability the obtained r value happen by chance.}} With a p-value of {0.17 and 0.39, we found the rank coefficients for the 15 sources in Chameleon and Lupus to be  $r = 0.37$ and $r = -0.24$ for {{strength versus $\alpha$ and shape versus $\alpha$}} respectively}
~-- see Table~\ref{tab-Spearman}. Thus we conclude that the {{correlations between 10~$\mu$m strength versus $\alpha$ and the 10~$\mu$m shape versus $\alpha$ {cannot be confirmed}}} with this sample.
{Spearman statistics was {{then}} performed on the sample of 36 sources from the four star forming regions and on each individual star forming region {{listed in Table~\ref{tab-10micronresults}.

The results suggest no correlation is present for each individual star forming regions (Lupus 1 and 3 only have three sources, thus were not included in the individual analysis) -- see Table~\ref{tab-Spearman}. A weak correlation between the strength of the {{10~$\mu$m}} silicate feature and $\alpha$ is found for the whole sample, while no correlation appears present between the shape of the {{10~$\mu$m}} silicate feature and $\alpha$.

There are multiple processes affecting the shape of the 10~$\mu$m silicate feature (grain growth, crystallisation and opacity) that would make the detection and interpretation of a possible correlation difficult. For example both RXJ1615.3-3255 and WSB~52 have a 10~$\mu$m shape of 1.02 and 1.01 respectively, {{suggesting}} a flat shape structure for both sources. However, this is not the case RXJ1615.3-3255 has a strong, boxy {{ 10~$\mu$m}} feature, while WSB~52 has a very flat structure -- see Fig.~\ref{fig-fits}. It was only when Spitzer allowed for sufficient statistics that such ambiguities were cancelled out and the correlation between the strength and the shape of the 10~$\mu$m feature was confirmed \citep[e.g.][]{2006ApJ...639..275K}.

{{\citet{2006ApJ...639..275K}  have shown a similar correlation exists between the 10 and 20~$\mu$m strengths, with some outliers with strong 10~$\mu$m features but weak 20~$\mu$m features. To reproduce these results, they had to increase the size of the grains emitting at 20~$\mu$m to a minimum radius of 2--3~$\mu$m, implying the emission at 20~$\mu$m is from a region in the disc deeper than the grains emitting at 10~$\mu$m with a minimum radius of 0.1~$\mu$m. A recent study performed on brown dwarfs by \citep{2012MNRAS.420.2603R} has found the flattening and shift of the peak position of the 10~$\mu$m silicate feature to have no correlation with an increase in the grain sizes or a higher degree of crystallinity in the disc. The four outliers identified in the study have a mix of small, large and crystalline grains, with a weak 20~$\mu$m. Other studies have found similar outliers in T~Tauri stars which were explained by enhanced dust settling in the outer disc region \citep{2007ApJ...659..680K}. \citet{2012MNRAS.420.2603R} suggest that the presence of these outliers indicates that the timescales for dust processing are different for the inner and outer regions of discs.}}

Compared to previous analysis of this correlation, our sample is the largest  (36 sources compared to 7 sources in \citet{Lommen07}, 30 sources in \citet{Lommen10} and 21 sources in \citet{Ricci10b}), excludes sources for which only upper limits on $\alpha$ are known, contains no upper limits for 1--3~mm spectral slope $\alpha$, and all the IR data were process and analysed consistently. However, the sources used may have several intrinsic differences. In particular, Cha~I and $\rho$ Ophiucus are very different environments, with Cha~I sources being fairly isolated \citep[e.g.][]{2007ApJS..173..104L} {{while}} $\rho$ Ophiucus {{sources}} are deeply embedded \citep[e.g.][]{2010ApJS..188...75M}. Although the $\rho$ Ophiucus  values in Table~\ref{tab-10micronresults} were not corrected for extinction, similar results were obtained by \citet{Ricci10b} using extinction corrected values for the {{10~$\mu$m}} shape. 
Ideally we would {{like}} the same sample number per star forming region for this type of analysis. A larger sample may also mitigate the intrinsic differences in the 10~$\mu$m silicate feature, allowing for a more definite conclusion on the presence or absence of a correlation with $\alpha$.

\begin{table*}
	 \centering
	 \caption{Spearman rank correlation coefficient values and percent confidence levels (p-values) for each {{star forming region in Table~\ref{tab-10micronresults}}} used in the correlation analysis for the 10~$\mu$m silicate feature strength and shape with the 1-3 mm spectral slope $\alpha$.} 
	 \begin{tabular}{cccccc}
	 \hline \hline
			&   &     \multicolumn{2}{c}{Strength vs. $\alpha$} & \multicolumn{2}{c}{Shape vs. $\alpha$} \\
			&      $\#$ of sources    & rank coeff. &  p-value & rank coeff. & p-value  \\			
	\hline
Cham \& Lupus	&	15	&	0.37	&	0.17	&	-0.24 	&	0.39	\\
All	&	36	&	0.40	&	0.02	&	0.02	&	0.90	\\
\hline															
Cha~I	&	7	&	-0.22	&	0.64	&	0.36	&	0.43	\\
\hline
$\rho$ Ophiucus	&	7	&	0.02	&	0.97	&	-0.04	&	0.93	\\
Taurus	&	11	&	0.32	&	0.34	&	-0.05	&	0.87	\\
	\hline
	 \label{tab-Spearman}	
	 \end{tabular}	
\end{table*}
	
\section{Conclusions}
\label{sec-summary}

Continuum observations were carried out with ATCA at  3, 7, 15~mm, and 3+6~cm for 20 T Tauri stars located in Chamaeleon and Lupus. We analysed the mm fluxes in order to determine the spectral slopes, maximum grain sizes, and dust masses in these discs. Using supplementary data from the literature we conducted temporal monitoring of the {{fluxes of a subsample of our sources}} over short (less than one day) and long (months to years) timescales to help {{constrain}} the emission mechanisms present in {{these}} discs. We also analysed the potential correlation between the millimetre spectral slope and the strength {{and shape}} of the 10~$\mu$m silicate feature suggested by \citet{Lommen07,Lommen10}.

\begin{enumerate}
\item{Our  3 and 7~mm continuum fluxes {{show}} that 11 sources do not have a break in the spectral slope at 7~mm, suggesting thermal dust emission is dominant to wavelengths as long as 7~mm. We found that 6 of those sources (all in Lupus) have a dust opacity index less than unity (assuming $\beta \sim \alpha-2$), suggesting grain growth up to at least mm sizes. }

\item{We obtained dust disc mass estimates ranging from {{$10^{-5}$--$10^{-3}~$M$_\odot$}}. Assuming a gas-to-dust ratio of 100, we find eight sources have {{a total disc mass (gas+dust)}} greater than 0.01~M$_{\odot}$, the minimum mass solar nebula according to \citet{1977MNRAS.180...57W}; and six have a {{total disc mass}} greater than 0.02~M$_{\odot}$, the minimum mass solar nebula according to \citet{1981PThPS..70...35H}.}

\item{All six Chamaeleon sources observed at 15~mm, have excess emission above thermal dust. At 3+6~cm, DK Cha, T Cha and Sz 32 have flux upper limits which suggest the emission at 3+6~cm is due to an ionised wind and/or chromospheric emission.}

\item{The monthly and yearly temporal flux monitoring revealed no excess emission {at 7~mm for MY Lup, Sz 111 and CS Cha, indicating thermal dust emission dominates in these sources and hence that they have grains up to $\sim$1~cm in size. 
At 15~mm the excess emission for CR Cha, CS Cha, DI Cha, T Cha, Sz 32 and DK Cha is not from a fast varying chromospheric emission, but most likely from thermal free-free emission consistent with the spectral slope of 0.6.}}

\item{Supplementing our sample with 16 other sources from Taurus-Auriga and $\rho$ Ophiucus, no correlation between the shape of the 10~$\mu$m silicate feature and the 1-3 mm spectral slope $\alpha$ was found, while the strength of the 10~$\mu$m feature appears to correlate weakly with $\alpha$.}

\end{enumerate}

ALMA will be vital in obtaining high resolution maps at high sensitivities of protoplanetary discs in the southern hemisphere at (sub)mm wavelengths, providing a better understanding of disc sizes and further analysis of mm grain growth signatures. However, at 7 and 15~mm, the ATCA {and VLA}
~{{will continue to provide invaluable}} information on other emission mechanisms present in protoplanetary discs. 

\section*{Acknowledgements}
Special thanks the ATCA staff, the Spitzer Science Center Helpdesk {{and}} Leonardo Testi, Lucca Ricci and Fran\c{c}ois Menard for useful discussions. Thanks also to Philip Edwards for awarding some discretionary time to help complete this project. {{We would also like to thank the referee for the useful comments and suggestions.}} This research was supported in part by a CSIRO OCE Postgraduate Top Up Scholarship. CMW acknowledges support from the Australian Research Council through Discovery Grant DP0345227 and Future Fellowship FT100100495.Ó This research has made use of the NASA/IPAC Infrared Science Archive, which is operated by the Jet Propulsion Laboratory, California Institute of Technology, under contract with the National Aeronautics and Space Administration, and data products from the Two Micron All Sky Survey, which is a joint project of the University of Massachusetts and the Infrared Processing and Analysis Center/California Institute of Technology, funded by the National Aeronautics and Space Administration and the National Science Foundation. 

\bibliographystyle{mn2e}
\bibliography{biblio}{}

\appendix
\label{app1}

\section{Comments on individual sources}
\label{app-individual}

\subsection{Chamaeleon}
\label{app-sub-cham}

Chamaeleon comprises of three clouds -- Cha~I, II and III -- {{located at a distance}} of $\sim$150--160~pc \citep{2008hsf2.book..169L,2009ApJ...703.1964F}. This survey focused on T Tauri stars on the Cha~I cloud (with the exception to DK Cha in Cha~II). Cha~I has two subclusters, with ages ranging from 3--4 and 5--6 Myr in the southern and northern subcluster respectively \citep{2007ApJS..173..104L}.

\textit{\textbf{CS Cha}} is a close binary with a $\lesssim$ 0.1$^{\prime\prime}$ separation \citep{2007ApJ...664L.111E}. {{A near-infrared}} flux deficit was detected, suggestive of an inner disc hole, modelled to be $\sim$43~AU having $\sim 5 \times 10^{\--5}$ lunar masses of dust from 0.1 to 1~AU \citep{2007ApJ...664L.111E}. \citet{2009ApJ...700.1017K} observed an excess at 5--8~$\mu$m higher than typical T Tauri stars.

\textit{\textbf{DI Cha}} is a binary system with $\sim4^{\prime\prime}$.6 separation \citep{1997ApJ...481..378G,2006A&A...459..909C}. The binary is yet to be resolved in (sub)mm wavelengths \citep{H1993,Lommen07}.

\textit{\textbf{T Cha}} is an isolated source. \citet{2007A&A...476..279G} detected a weak 11.2~$\mu$m feature and a marginal 3.3 $\mu$m feature. Both features were unresolved, corresponding to a distance {{$\geq$ 13 \rm and $< 22$ AU}} respectively. \citet{2007ApJ...664L.107B} detected a possible {{gap in the disc at 66~AU}}. 

\textit{\textbf{Glass I}} is {{a binary}} with separation $\sim 2^{\prime\prime}.5 \pm 0.5$ \citep{1989ApJ...338..262F,1993A&A...278...81R, 1997ApJ...481..378G,2004ApJ...602..816L}. Glass Ia is a K4 WTTS \citep{1988A&A...207...46C} and Glass Ib an IR G-type source \citep{1989ApJ...338..262F}.  The 10~$\mu$m feature was assigned to Glass Ib, no detection was obtained for Ia \citep{2000IAUS..200P..38S}.  The binary system has not been resolved at mm wavelengths.

\textit{\textbf{SZ Cha}} was classified by \citet{2011ApJS..193...11M} as {an object with an inner and outer optically thick disc separated by optically thin gap} with the 10~$\mu$m silicate feature in emission. It is an accreting source \citep{2004ApJ...602..816L} with two potential companions 2M J0581804-777197 at $5^{\prime\prime}.2$ and 2M J10581413-7717088 at $12^{\prime\prime}.5$ \citep{1997ApJ...481..378G}, which are not confirmed members of the Cha~I association \citep{2007ApJS..173..104L, 2008hsf2.book..169L}. The IRS spectra from \citet{2009ApJ...700.1017K} detected polycyclic aromatic hydrocarbon emission at 11.3~$\mu$m.

\textit{\textbf{DK Cha}} is the brightest IR and submm source in Cha~II \citep{1997A&A...327.1194W}, found to be transitioning from {an embedded disc to a protoplanetary disc} through modelling \citep{2010A&A...518L.128V}, and still surrounded by a protostellar envelope and molecular outflow \citep{2006A&A...454L..75V,2009A&A...508..259V}. 

\subsection{Lupus}
\label{app-sub-lup}

The Lupus southern star forming region is composed of 8 clouds (Lupus 1 trough 8) \citep{1999PASJ...51..895H}. This work focused on sources in Lupus 1, 3 and 4 at distances of 150$\pm$20~pc, 200$\pm$20~pc and 165$\pm$15~pc respectively \citep{2008hsf2.book..295C}.

\textit{\textbf{IK Lup and Sz 66}} are a binary pair with a 6$^{\prime\prime}$.4 (960~AU) separation \citep{1977ApJS...35..161S, 1993A&A...278...81R,2006A&A...446..201M}. {{Sz 66 was not detected for the 1.2 and 3.2~mm observations by \citet{Lommen10}, in both cases the beam size was sufficient to resolved the binary system. Sz 66 was detected with Spitzer and has a 10~$\mu$m silicate feature -- see {\citet{Lommen10}. However, the lack of cold dust from 1.2 to 7~mm would suggest there is no circumbinary disc, and if a cold dust disc is present its mass is too low to have been detected by our observations.}}

\textit{\textbf{HT Lup}} is a triple system (Ghez et al., 1997), with one companion at 2$^{\prime\prime}.8\pm0.1$ (Reipurth \& Zinnecker 1993; Brandner et al., 1996) and another at 0$^{\prime\prime}.107\pm$0.007 (Ghez et al. 1997). Heyer \& Graham (1989) found a reflection nebula near HT Lup and H$\alpha$ emission possibly associated with an HH jet. 

\textit{\textbf{GQ~Lup}} has a known companion, GQ~Lup B, at a distance of $0.7^{\prime\prime}$, which may be a brown dwarf \citep{2005AN....326..958G,2008poii.conf..539N} or a planet \citep{2008A&A...484..281N}. The system was observed at 1.3~mm with the Submillimetre Array by \citet{Dai10} who did not detect thermal dust emission from QQ~Lup~B and model a compact disc of mass $\sim$M$_{\rm Jup}$ around GQ~Lup{{, which is consistent with our disc mass estimate.}} 

\textit{\textbf{Sz 111}} was identified as a cold disc by \citet{2008ApJS..177..551M} who suggests the fluxes from the IRAC and MIPS band 1 were more representative of a debris disc with no gas present (where $\alpha_{\rm K-24\mu\rm m} < -1.6$), while showing an excess at 70~$\mu$m typical of a classical T Tauri star. Cold discs were defined by \citet{2005ApJ...630L.185C} and \citet{2007ApJ...664L.107B} as optically thick discs with large inner holes, where potential large grain growth may occur. Sz 111 inner hole was determined to be $\sim70$ AU \citep{2008ApJS..177..551M}.

\textit{\textbf{EX Lup}} is a highly variable M-type star with no known companion \citep{2007AJ....133.2679H,2008IBVS.5819....1K}. It is located at the edge of a gap between Lupus 3 and 4 \citep{1999A&A...345..965C}. It has a quiescent accretion rate of $\sim 4 \times 10^{-10} \rm M_{\odot} /  \rm yr$ \citep{2004ASPC..323..279N,2009A&A...507..881S}{{, though five}} flares-ups have been observed from 1995--2008 \citep{2001PASP..113.1547H,2007AJ....133.2679H,2010A&A...522A..56G,2010ApJ...719L..50A,2011ApJ...728....5G}. In February of 2008, \citet{2008IBVS.5819....1K} observed the extreme outburst of EX Lup. They found there a wealth of metallic lines with no added absorption features, which lead to the conclusion the brightness increase was due to an increase in accretion and possibly stellar winds. {{Modelling by}} \citet{2009A&A...507..881S} found the 10 $\mu$m feature could be recreated with amorphous silicates with olive and pyroxene stoichiometry, and the disc an inner and outer radius of 0.2, 150~AU respectively. They concluded the inner hole may be the cause of the outburst, however more evidence {{is required to confirm this hypothesis.}}

\section{Observing details}
\label{ap-sec-ob-details}

{The Australia Telescope Compact Array (ATCA) was used to observed 9 sources in Chamaeleon and 11 source in Lupus star forming regions at 3, 7, 15 mm and 3+6 cm. A log of the ATCA observations is given in Table~\ref{tab-obdetails}.}

 \begin{table*}
 \caption{ATCA observation log. (1) Observation date. (2) Sources observed. (3) Total integration time used in analysis in this paper. (4) Central frequency pair of observations. (5) Array configuration$^*$. (6) Notes.}
 \begin{tabular}{cccccc}
 \hline \hline
	Observation dates	&	Sources	&	T$_{\rm{int}}$ 	&	Frequency pairs &	Array &	Notes	\\
	 & & (min) & (GHz) & Config. & \\
	 \hline
	May-09$^a$	&	SZ Cha	&	200	&	43, 45	&	 H214 & \\
		&	CS Cha	&	80	& & \\
		&	CR Cha	&	80	& & \\
		&	Glass I	&	150	& & \\
		&	Sz 32	&  	130	& & \\
		&	DI Cha	&	100 & & 	\\
		&	T Cha	&	60	& & \\
	May-09$^b$	&	Sz 65-Sz 66	&	70	&	43, 45	&	 H214  &	\\
		&	HT Lup	&	40	&		&		\\
		&	GW Lup	&	60	&		&		\\
		&	RY Lup	&	60	&		&		\\
		&	EX Lup	&	60	&		&		\\
		&	MY Lup	&	50	&		&		\\
		&	Sz 111	&	60	&		&		\\
		&	RXJ1615.3-3255	&	50	&		&		\\
		&	HK Lup	&	50	&		&		\\
				 \hline
	May-10$^a$	&	Glass I	&	50	&	93, 95	 &	H214 &	\\
		&	DI Cha	&	104	&		&		\\
		&	Sz 32	&	180	&		&		\\
		\hline
	Jul-10$^a$ &	DI Cha	&	80	&		&	EW352 &	\\
		&	DK Cha	&	43	&		&		\\
		&	CS Cha	&	60	&		&		\\
	Jul-10$^a$ &	DK Cha	&	54	&	43, 45	&	EW352 &	\\
		\hline
	Aug-10$^b$	&	EX Lup	&	70	&	93, 95 	&	H168. & Antenna 2 offline.  \\ 
		&	GQ Lup	&	44	&		&		\\
		&	ATCA160043-415442	&	73	&		&		\\
	Aug-10$^b$	&	RXJ1615.3-3255	&	60	&	43, 45 	&       H168 &	\\
		&	GQ Lup	&	80	&		&		\\
		&	SY Cha	&	30	&		&		\\
		\hline
	Jul-11$^c$ & DK Cha & 18& 17, 19 & H214 & Errors with antenna 6 polarization. \\
		   & Sz 32 and WW Cha & 42 &  &  \\
		   & T Cha &  39 & & \\
		   & DI Cha  & 100 & & \\
		   & CR Cha &  42 & & \\
		   & CS Cha &  105 & & \\
	Jul-11$^c$ & DK Cha & 180 & 5.5, 9 &EW352 & July 28 -- antenna 5 offline. \\
		   &  &  & & & July 29 -- antenna 3 offline for last hour.  \\
		   &  &  & & & Flux calibrator {{bootstrapped}} from gain calibrator. \\
		  & Sz 32 and WW Cha & 125 & & &\\
		  & T Cha &  170 & & &\\
	\hline	 
\end{tabular}
	\begin{tablenotes} 
	         \item[1] $^*$  ATCA array configuration information: http://www.narrabri.atnf.csiro.au/operations/array\_configurations/upcoming\_configs.html
		\item[2] $^a$ Phase calibrator QSO~B1057-797, primary flux calibrator Uranus.
		\item[3] $^b$ Phase calibrator QSO~B1600-44, primary flux calibrator Uranus.
		\item[4] $^c$ Phase calibrator QSO~B1057-797, primary flux calibrator QSO~B1934-638.
	\end{tablenotes}
 \label{tab-obdetails}
 \end{table*}

\section{Result details}
\label{app-resultdetails}

{The complete results used in Section~\ref{subsec-resolved} analysis to determine if a source was extended are given in Table~\ref{tab-results} and the full spectral energy distribution plots are given in Figures~\ref{fig-chamSED}~and~\ref{fig-lupSED}.}

 \begin{landscape}
 \begin{table}
 \caption{{Summary of 3 and 7~mm flux fittings. (1) Source name. (2), (3)}
 ~Point source and Gaussian fit fluxes. (4) Gaussian size obtained from the Gaussian fit. (5) Synthesised beam size using natural weighting. (6) Image RMS (natural weighting). (7) Factor of $\sigma$ the point source fit is below the Gaussian fit (F$_{\rm g}$=F$_{\rm p}$+n$\sigma$). (8) {Description of the amplitude as a function of u-v distance plots (UVAMP) presented in Figs.~\ref{fig-uvamps-3mm}~and~\ref{fig-uvamp-lup-7mm}.}
 ~(9) If the emission was considered resolved or not.}	
 \begin{tabular}{ccccccccc}
 \hline \hline
	Sources	&	F$_{\rm P}$	&	F$_{\rm G}$	&	Gaussian size	&	Beam size	&	RMS	&	n	&	UVAMP$^a$	&	Resolved$^b$	\\
		&	(mJy)	&	(mJy)	&	(arcsec)	&	(arcsec)	&	(mJy/beam)	&	F$_{\rm g}$=F$_{\rm p}$+n$\sigma$	&	F,D,S	&	Y,N	\\
	\hline	
	\multicolumn{9}{c}{3 mm} \\
	\hline
	CS Cha	&	$7.8\pm0.8$	&	$9.4\pm0.9$	&	$14.3\pm1.9\times1.8\pm0.2$	&	13.2$\times$1.4	&	0.6	&	2.7	&	D	&	Y	\\
	DI Cha	&	$2.3\pm0.5$	&	$2.3\pm0.4$	&	4.2$\pm1.4\times1.9\pm0.6$	&	4.6$\times$1.8	&	0.3	&	0.0	&	F	&	N	\\
	Glass I	&	$4.1\pm0.1$	&	$4.4\pm0.1$	&	2.5$\pm0.2\times2.0\pm0.1$	&	2.4$\times$1.8	&	0.1	&	3.0	&	F	&	N	\\
	Sz 32	&	$3.1\pm0.3$	&	$3.1\pm0.2$	&	2.4$\pm0.6\times1.9\pm0.5$	&	2.4$\times$1.8	&	0.2	&	0.0	&	F	&	N	\\
	DK Cha	&	$33.0\pm3.0$	&	$49.8\pm1.3$	&	$7.7\pm0.5\times2.4\pm0.2$	&	7.1$\times$1.3	&	1.3	&	12.9	&	D	&	Y	\\
	GQ Lup 	&	$3.6\pm0.3$	&	$3.5\pm0.8$	&	2.9$\pm0.9\times1.4\pm1.1$	&       2.8$\times$2.0	&	0.4	&	-0.3	&	F	&	N	\\
	EX Lup    	&	${{2.0\pm0.2}}$	&	$1.9\pm0.3$	&	${{3.2\pm0.4\times1.7\pm0.2}}$	&	${{2.9\times1.9}}$	&	0.3	& -0.3	&	F	&	N	\\
	\hline \hline
	\multicolumn{9}{c}{7 mm} \\		
	\hline																	
	\multicolumn{9}{c}{Chamaeleon}																					\\
	\hline
	SY Cha    	&	$<$1.0$^c$	&	---	&	---	&	16.0$\times$5.4	&	0.3	&	---	&	---	&	---	\\
	CR Cha	&	$1.7\pm0.1$	&	$1.8\pm0.1$	&	5.1$\pm0.4\times4.8\pm0.4$	&	5.0$\times$4.3	&	0.1	&	1.0	&	D	&	N	\\
	CS Cha	&	$1.5\pm0.1$	&	$1.7\pm0.1$	&	$6.0\pm0.6\times4.8\pm0.5$	&	5.0$\times$4.3	&	0.1	&	2.0	&	D	&	N	\\
	DI Cha	&	$0.9\pm0.1$	&	$0.9\pm0.1$	&	4.7$\pm0.5\times4.4\pm0.4$	&	4.8$\times$4.2	&	0.1	&	0.0	&	F	&	N	\\
	T Cha	&	$3.0\pm0.1$	&	$3.0\pm0.1$	&	5.2$\pm0.2\times4.1\pm0.1$	&	5.1$\times$4.1	&	0.1	&	0.0	&	F	&	N	\\
	Glass I	&	$0.7\pm0.1$	&	$0.7\pm0.2$	&	5.2$\pm0.9\times4.5\pm0.8$	&	4.9$\times$4.1	&	0.1	&	0.0	&	F	&	N	\\
	SZ Cha	&	$0.7\pm0.1$	&	$0.8\pm0.1$	&	5.8$\pm0.9\times4.7\pm0.8$	&	5.0$\times$4.1	&	0.1	&	1.0	&	F	&	N	\\
	Sz 32	&	$1.2\pm0.1$	&	$1.2\pm0.1$	&	5.3$\pm2.0\times3.8\pm1.4$	&	5.0$\times$4.0	&	0.2	&	0.0	&	S	&	N	\\
	DK Cha	&	$6.6\pm0.6$	&	$8.4\pm0.2$	&	$54.9\pm4.5\times3.1\pm0.3$	&	47.4$\times$2.5	&	0.7	&	2.6	&	D	&	Y	\\						
	\hline			
	\multicolumn{9}{ c}{Lupus}																					\\
	\hline
	Sz 65-66	&	$0.9\pm0.7$	&	$1.0\pm0.1$	&	4.9$\pm0.5\times3.5\pm0.4$	&	4.5$\times$3.5	&	0.1	&	1.0	&	F	&	N	\\
	HT Lup	&	$3.4\pm0.1$	&	$3.5\pm0.1$	&	4.6$\pm0.1\times3.5\pm0.1$	&	4.4$\times$3.5	&	0.1	&	1.0	&	F	&	N	\\
	GQ Lup	&	$0.6\pm0.1$	&	$0.6\pm0.1$	&	4.2$\pm1.3\times1.1\pm2.8$	&	6.7$\times$4.4	&	0.1	&	0.0	&	F	&	N	\\
	GW Lup	&	$0.6\pm0.1$	&	$0.6\pm0.1$	&	5.2$\pm1.0\times3.6\pm0.7$	&	4.8$\times$3.5	&	0.1	&	0.0	&	F	&	N	\\
	RY Lup	&	$1.0\pm0.1$	&	$1.0\pm0.3$	&	4.9$\pm0.3\times3.5\pm0.2$	&	4.7$\times$3.5	&	0.1	&	0.0	&	F	&	N	\\
	HK Lup	&	$1.0\pm0.1$	&	$1.0\pm0.1$	&	5.1$\pm0.5\times3.4\pm0.3$	&	5.2$\times$3.5	&	0.1	&	0.0	&	F	&	N	\\
	Sz 111	&	$0.5\pm0.1$	&	$0.5\pm0.1$	&	5.8$\pm1.4\times3.3\pm0.8$	&	5.2$\times$3.5	&	0.1	&	0.0	&	F	&	N	\\
	EX Lup	&	$1.4\pm0.1$	&	$1.4\pm0.1$	&	5.1$\pm0.5\times3.5\pm0.9$	&	4.8$\times$3.5	&	0.1	&	0.0	&	F	&	N	\\
	MY Lup	&	$1.1\pm0.1$	&	$1.1\pm0.3$	&	4.8$\pm0.7\times3.7\pm0.5$	&	4.8$\times$3.4	&	0.2	&	0.0	&	S	&	N	\\	
	RXJ 1615.3-3255	&	${{0.8\pm0.2}}$	&	${{0.8\pm0.2}}$	&	${{7.2\pm4.4\times6.2\pm4.9}}$	&	${{9.5\times4.1}}$	&	0.2	& 0.0 	&	D	&	N	\\
	\hline
 \end{tabular}
 \label{tab-results}	
   	\begin{tablenotes}
	   	\item[1] $^a$ {UVAMP: F -- flat, D -- drop, S -- sinusoidal.}
		\item[2] $^b$ {Resolved: Y -- yes, N -- no. }
		\item[3] $^c$ A 3$\sigma$ upper limit is given for non-detections.
	\end{tablenotes}	
\end{table}
\end{landscape}

\begin{figure*}
\subfloat{\label{fig:sycha-fullsed}\includegraphics[width=.68\columnwidth]{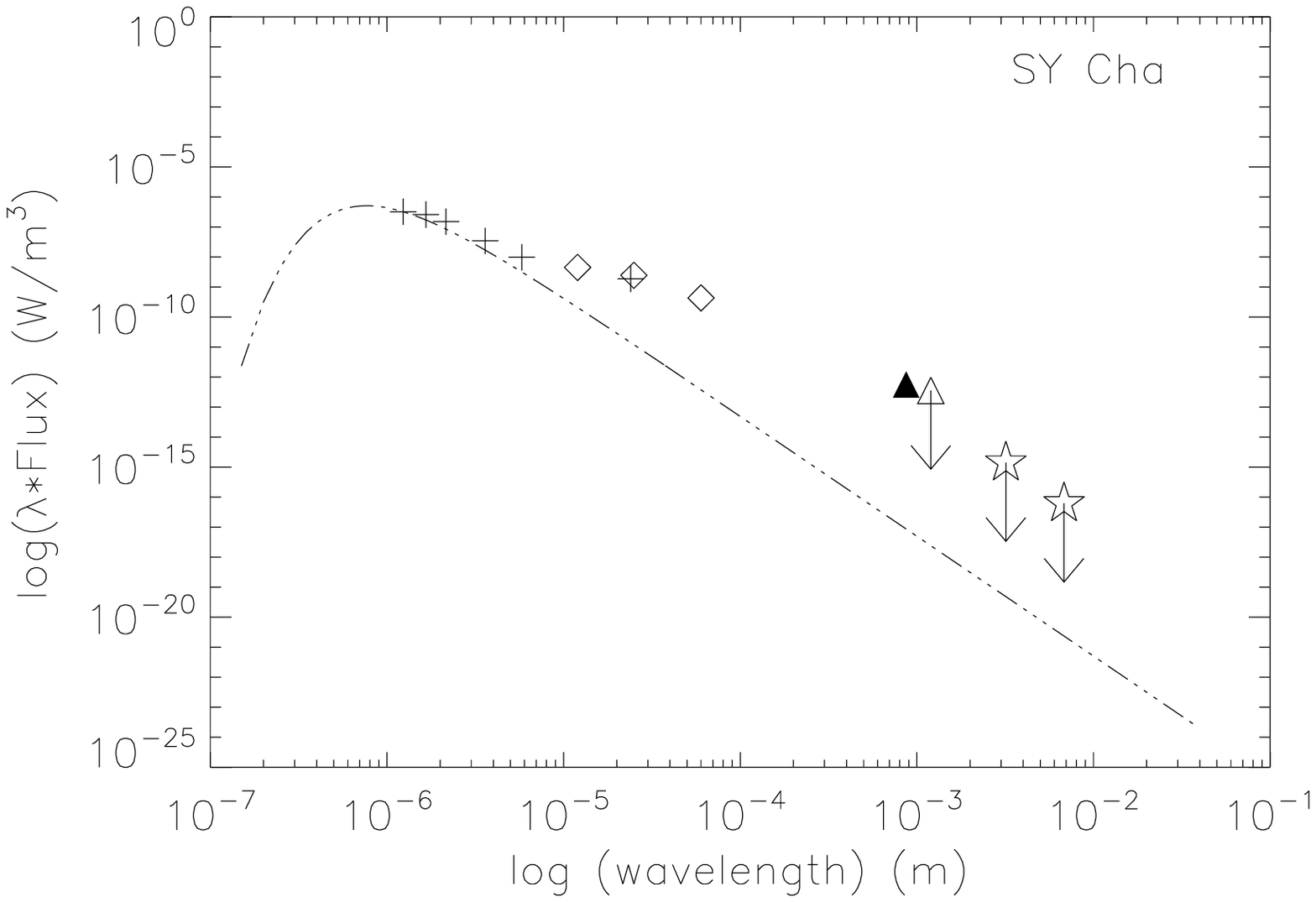}}
\subfloat{\label{fig:crcha-fullsed}\includegraphics[width=.68\columnwidth]{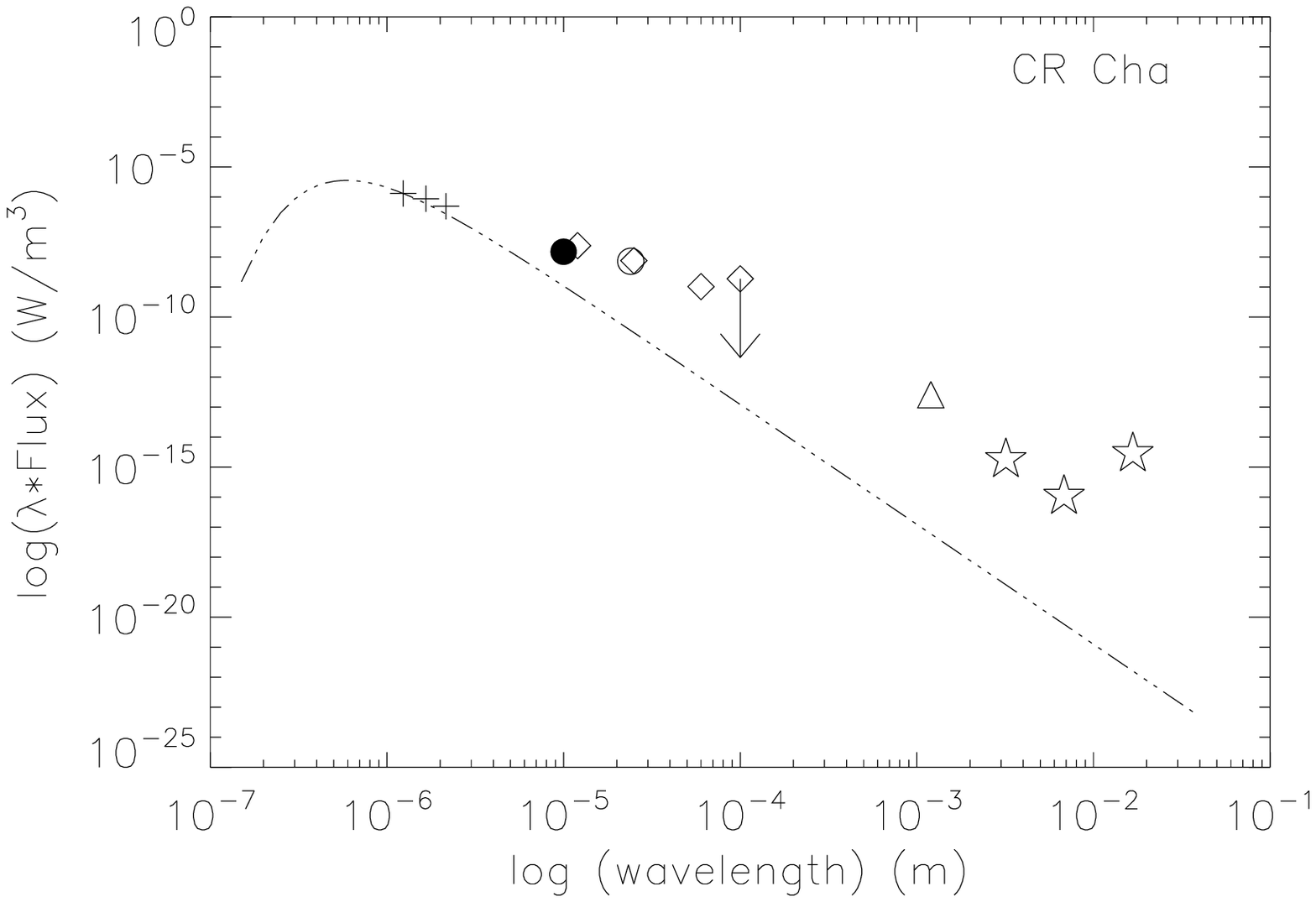}}
\subfloat{\label{fig:cscha-fullsed}\includegraphics[width=.68\columnwidth]{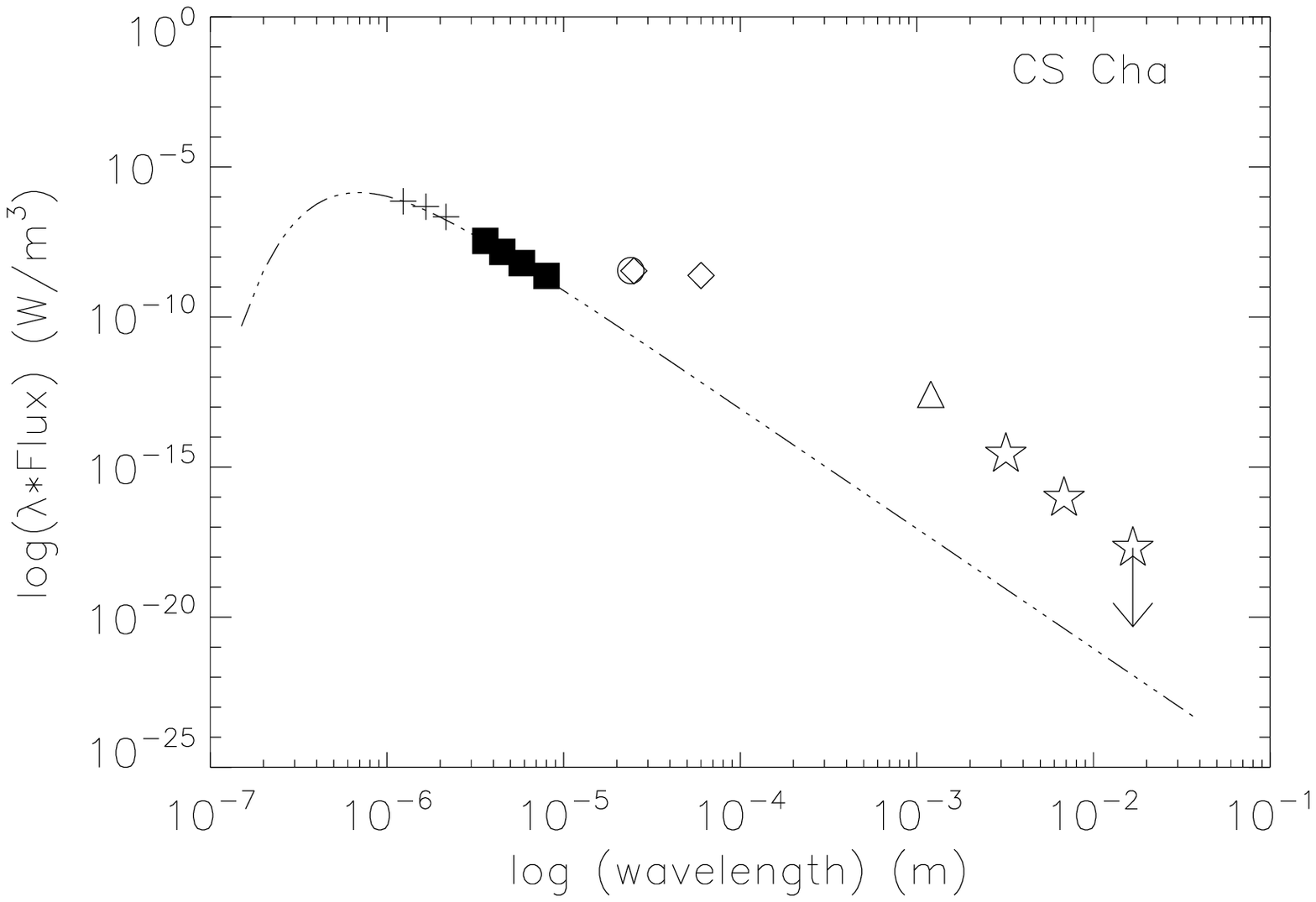}} \\     
\subfloat{\label{fig:dicha-fullsed}\includegraphics[width=.68\columnwidth]{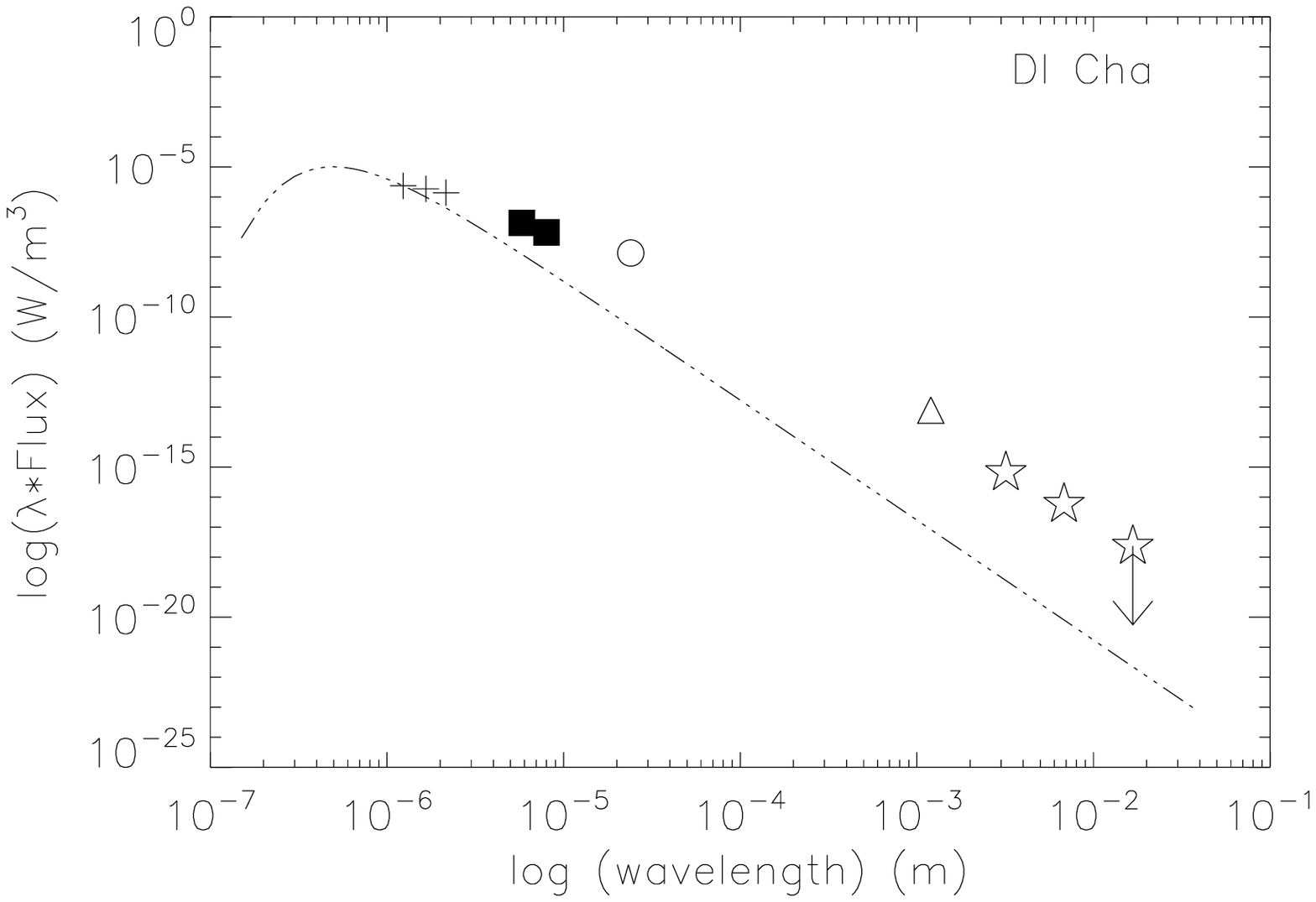}} 
\subfloat{\label{fig:tcha-fullsed}\includegraphics[width=.68\columnwidth]{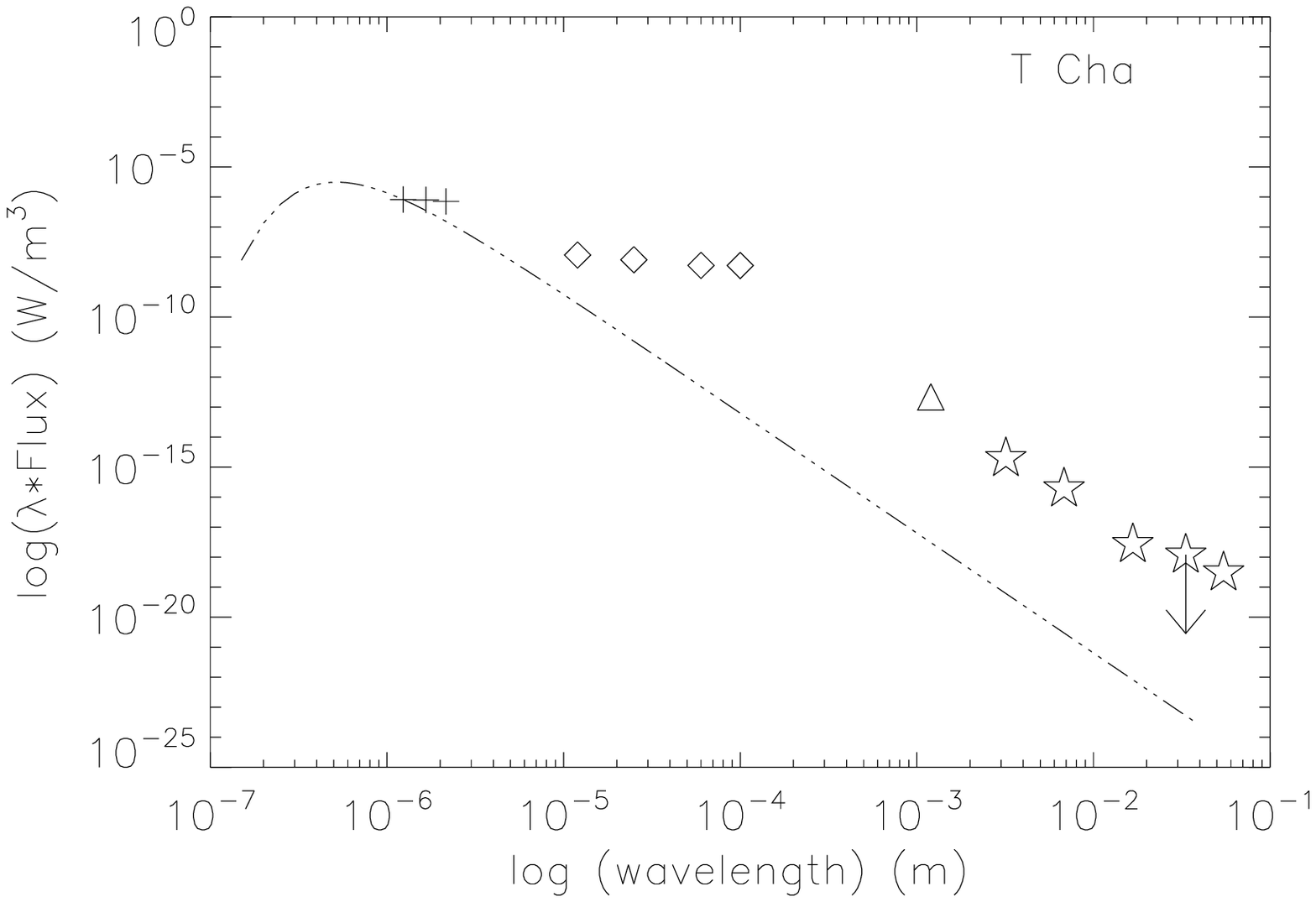}}
\subfloat{\label{fig:glassi-fullsed}\includegraphics[width=.68\columnwidth]{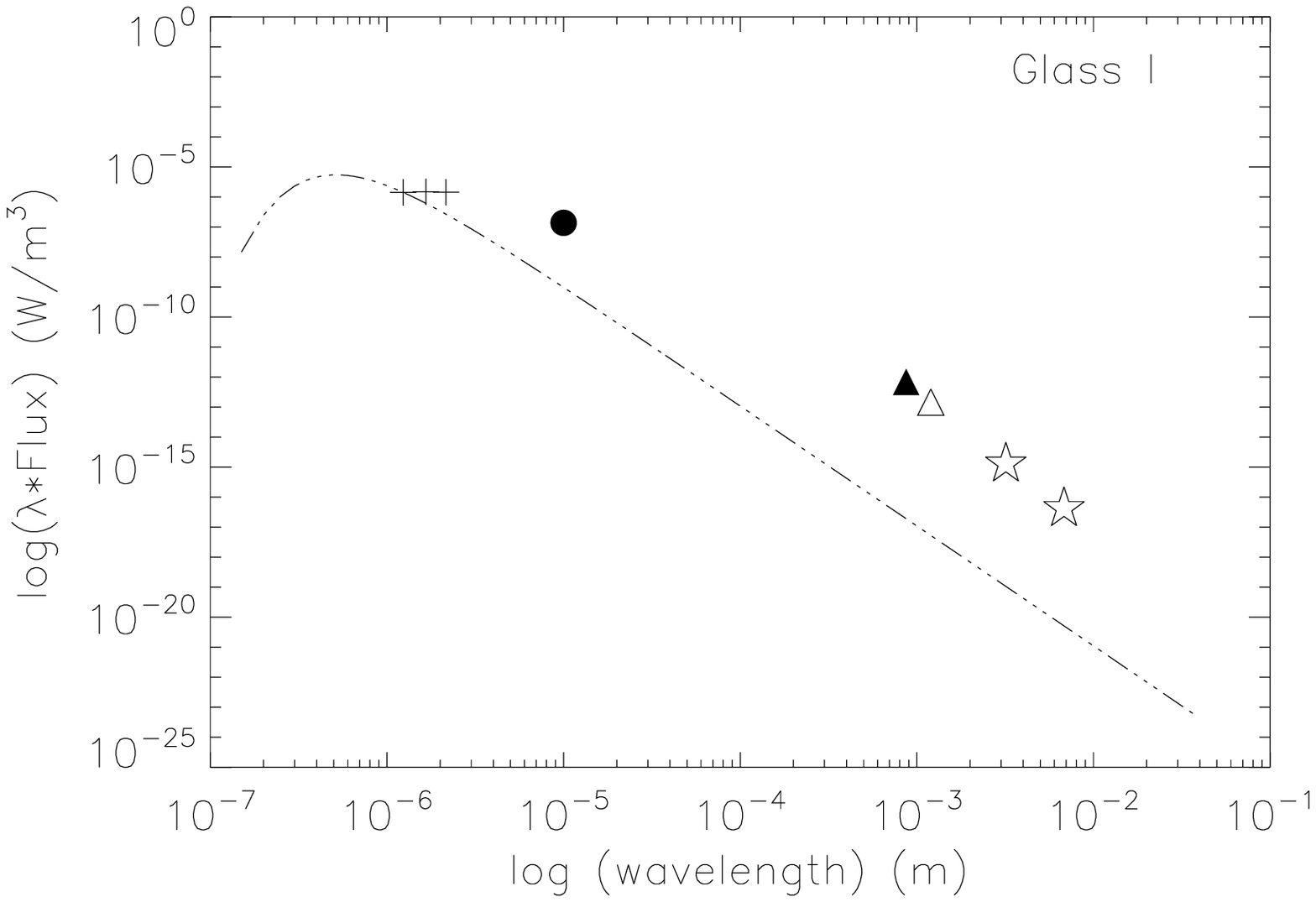}} \\
\subfloat{\label{fig:szcha-fullsed}\includegraphics[width=.68\columnwidth]{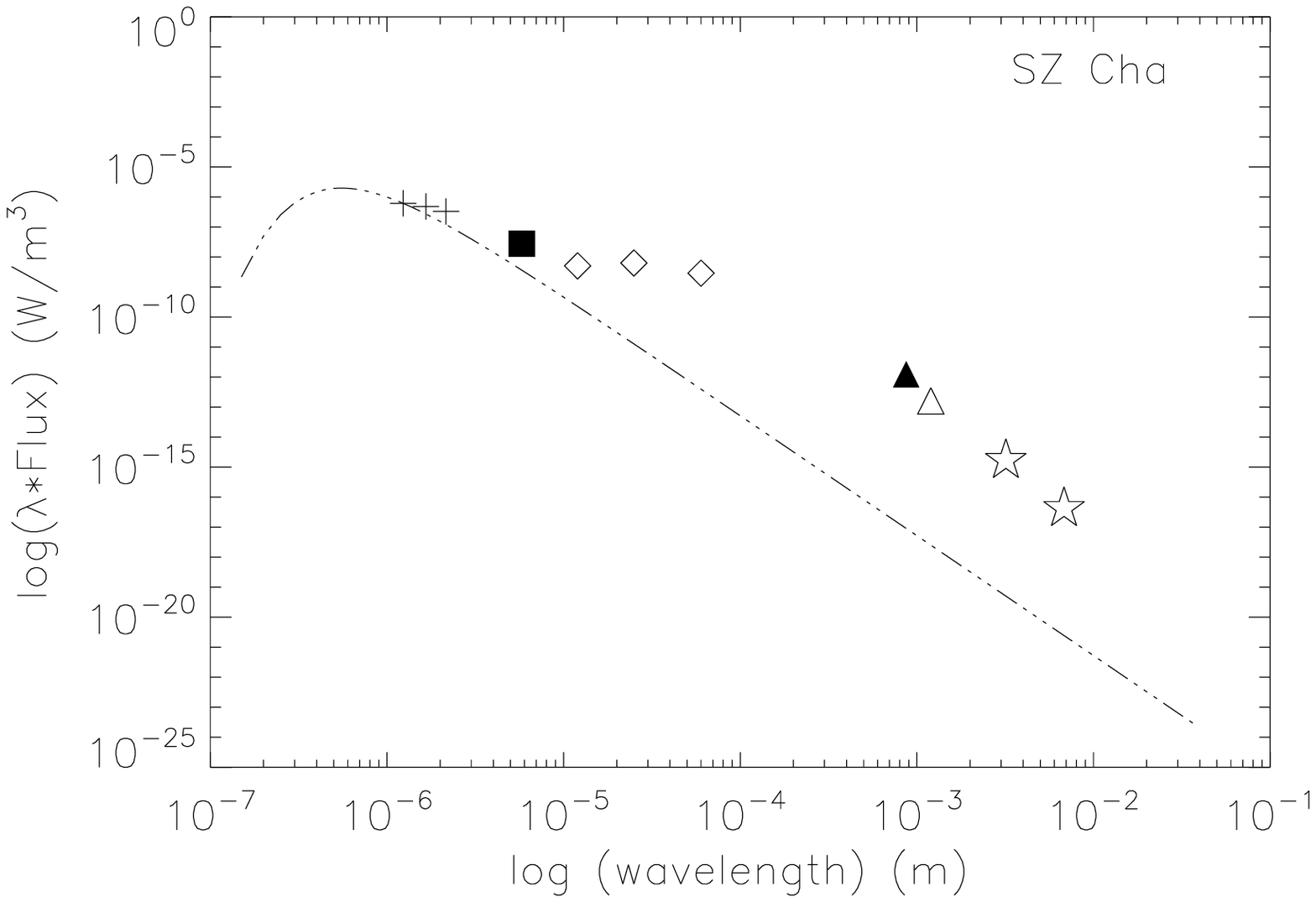}} 
\subfloat{\label{fig:sz32-fullsed}\includegraphics[width=.68\columnwidth]{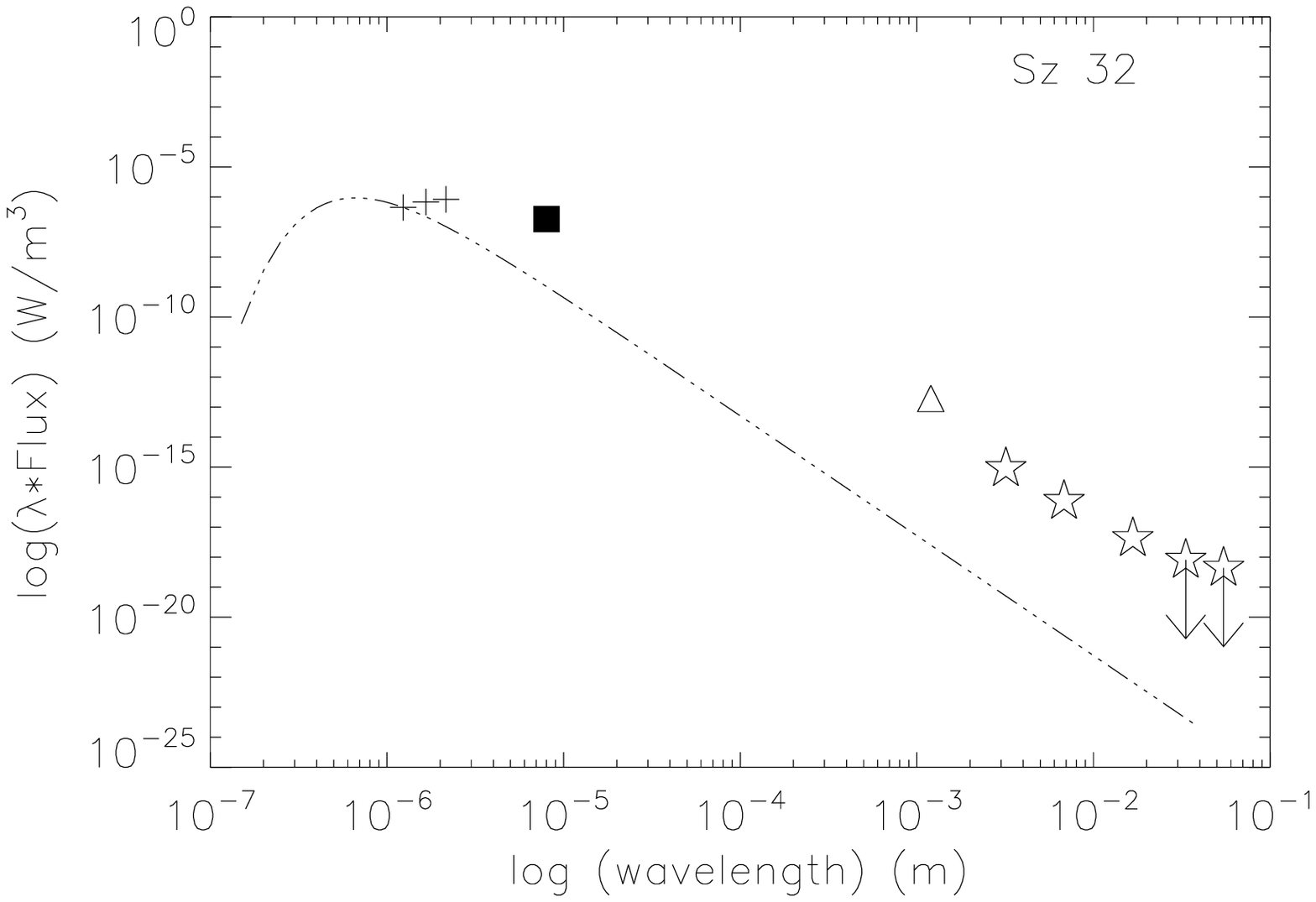}} 
\subfloat{\label{fig:dkcha-fullsed}\includegraphics[width=.68\columnwidth]{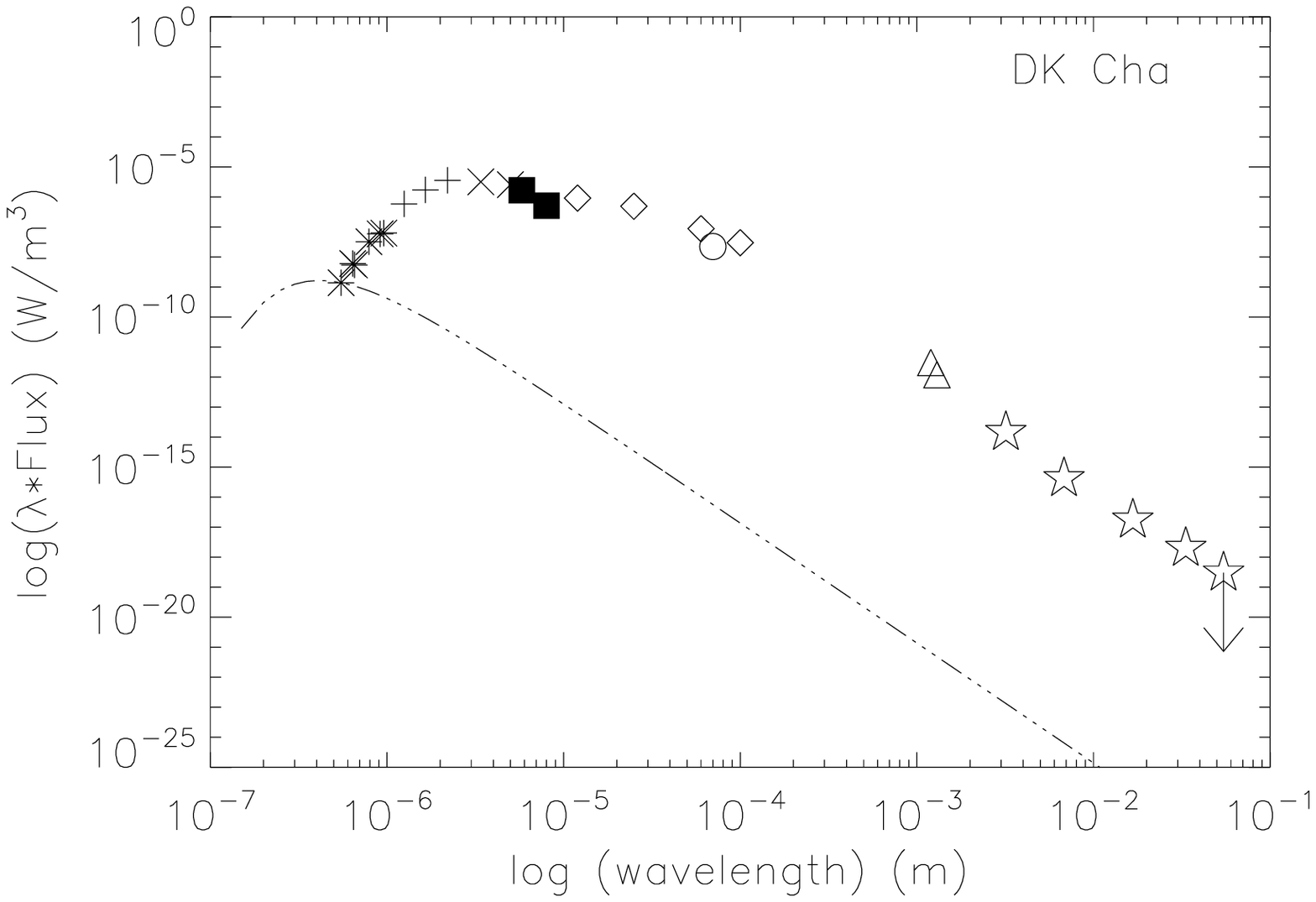}} \\
\caption{Full SEDs for Chamaeleon sources. Data obtained from the literature and this work. The stellar photosphere was obtained using the method of \citet{Furlan06} normalising sources of spectral type later than G by their J-band flux, and earlier spectral types by their V-band flux. The photometry data were obtained from published catalogs. Legend:
{* WFI (Wide Field Imager); $+$ 2MASS (Two Micron All Sky Survey); $\times$ CTIO (Cerro Tololo Inter-American Observatory); $\triangledown$ USNO (U.S. Navel Observatory);
$\blacksquare$ IRAC (Infrared Array Camera on Spitzer Space Telescope);
$\diamond$ IRAS (Infrared Astronomy Satellite); $\circ$ MIPS (Multiband Imaging Photometer for Spitzer); $\bullet$ ESO-TIMMI2 (Thermal Infrared MultiMode Instrument 2) on the ESO 3.6-m telescope at La Silla;
$\blacktriangle$ LABOCA (LArge BOlometer CAmera) on the 12-m APEX telescope, \citep{2011A&A...527A.145B};
$\triangle$ SEST (Swedish-ESO 15m Submillimeter Telescope), \citet{H1993}; $\square$ SMA (Submillimeter Array), \citet{Lommen10};
$\medstar$ ATCA (Australia Telescope Compact Array), \citet{Lommen07,Lommen10} and this work.}
~}
\label{fig-chamSED}
\end{figure*}

\begin{figure*}
  \centering
  \subfloat{\label{fig:iklupfull-sed}\includegraphics[width=.68\columnwidth]{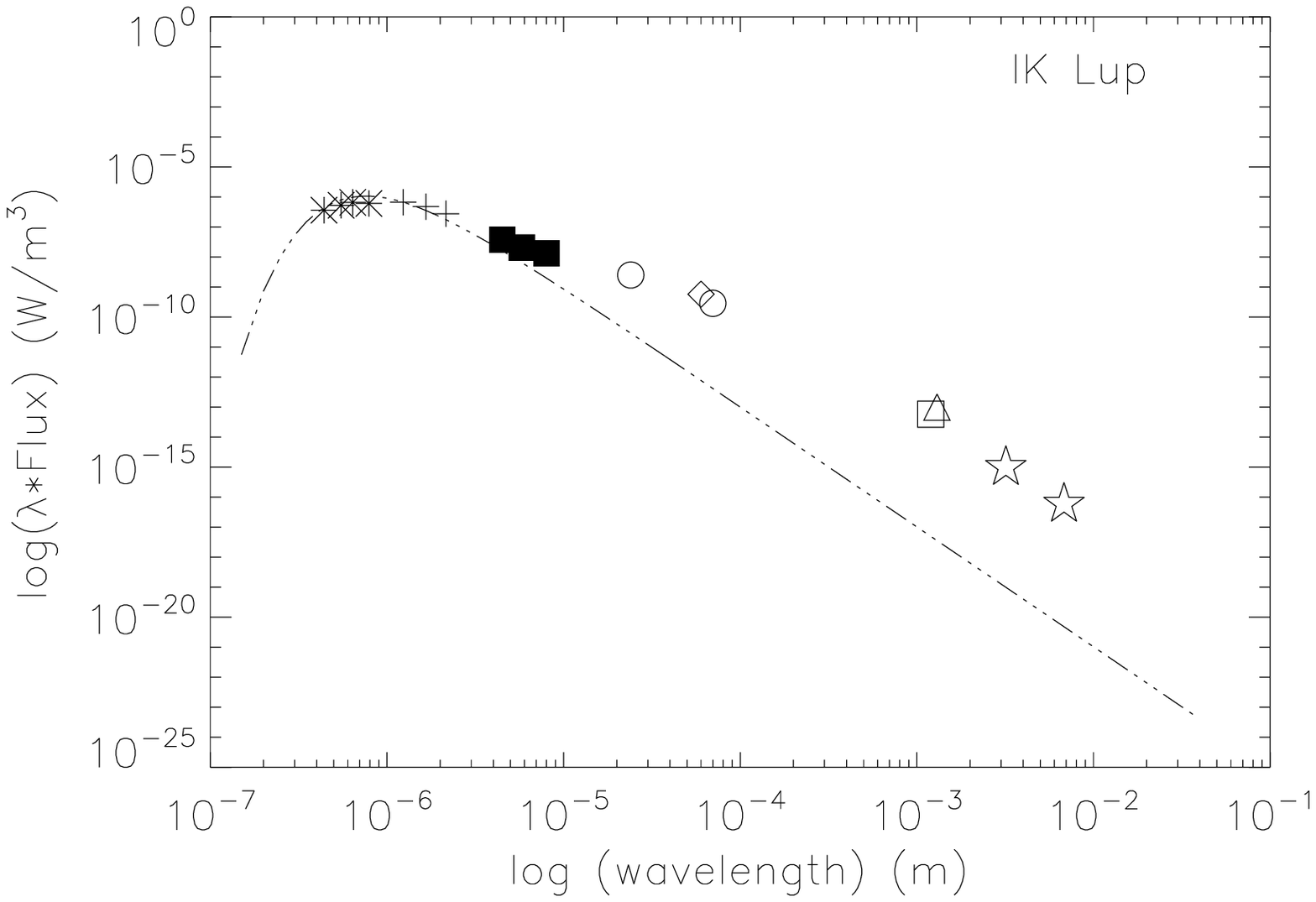}}    
  \subfloat{\label{fig:sz66full-sed}\includegraphics[width=.68\columnwidth]{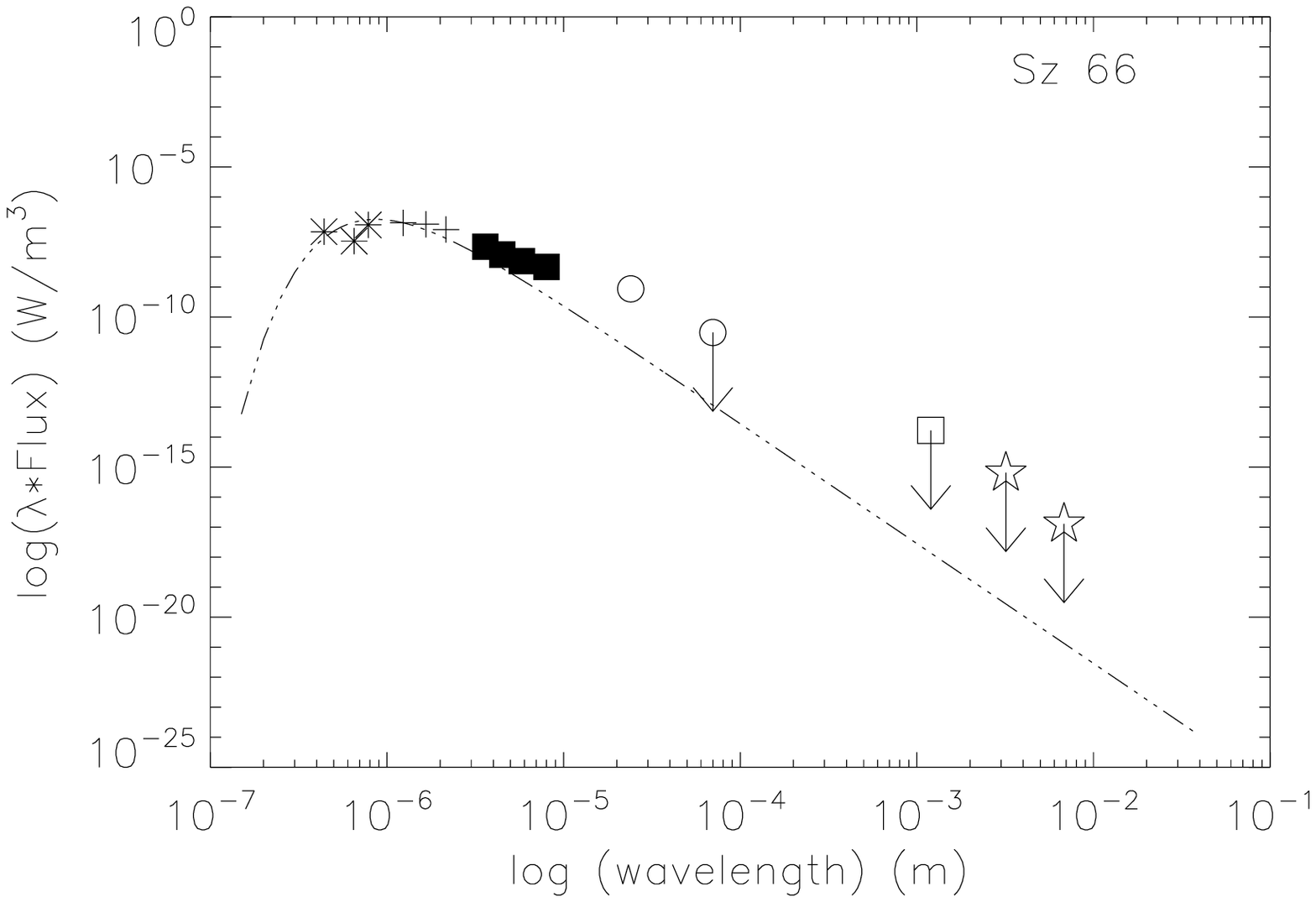}}  
   \subfloat{\label{fig:htlupfull-sed}\includegraphics[width=.68\columnwidth]{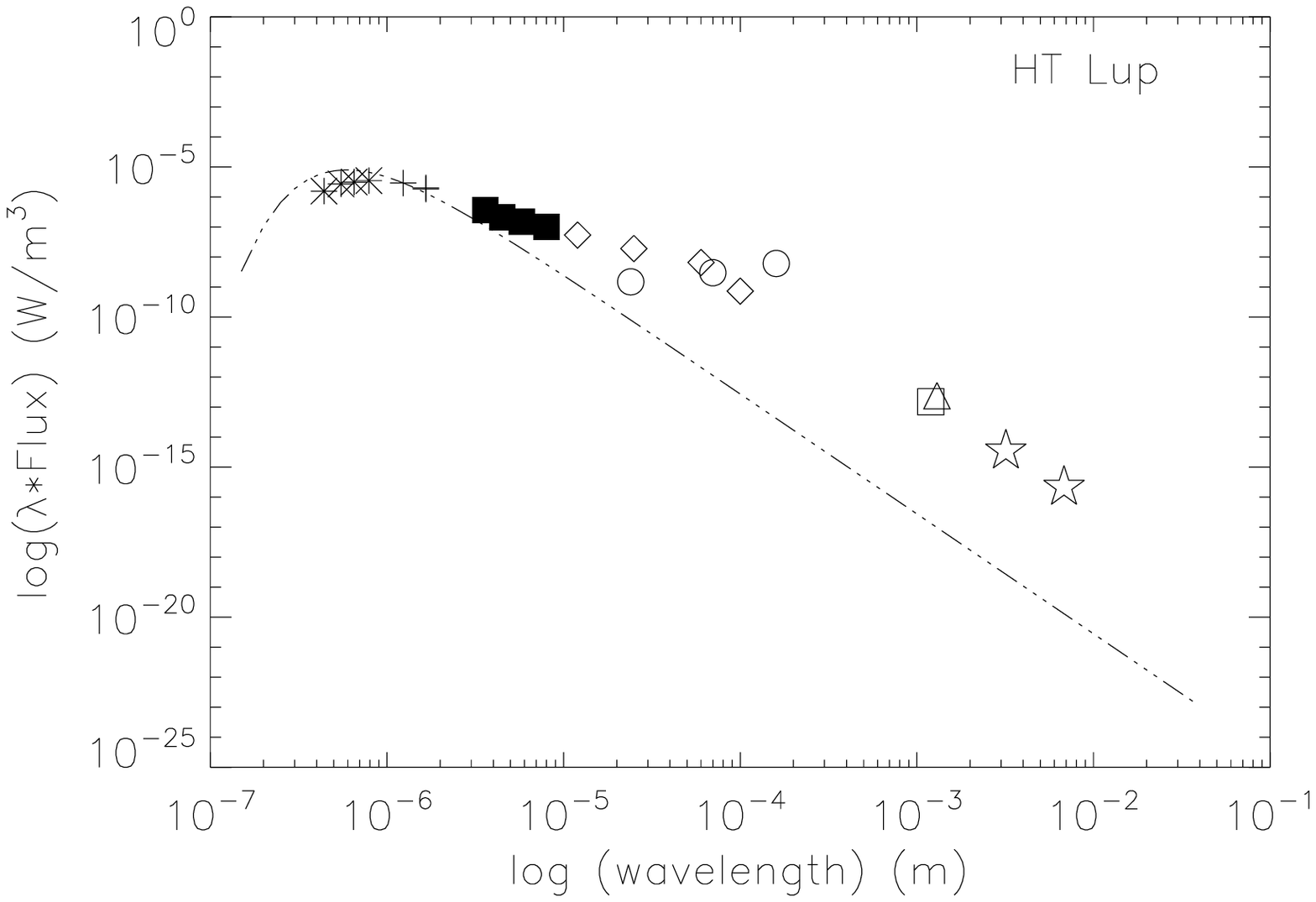}} \\
  \subfloat{\label{fig:gwlupfull-sed}\includegraphics[width=.68\columnwidth]{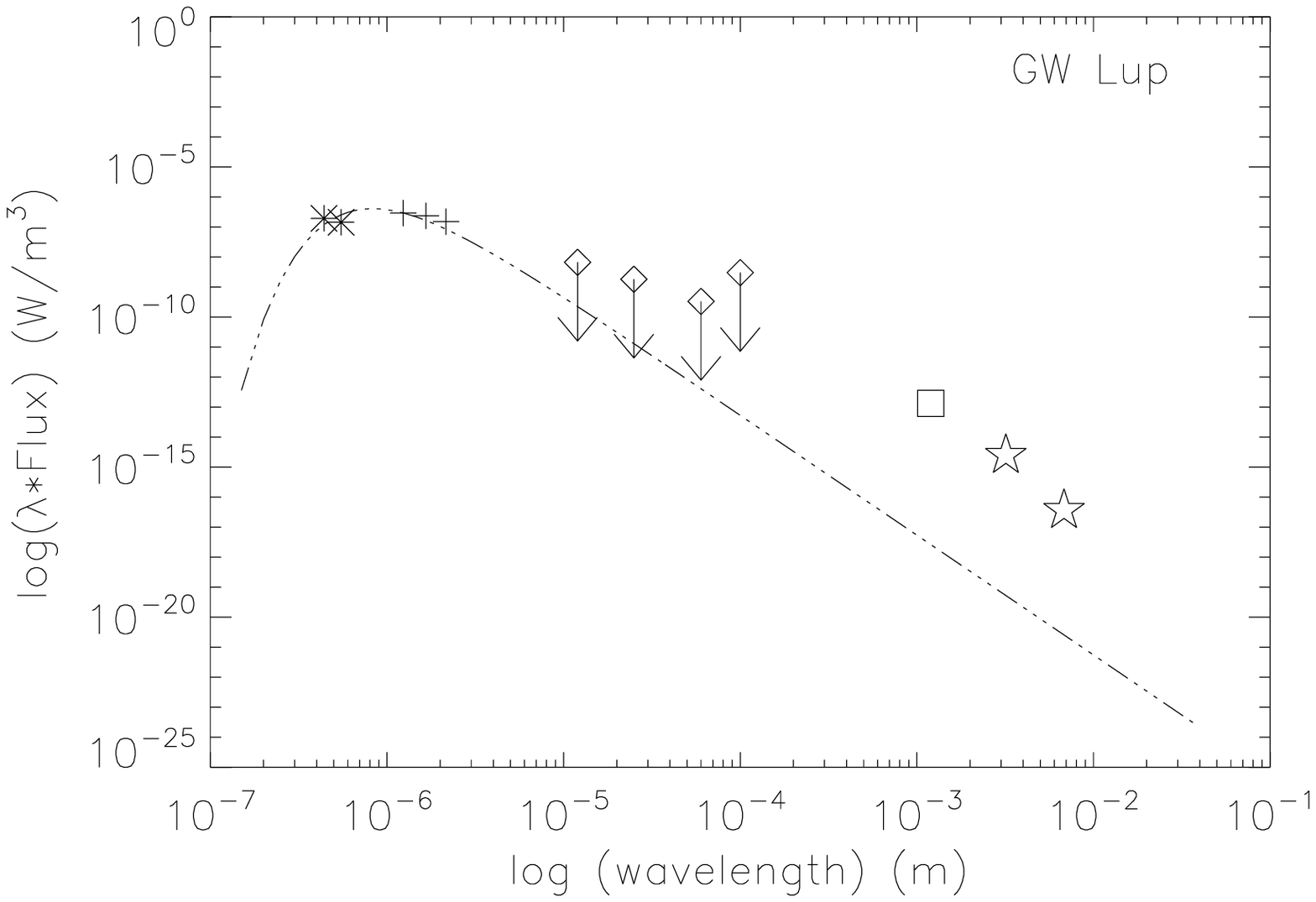}} 
    \subfloat{\label{fig:gqlupfull-sed}\includegraphics[width=.68\columnwidth]{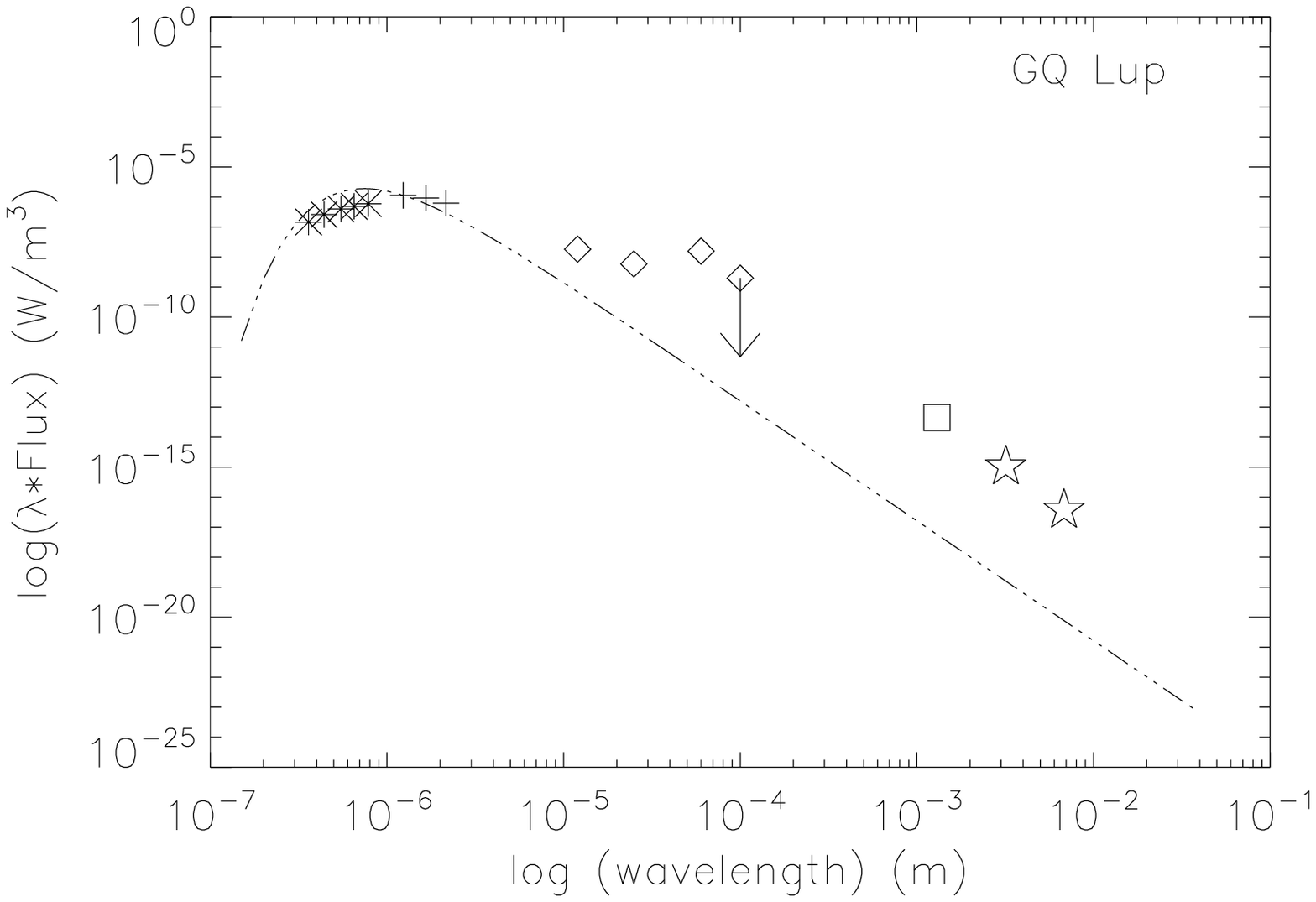}} 
  \subfloat{\label{fig:rylupfull-sed}\includegraphics[width=.68\columnwidth]{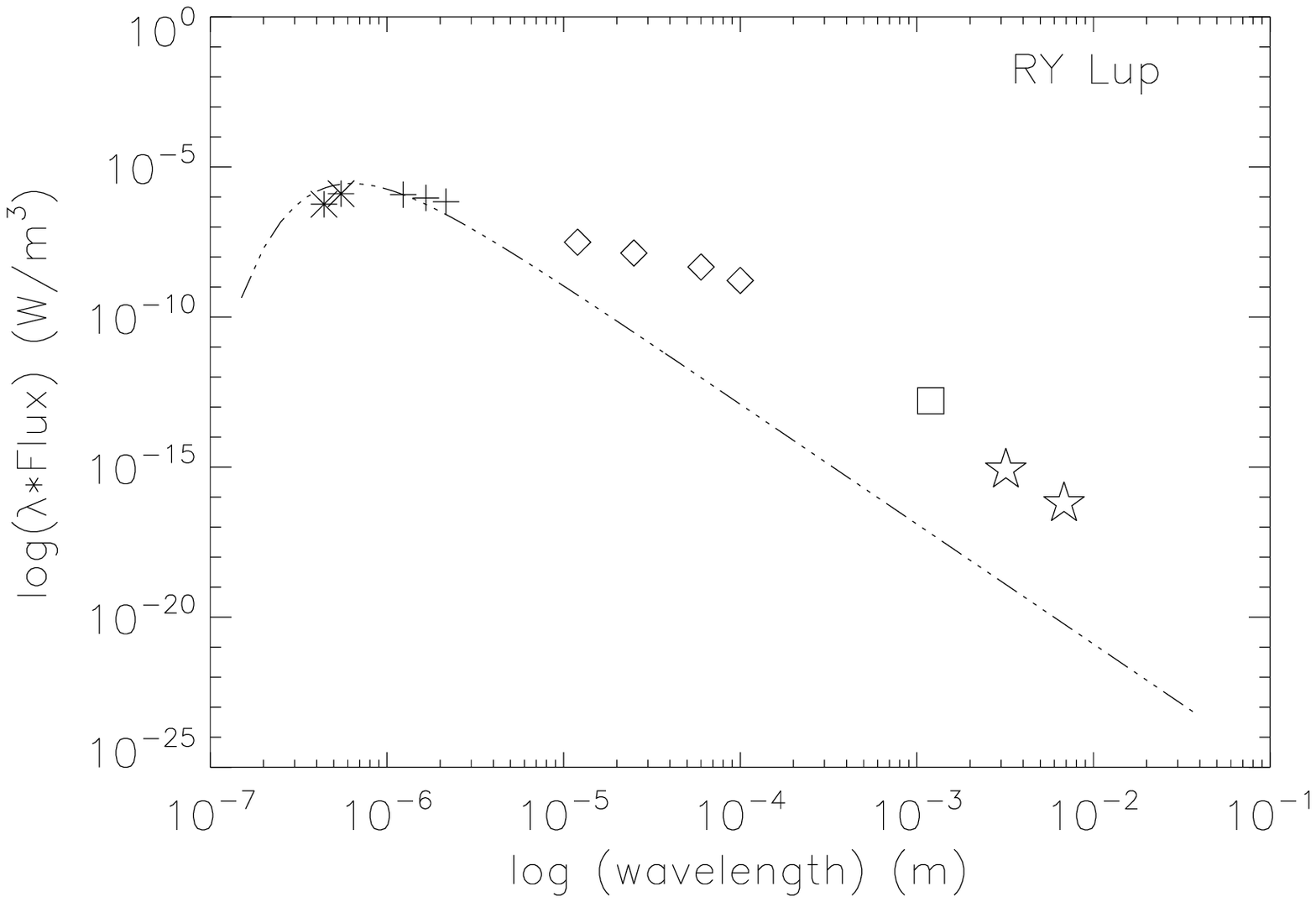}} \\
    \subfloat{\label{fig:hklup1full-sed}\includegraphics[width=.68\columnwidth]{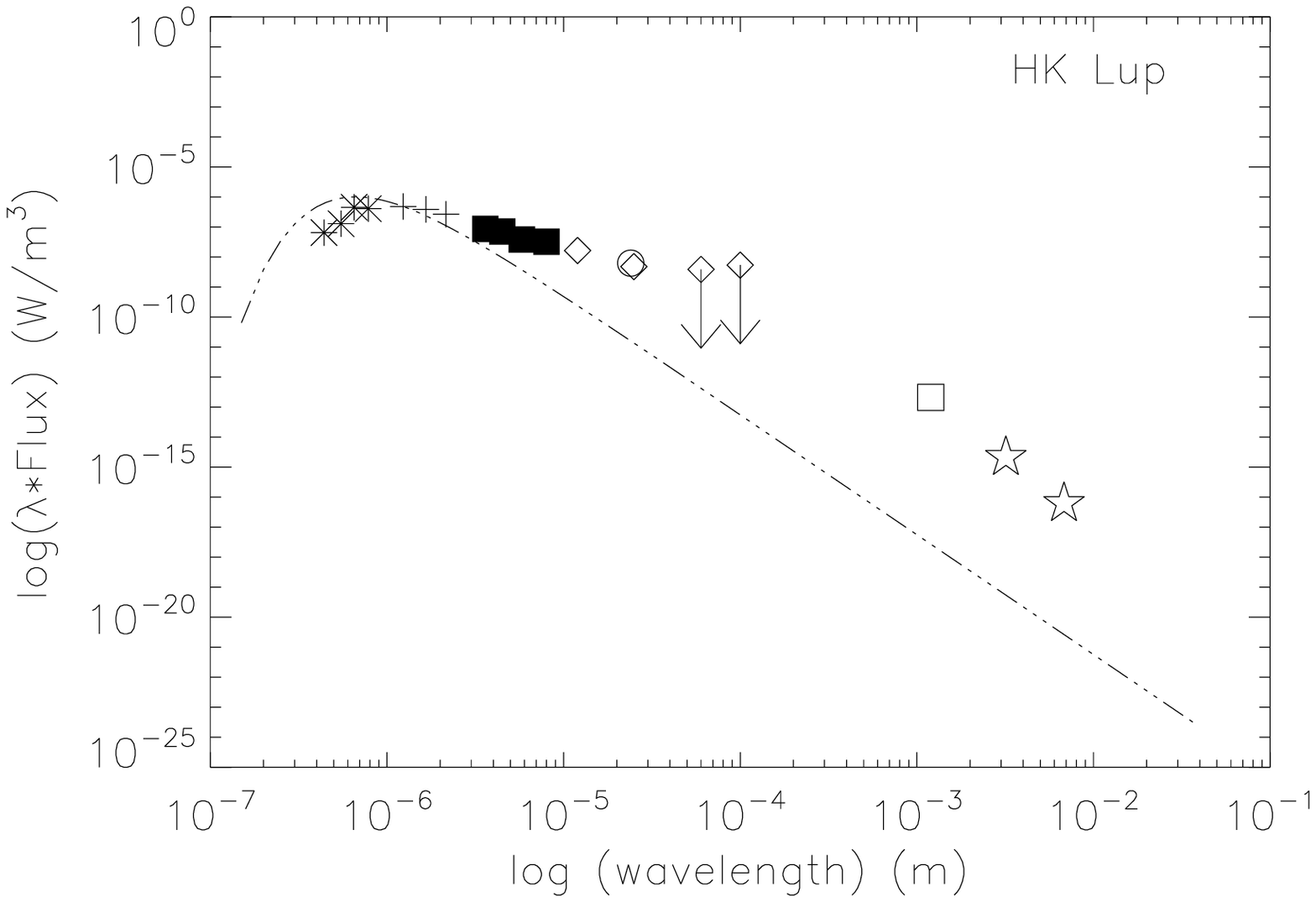}} 
  \subfloat{\label{fig:sz111full-sed}\includegraphics[width=.68\columnwidth]{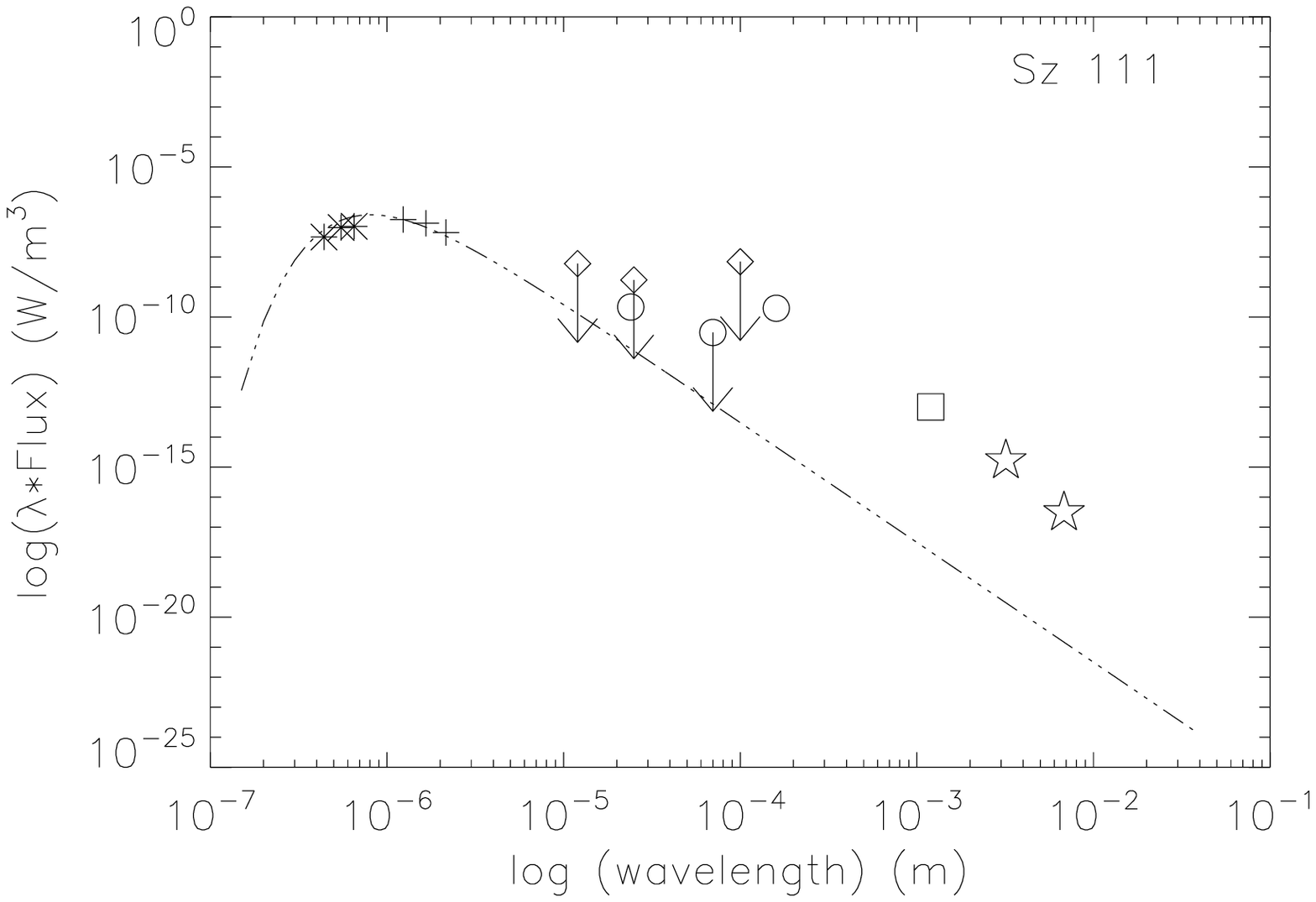}}
  \subfloat{\label{fig:exlupfull-sed}\includegraphics[width=.68\columnwidth]{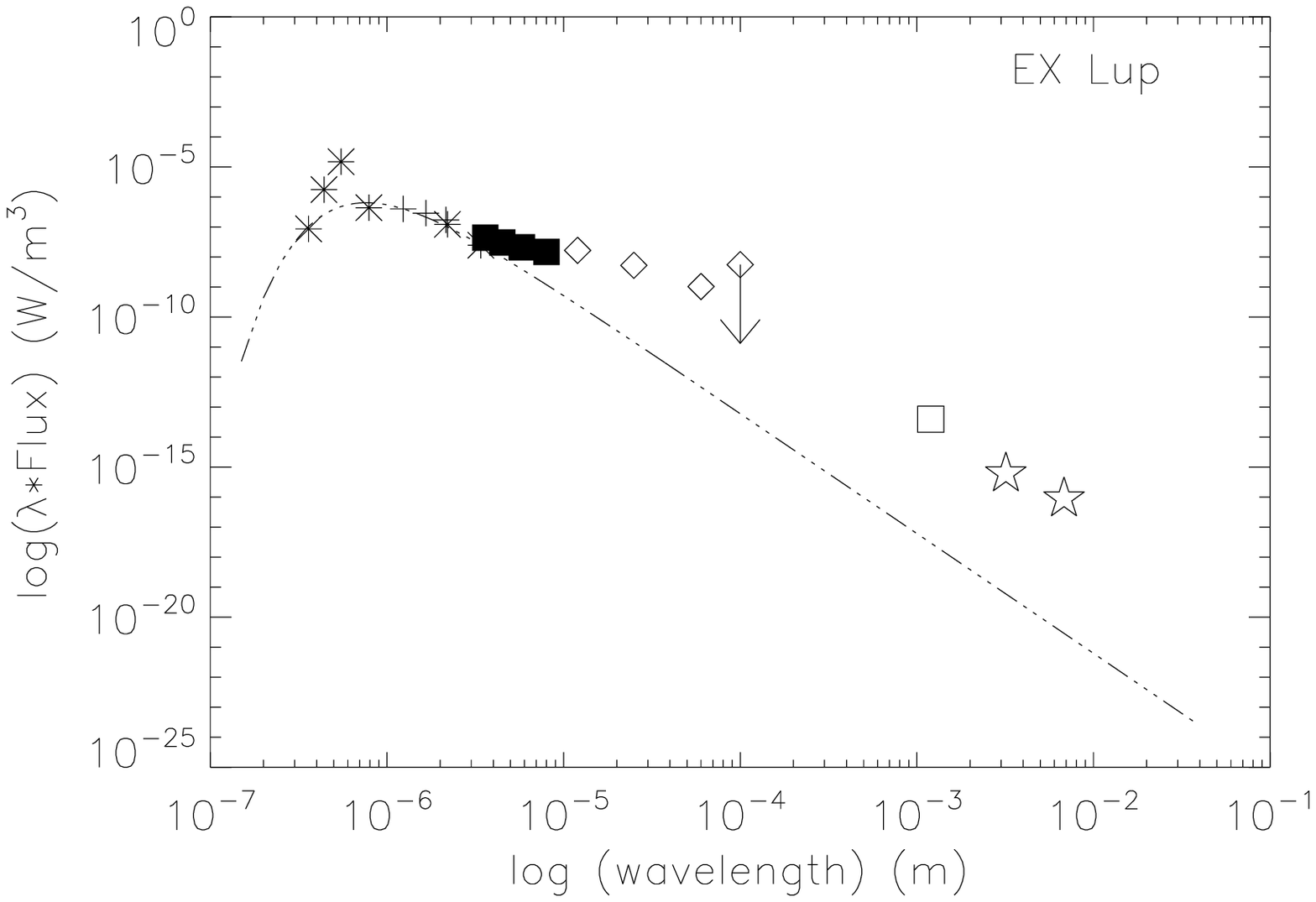}} \\
  \subfloat{\label{fig:mylupfull-sed}\includegraphics[width=.68\columnwidth]{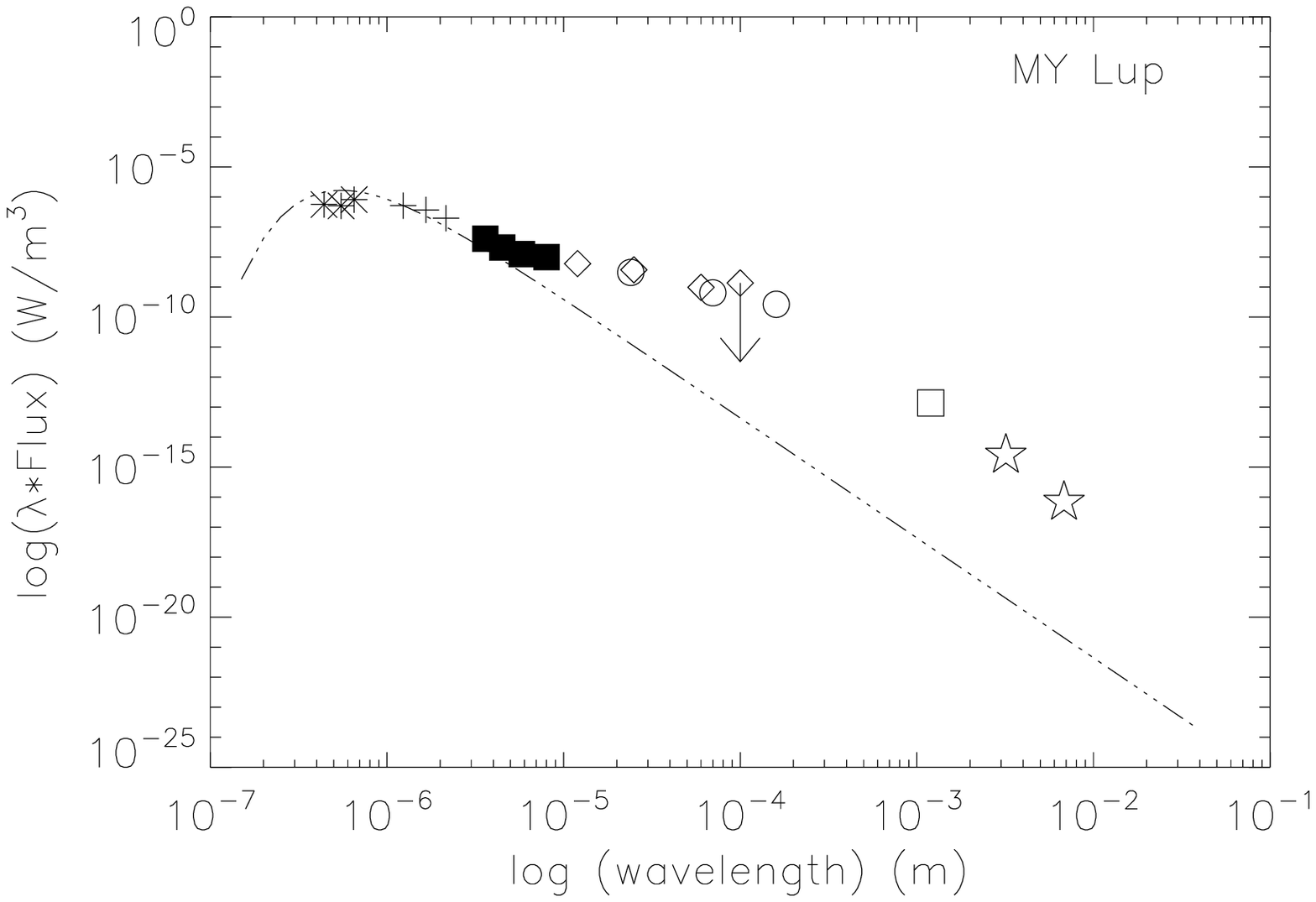}} 
 \subfloat{\label{fig:rxj1615full-sed}\includegraphics[width=.68\columnwidth]{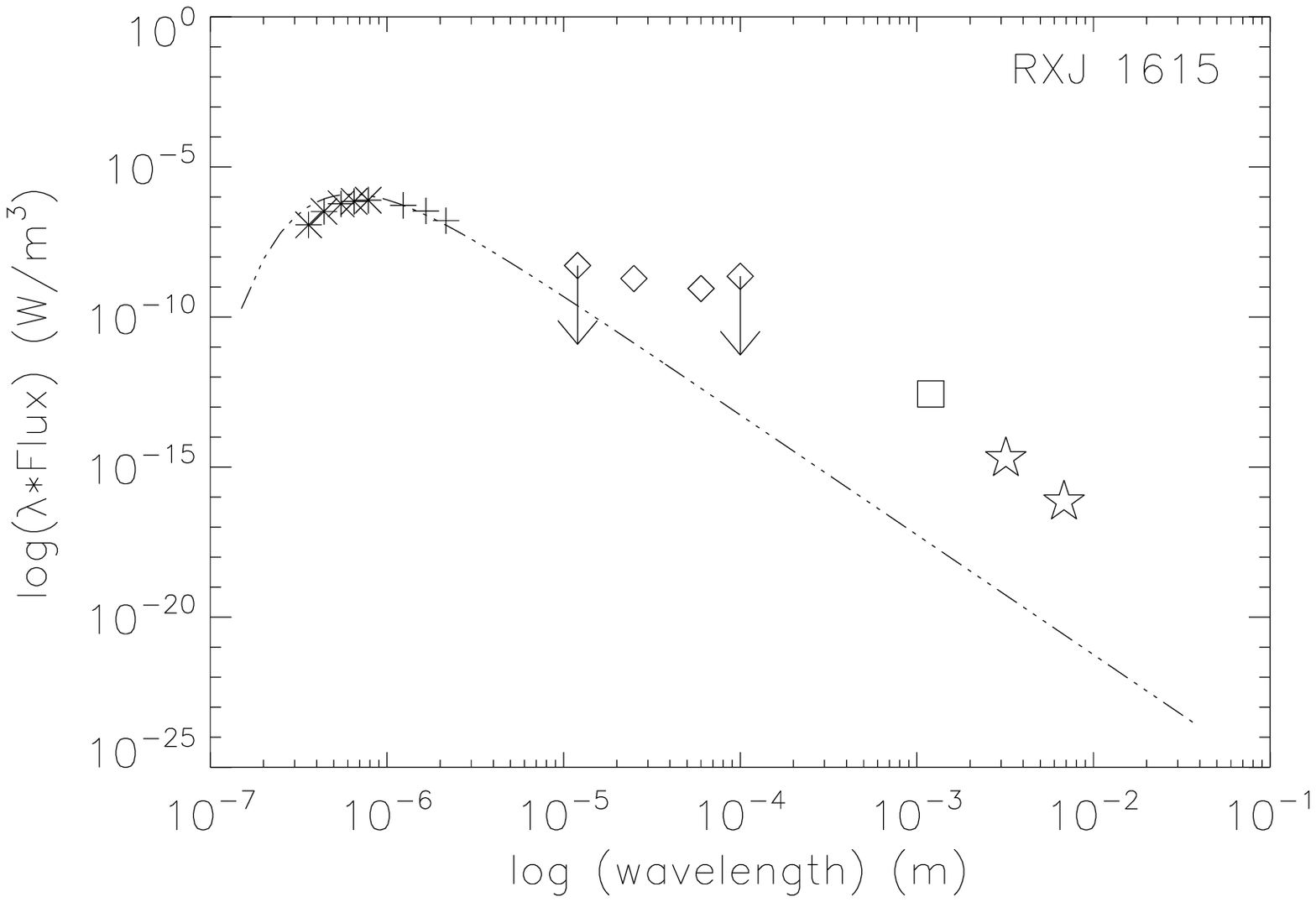}} 
\caption{Same as Fig.~\ref{fig-chamSED}, but for Lupus sources. Note the stellar photosphere of MY Lup was normalised to the J-band flux since its spectral type is unknown.}
  \label{fig-lupSED}
\end{figure*}

\section{Monitoring details}
\label{app-monitoringdetails}

{The complete results used for the monitoring analysis presented in Section~\ref{subsec-discussion-emission}. The 7~mm fluxes are given in Table~\ref{tab-monitoring2}, 15~mm in Table~\ref{tab-15mm} and 3+6~cm in Table~\ref{tab-cm}.}

\begin{table*}
\caption{Results of 7~mm flux monitoring of RY Lup, Sz 111, MY Lup, RXJ1615.3-3255, CS Cha and Sz 32 from this work and the literature. (1) Source name and dust opacity index $\beta$ (see Section~\ref{subsec-discussion-index}). (2) Date of observation in YYYYMMDD format.  (3) ATCA array configuration$^a$. (4) References: pre-CABB data from \citet{Lommen09} and \citet{Lommen10}, and CABB data from this work. (5), (6) continuum fluxes from point source fits obtained at approximately 6.7 and 7.0~mm respectively with the RMS given in parenthesis in units of mJy/beam.}
\begin{tabular}{cccccc}
	 \hline \hline
Source	&	Date	&	Array	&	Reference	&	F(6.7~mm)$^b$ 	&	F(7~mm)$^b$ 	\\
	&		&		&		&	(mJy) 	&	(mJy)	\\		
\hline									
RY Lup	&	20080801	&	H214	&	\citet{Lommen10}	& ---	& $<0.6$(0.2)			\\
$\beta = 1. 9$		& 20090529	&	H214	&	This work &	$<1.2$(0.4)	&	$<0.9$(0.3)	\\
		& 20090530	&	H214	&	This work	&	$1.1\pm0.1$(0.1)	&	$0.8\pm0.1$(0.1)	\\
		\hline									
MY Lup	&	20080805	&	H214	&	\citet{Lommen10}	&	1.3(0.1)	&	---	\\
$\beta = 0.3$	&	20090529	&	H214	&	This work	&	$<1.2$(0.4)	&	$<1.2$(0.4)	\\
	&	20090530	&	H214	&	This work.	&	$1.0\pm0.1$(0.1)	&	$1.0\pm0.1$(0.1)	\\
			\hline								
RXJ1615.3-3255	&	20080801	&	H214	&	\citet{Lommen10}	&	$<0.5$(0.2)	&	---	\\
$\beta = 1.6$	&	20100821	&	H168	&	This work 	&	${{0.8\pm0.2(0.2)}}$	&	${{0.8\pm0.2(0.2)}}$	\\
				\hline							
Sz 111	&	20080801	&	H214	&	\citet{Lommen10}	&	$<0.6$(0.2)	&	---	\\
$\beta = 0.5$	&	20090529	&	H214	&	This work 	&	$<0.9$(0.3)	&	$<0.9$(0.3)	\\
	&	20090530	&	H214	&	This work 	&	$0.3\pm0.1$(0.1)	&	$0.3\pm0.1$(0.1)	\\
\hline
CS Cha	&	20080426	&	6A	&	\citet{Lommen09}		& ---	&	1.0$\pm$0.3(0.1)		\\
$\beta = 0.9$	&	20080705	&	1.5B	&	\citet{Lommen09}	&	$<0.8$(0.3)	&	$<0.7$(0.2)			\\
	&	20080706	&	1.5B	&	\citet{Lommen09}	&	$<1.1$(0.4)	&	1.4$\pm$0.3(0.2)			\\
	&	20090528	&	H214	&	This work 	&	1.7$\pm$0.4(0.3)	&	1.2$\pm$0.3(0.3)			\\
	&	20090529	&	H214	&	This work 	&	2.0$\pm$0.3(0.3)	&	1.9$\pm$0.2(0.2)			\\
\hline
Sz 32 	&	20080331$^c$ 	&	H168	&	\citet{Lommen09}	&	---	&	0.8$\pm$0.1(0.2)			\\
$\beta = 1.8$	&	20090528	&	H214	&	This work &	$<0.5$(0.2)	&	$<0.5$(0.16)			\\
	&	20090529	&	H214	&	This work 	&	1.5$\pm$0.5(0.3)	&	1.0$\pm$0.4(0.2)		\\
\hline
\end{tabular}
	    	\begin{tablenotes}
		\item[1]  $^a$ ATCA array configuration: http://www.narrabri.atnf.csiro.au/operations/array\_configurations/upcoming\_configs.html
		 \item[2] $^b$ A 3$\sigma$ upper limit is given for non-detections.
		 \item[3] $^c$ Sz 32 was detected in the field of view of WW Cha observations.
		\end{tablenotes}
	 \label{tab-monitoring2}
 \end{table*}

\begin{table}
\caption{Results of the 15~mm flux monitoring of 6 Chamaeleon sources, listing the highest and lowest fluxes obtained from a point source fit in the u-v plane. A 3$\sigma$ value is given for non-detections. RMS = 0.1 mJy/beam.}  
\centering
 \begin{tabular}{cc|cc|cc}
	\hline
	\hline 										
		  &  &\multicolumn{2}{c}{17~GHz}		& \multicolumn{2}{c}{19~GHz}			\\					
	Sources	&	Time &F$_{\rm high}$ 	&	F$_{\rm low}$	&	F$_{\rm high}$	&	F$_{\rm low}$ 	\\
		&	(minutes) & (mJy) 	&	(mJy) &	(mJy) &	(mJy) 	\\
	\hline
	CR Cha	& 3 &	$0.6\pm0.1$ 	&	$<0.3$ 	&	$0.6\pm0.2$	&	$<0.2$ 	\\
	CS Cha	& 5 &	$0.5\pm0.1$	&	$<0.3$  &	$0.4\pm0.1$	&	$<0.2$ 	\\
	DI Cha	& 5 &	$<0.3$ & 	$<0.3$ 	&	$0.3\pm0.1$	&	$<0.3$ 	\\
	T Cha	& 3 &	$0.9\pm0.2$ 	&	$<0.3$ 	&	$0.5\pm0.2$	&	$<0.3$ 	\\
	Sz 32	& 5 &	$0.8\pm0.1$	&	$<0.2$ 	&	$0.7\pm0.1$ 	&	$<0.2$ \\
	DK Cha	& 3 &	$2.0\pm0.2$	&	$1.5\pm0.2$ 	&	$1.8\pm0.2$	&	$1.3\pm0.2$  \\
	\hline
\end{tabular}
\label{tab-15mm}
\end{table}

\begin{table}
\caption{Results of the 3+6~cm flux monitoring of 3 Chamaeleon sources, listing the highest and lowest fluxes obtained from a point source fit in the u-v plane. A 3$\sigma$ value is given for non-detections. RMS$_{5.5\rm{GHz}} =$ 0.1 mJy/beam and RMS$_{9\rm{GHz}} \leq 0.2$~mJy/beam. All sources had a scan length of 15~minutes.}  
\centering
 \begin{tabular}{c|cc|cc}
	\hline 
	\hline 										
			    &\multicolumn{2}{c}{5.5~GHz}		& \multicolumn{2}{c}{9~GHz}			\\	
	Sources	&	F$_{\rm high}$ 	&	F$_{\rm low}$&	F$_{\rm high}$	&	F$_{\rm low}$	\\
		&	(mJy) & 	(mJy)  & 	(mJy) & 	(mJy) 	\\
	\hline
	T Cha	&	$0.9\pm0.1$ 	&	$<0.3$	&	$0.5\pm0.1$	&	$<0.5$	\\
	Sz 32	&	$0.8\pm0.1$ 	&	$<0.3$	&	$0.7\pm0.1$	&	$<0.5$	\\
	DK Cha	&	$0.7\pm0.1$	&	$<0.3$	&	$1.1\pm0.1$	&	$<0.4$	\\
	\hline
\end{tabular}
\label{tab-cm}
\end{table}

\section{Strength and Shape}
\label{app-SS}

The third-order polynomial fits to the infrared continuum obtained from the Heritage Infrared Archive of each source are presented. Spitzer data has been published in previous works \citep{Furlan06,2006ApJ...639..275K,Lommen10}.

\begin{figure*}
  \subfloat{\includegraphics[width=1.5\columnwidth]{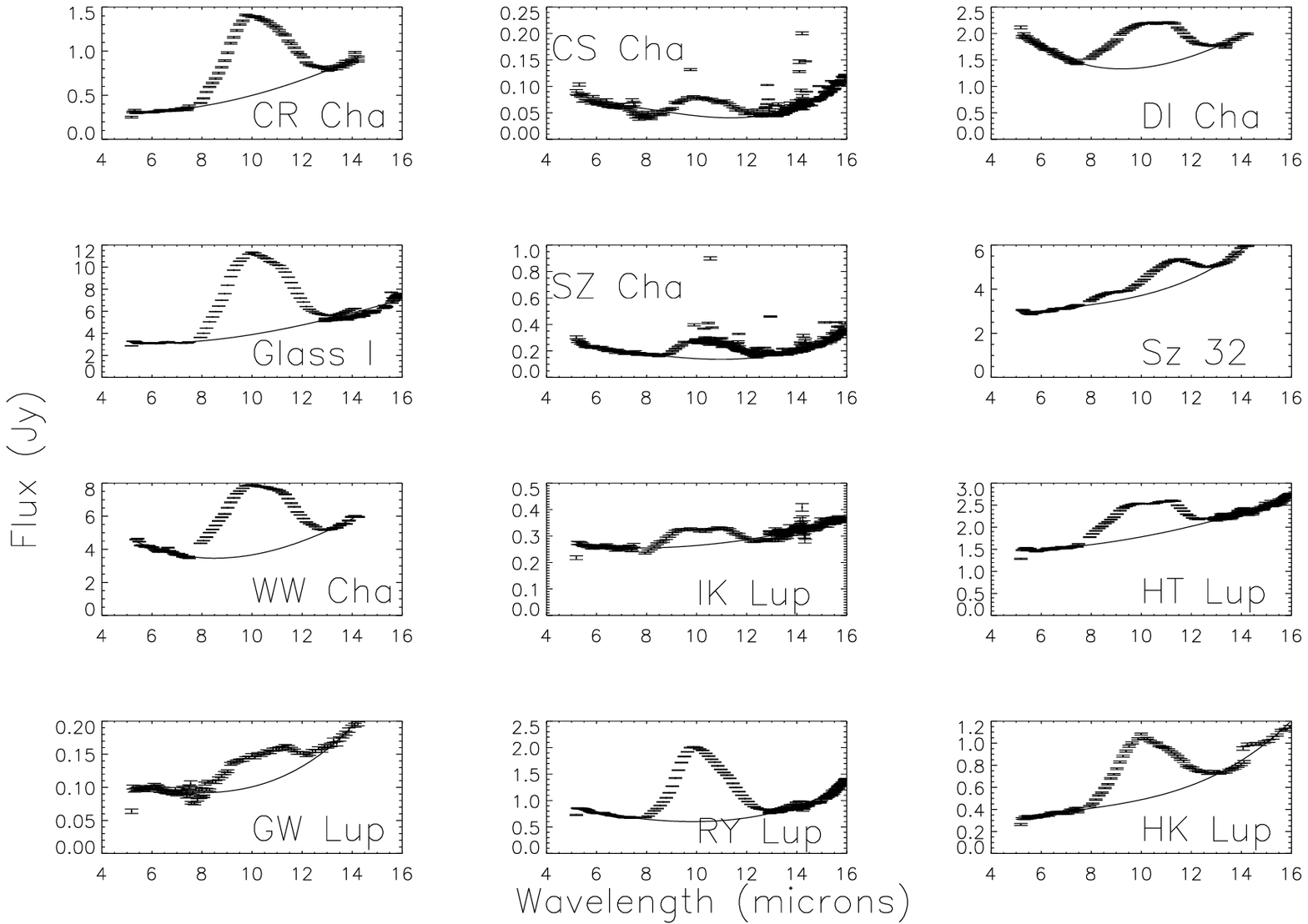}}\\
  \subfloat{\includegraphics[width=1.5\columnwidth]{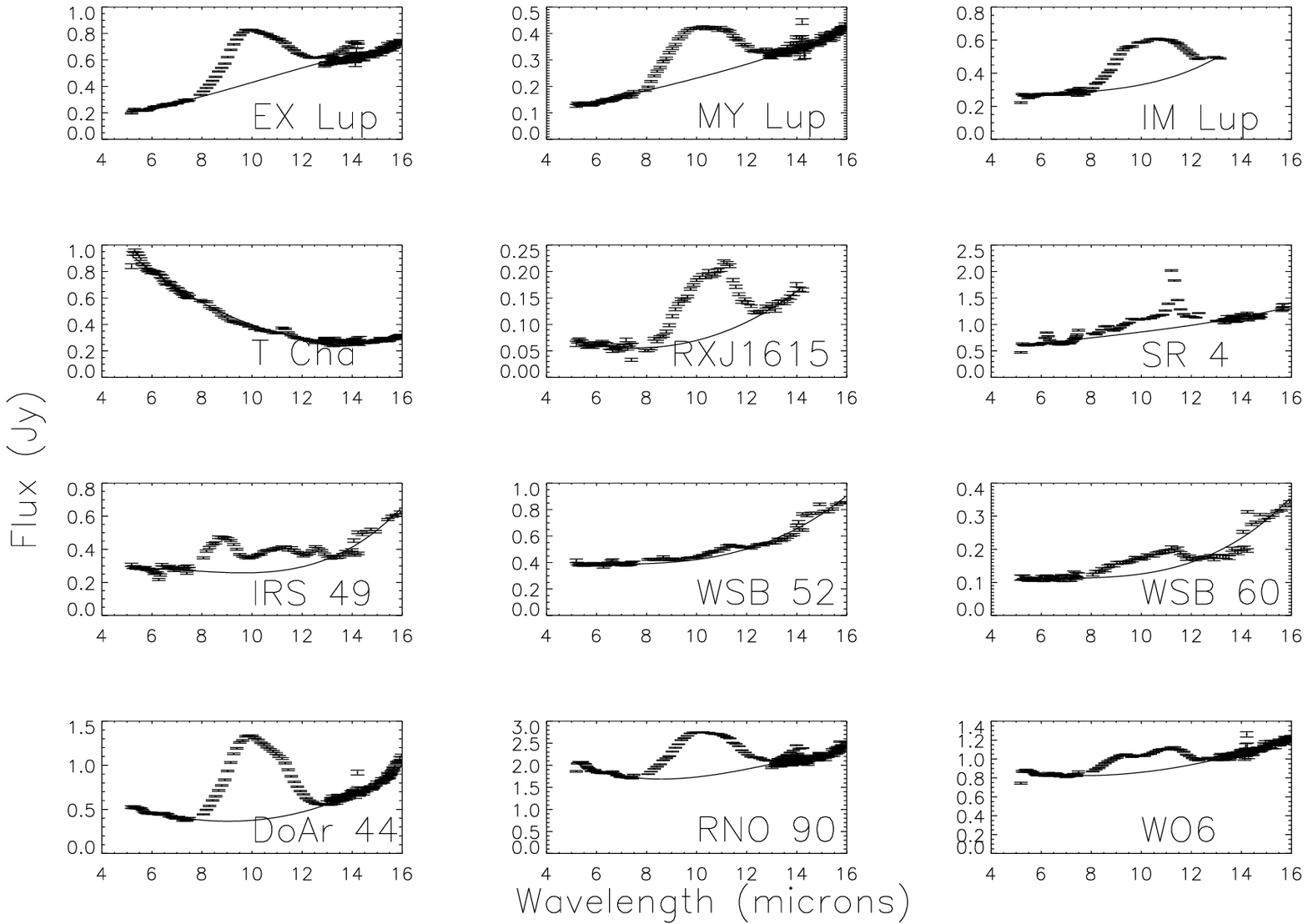}}
   \caption{{The third-order polynomial fit to the infrared continuum used to obtain the strength of the 10$\mu$m silicate feature. The Spitzer infrared data were obtained from the Heritage Archive.}}
 \end{figure*}

\begin{figure*}
  \ContinuedFloat  
  \subfloat{\includegraphics[width=1.5\columnwidth]{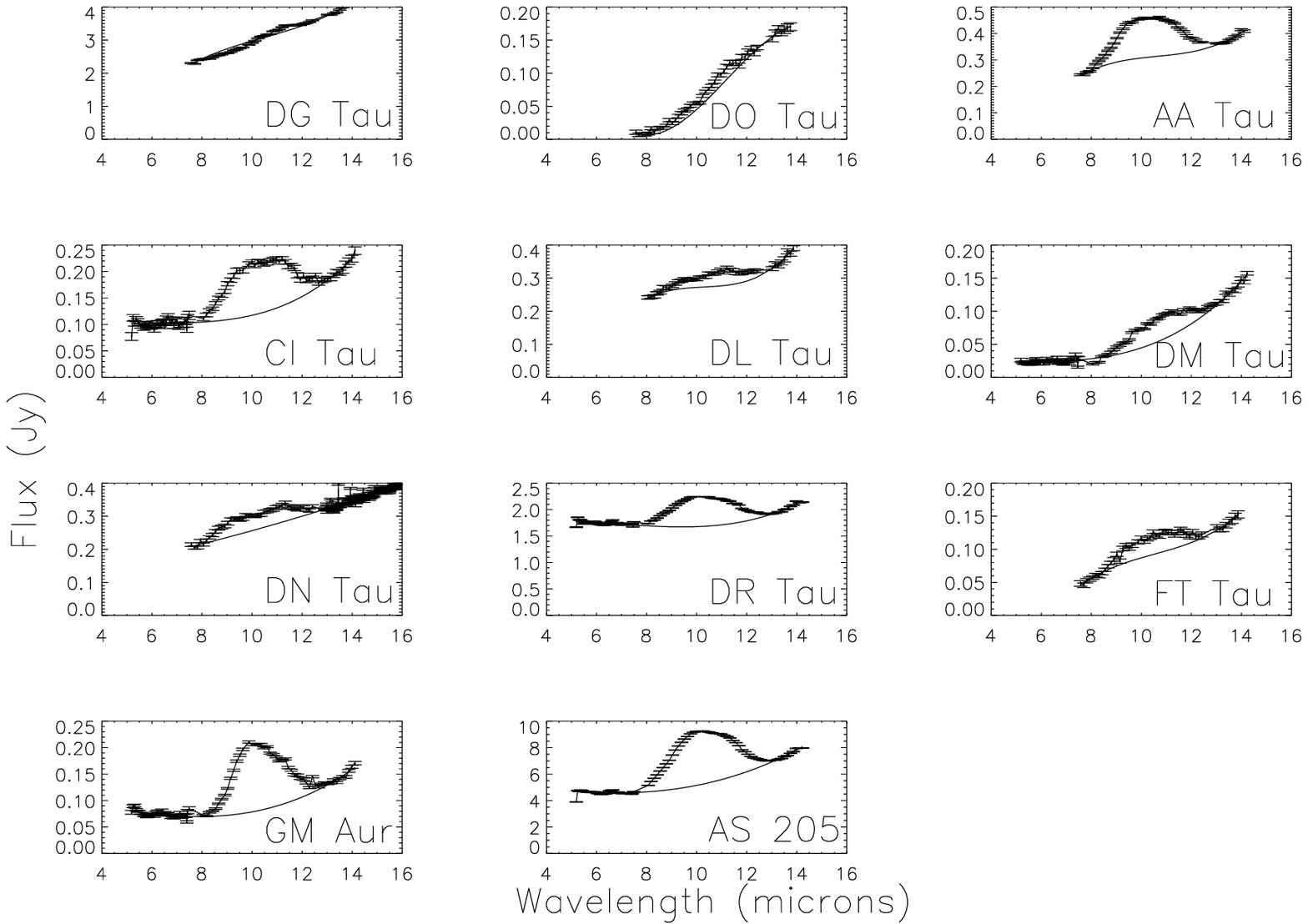}}
   \caption{{Continued.}}
   \label{fig-fits}
 \end{figure*}

\end{document}